\documentclass[11pt]{article}
\usepackage[margin=1in]{geometry}
\usepackage{graphicx} 
\usepackage{setspace}

\usepackage[colorlinks=True,linkcolor=blue,citecolor=magenta,urlcolor=black,pagebackref=true,backref=true]
{hyperref}
\usepackage{amsmath,amsfonts}
\usepackage[mathscr]{euscript}
\usepackage{enumerate,enumitem}
\usepackage{float,multirow}
\usepackage{placeins}
\usepackage{mathtools}
\renewcommand*{\backrefalt}[4]{%
    \ifcase #1 \footnotesize{(Not cited.)}%
    \or        \footnotesize{(Cited on page~#2.)}%
    \else      \footnotesize{(Cited on pages~#2.)}%
    \fi}
\usepackage[table]{xcolor}
\usepackage[export]{adjustbox}

\usepackage{multirow}
\usepackage{booktabs,tabularx}
\usepackage{caption}
\usepackage{subcaption}
\usepackage{tabularx}
\usepackage[colorinlistoftodos]{todonotes}

\usepackage[capitalize]{cleveref}  
\usepackage{datetime}
\crefname{nlem}{Lemma}{Lemmas}
\crefname{nprop}{Proposition}{Propositions}
\crefname{ncor}{Corollary}{Corollaries}
\crefname{nthm}{Theorem}{Theorems}
\crefname{exa}{Example}{Examples}
\crefname{assumption}{Assumption}{Assumptions}
\crefname{equation}{equation}{equations}
\crefname{nclm}{Claim}{Claims}

\newcommand{\braces}[1]{\left\{ #1 \right\}}
\DeclarePairedDelimiter{\paren}{(}{)}
\newcommand{\parenth}[1]{\left( #1 \right)}
\newcommand{\E}{\mathbb{E}}
\newcommand{\brackets}[1]{\left[ #1 \right]}
\newcommand{\Var}{\textnormal{Var}}

\newcommand{\trm}[1]{\textrm{#1}}
\newcommand{\qtext}[1]{\quad\text{#1}\quad}

\newcommand{\wtilde}[1]{\widetilde{#1}}
\newcommand{\what}[1]{\widehat{#1}}

\newcommand{\mc}[1]{\mathcal{#1}}
\newcommand{\mb}[1]{\mathbb{#1}}
\newcommand{\mbf}[1]{\mathbf{#1}}
\newcommand{\mf}[1]{\mathfrak{#1}}

\newcommand{\fspace}{\mc {X}}
\newcommand{\group}{\mbf G}
\newcommand{\ate}{\tau_{\trm{ATE}}}
\newcommand{\cate}[1][x]{\tau(#1)}
\newcommand{\gate}[1][\group]{\tau_{#1}}

\newcommand{\atehat}{\what{\tau}_{\trm{ATE}}}

\newcommand{\gatehat}[1][\group]{\what{\tau}_{#1}}

\newcommand{\defn}{\coloneqq}

\newcommand{\real}{\mb R}

\newcommand{\zscore}{{\mb T}}

\newcommand{\varhat}{\what{\Var}}
\newcommand{\treat}{\mbf T}
\newcommand{\control}{\mbf C}
\newcommand{\abs}[1]{\left \vert #1 \right \vert}

\newcommand{\qvalue}{\mf{q}}
\newcommand{\tset}{\mbf S}
\newcommand{\trainset}{\tset_{\trm{TRAIN}}}
\newcommand{\testset}{\tset_{\trm{TEST}}}
\newcommand{\valset}{\tset_{\trm{VAL}}}

\newcommand{\fold}{\mf{f}}
\newcommand{\trainfold}[1][\fold]{\tset_{#1}}
\newcommand{\tfold}{\trm{TF}}
\newcommand{\vfold}{\trm{VF}}
\newcommand{\trainfolds}{\tset_{\tfold}}
\newcommand{\valfold}{\tset_{\vfold}}

\newcommand{\model}{\mbf M}
\newcommand{\bin}[1][j]{\mf m_{#1}}
\renewcommand{\j}{j}
\newcommand{\K}{K}
\newcommand{\score}{\textrm{Cal-Score}}

\newcommand{\rsquare}{\ensuremath{\mc{R}^2_{\trm{C}}}}

\newcommand{\cell}{\mb C}
\newcommand{\tbquant}{\group_{\trm{top}}}
\newcommand{\truepos}{\texttt{TP}}
\newcommand{\falsepos}{\texttt{FP}}

\newcommand{\stab}{\texttt{Stab}}
\newcommand{\cellsearch}{\texttt{CellSearch}}
\newcommand{\stabilizedcellsearch}{\texttt{StabilizedCellSearch}}
\newcommand{\tstattext}{\text{$t$-statistic}}

\newcommand{\lbr}{\ensuremath{\{}}
\newcommand{\rbr}{\ensuremath{\}}}
\newcommand{\textbrace}[1]{\lbr#1\rbr}
\newcommand{\cvorig}{\texttt{cv\_orig}}
\newcommand{\cvzero}{\texttt{cv\_0}}
\newcommand{\cvone}{\texttt{cv\_1}}

\newcommand{\grouplabels}{\mathcal{F}}

\begin{document}



\begin{center}

{\bf{\LARGE 
Stable discovery of interpretable subgroups via calibration in causal studies
}}

\vspace*{.2in}
 {\large{
 \begin{tabular}{ccc}
 Raaz Dwivedi$^{\star, 1}$, Yan Shuo Tan$^{\star, 2}$, Briton Park$^2$, Mian Wei$^2$, \\Kevin Horgan$^6$,  David Madigan$^{\Diamond, 5}$, Bin Yu$^{\Diamond,1, 2, 3, 4, 7}$
 \end{tabular}
 }}
\vspace*{.2in}

\begin{tabular}{c}
      \textsuperscript{1}Department of EECS,
      \textsuperscript{2}Department of Statistics \\
      \textsuperscript{3}Division of Biostatistics,
      \textsuperscript{4}Center for Computational Biology
      \\
      University of California,
      Berkeley
 \end{tabular}
 
\vspace*{.1in}
 \begin{tabular}{c}
    \textsuperscript{5}Khoury College of Computer Sciences,
    Northeastern University
  \end{tabular}

 \vspace*{.1in}
 \begin{tabular}{c}
    \textsuperscript{6}Protypia Inc, Nashville$^\dagger$
  \end{tabular}

 \vspace*{.1in}
 \begin{tabular}{c}
    \textsuperscript{7}Chan Zuckerberg Biohub, San Francisco
  \end{tabular}

\vspace*{.2in}

\today

\vspace*{.2in}
\end{center}

\begin{abstract}
     Building on Yu and Kumbier's PCS framework and for randomized experiments, we introduce a novel methodology for Stable Discovery of Interpretable Subgroups via Calibration (StaDISC), with large heterogeneous treatment effects. StaDISC was developed during our re-analysis of the 1999-2000 VIGOR study, an 8076 patient randomized controlled trial (RCT), that compared the risk of adverse events from a then newly approved drug, Rofecoxib (Vioxx), to that from an older drug Naproxen. Vioxx was found to, on average and in comparison to Naproxen, reduce the risk of gastrointestinal (GI) events but increase the risk of thrombotic cardiovascular (CVT) events. Applying StaDISC, we fit 18 popular conditional average treatment effect (CATE) estimators for both outcomes and use calibration to demonstrate their poor global performance. However, they are locally well-calibrated and stable, enabling the identification of patient groups with larger than (estimated) average treatment effects. In fact, StaDISC discovers three clinically interpretable subgroups each for the GI outcome (totaling 29.4\% of the study size) and the CVT outcome (totaling 11.0\%). 
    Complementary analyses of the found subgroups using the 2001-2004 APPROVe study, a separate independently conducted RCT with 2587 patients, provides further supporting evidence for the promise of StaDISC.

    \let\thefootnote\relax\footnotetext{$^\star$Raaz Dwivedi \& Yan Shuo Tan are joint first authors and contributed equally to this work.
    }
    \let\thefootnote\relax\footnotetext{
    $^\Diamond$Co-Senior Authors.
    }
    \let\thefootnote\relax\footnotetext{
    $^\dagger$Protypia, Inc., 111 10th Avenue South, Suite 102, Nashville, TN 37023.} \vspace{5mm}
    
    \noindent \textbf{Key words:} Causal inference, randomized experiments, subgroup discovery, CATE modeling, calibration, stability, PCS framework, VIGOR study
    \end{abstract}

\newpage
\begin{small}
\tableofcontents
\end{small}

\section{Introduction}

Since its inception, the field of statistics has aimed to produce tools to help scientists seek scientific truth. Scientific truth, however, is not of a singular quality. While some relations in physics like Hooke's law are made apparent using simple linear regression, questions dealing with complex, emergent phenomena such as the efficacy of drugs or job training programs seem to have more contingent answers. It was the urge to formalize and investigate such questions that begot and nurtured the field of causal inference in statistics 
over the past century. One of the two most influential frameworks for causal inference,
the Neyman-Rubin causal model~\cite{holland1986statistics}, has its roots in Fisher and Neyman's~\cite{fisher1936design,neyman1923translated,neyman1935statistical} work on randomized experiments for agriculture, and was later codified by Rubin~\cite{rubin1974estimating}, who was then interested in psychometrics.\footnote{With important extensions also by Cox~\cite{cox1958planning}.}

Historically, causal inference researchers have used traditional regression methods in their analyses, with econometricians in particular developing a comprehensive theory of drawing inference from linear models~\cite{angrist2008mostly}. This is rapidly changing, however, with recent works~\cite{Athey-survey,Kunzel2017,Chernozhukov2017,Molina2019} bringing in machine learning tools to tackle causal inference problems, one genre of which has been the investigation of heterogeneous treatment effects.

\subsection{Heterogeneous treatment effects} \label{sec:HTE}
In both randomized experiments as well as observational studies, apart from the treatment and response variables, additional pre-treatment information is often known about the study subjects. For instance, information on medical risk factors is collected in clinical trials, while demographic and socioeconomic data is collected in social science studies. Such side information has always been important because it allows us to adjust for confounding in observational studies, and also to create more efficient estimators in randomized experiments~\cite{Lin2013,imbens2015causal}.

In addition to these uses, researchers are also increasingly interested in drawing inference about how the effect of a treatment varies depending on an individual's observed covariates. The past decade in particular has witnessed a wave of innovation in the modeling and estimation of heterogeneous treatment effects. Underlying the hot topic of \emph{precision medicine}~\cite{precisionmedicine} is a realization that how a patient responds to a particular drug or treatment depends on the patient's genetics, lifestyle and environment, and that consequently, accounting for these differences will allow doctors to deliver better and more targeted care. Moreover, this emphasis on understanding and exploiting heterogeneity is not unique to the biomedical sciences, and has also arisen in economics~\cite{imbens2009recent}, political sciences~\cite{gerber2008social,feller2009beyond}, online advertising~\cite{michel2019targeting}, and many other fields~\cite{feller2009beyond}.

Broadly speaking, methodological research on heterogeneous treatment effects can be put into two categories: (i) conditional average treatment effect (CATE) function estimation~\cite{imbens2009recent,gerber2008social,feller2009beyond,cai2011analysis,foster2011subgroup,Tian2014,bloniarz2016lasso}, and (ii) subgroup analysis,~\cite{subgroup_survey,subgroup_survey_2,Athey2016,Lipkovich2011}
with the latter having a longer history. Here we attempt a brief review of the existing literature, and refer the readers to referenced papers for further background.

\paragraph{CATE Estimation:} For a binary treatment, the CATE is defined to be the expected difference between the potential outcome under treatment and that under no treatment, conditional on a subject's observed covariates (see \cref{sec:framework_and_notation} for formal definitions). While the average treatment effect (ATE) is a scalar quantity, the CATE is a function and thus far more challenging to estimate. Because one observes only one of the two potential outcomes for every individual---an issue referred to as the fundamental problem of missing data in causal inference~\cite{holland1986statistics}---one cannot directly solve this problem using the conventional supervised learning techniques.

Over the past decade or so, researchers have made tremendous progress with CATE estimation and proposed  numerous methods for it~\cite{imbens2009recent,gerber2008social,feller2009beyond,cai2011analysis,foster2011subgroup,Tian2014,bloniarz2016lasso}.
A large fraction of these~\cite{imbens2009recent,feller2009beyond,cai2011analysis,bloniarz2016lasso} fall under the framework of meta-learners. These are ``meta-algorithms [that] decompose estimating the CATE into several regression sub-problems that can be solved with any regression or supervised learning method''~\cite{Kunzel2017}. Some of these meta-algorithms are fairly obvious. For instance, the $T$-learner strategy~\cite{foster2011subgroup} comprises fitting models for the two response functions (the conditional expectation of each potential outcome), and then taking their difference. Others, such as the $X$-learner~\cite{Kunzel2017} and $R$-learner~\cite{Nie2017} strategies, are more sophisticated, and require more notation to explain (see \cref{sub:cate_estimators} for further details). 
Not all proposed algorithms follow a meta-learner strategy, the popular causal tree and causal forest algorithms~\cite{Athey2016, Wager2018} being prominent examples.

\paragraph{Concerns with model choice for CATE Estimation:}
With such a diverse range of estimators, most of which come with hyperparameters, model choice becomes a primary concern. 
Some researchers have used asymptotic efficiency~\cite{Nie2017, kennedy2020optimal} to establish when certain estimators can be definitely favored under (uncheckable) generative models. 
Such arguments, however, rely on smoothness assumptions and asymptotic data regimes that are typically hard to verify for the problems typically considered by causal inference researchers. 
Meanwhile, plug-in prediction accuracy on holdout test sets is frequently used to do model selection in supervised learning, but this is infeasible for CATE estimation due to the data missingness we alluded to earlier.
To circumvent this issue, researchers have formulated proxy loss functions~\cite{schuler2018} for data-driven model choice, with ideas including using nearest neighbor matching~\cite{rolling2014}, kernel-based local linear squares fit~\cite{cai2011analysis}, and influence functions~\cite{alaa19a}. These model choice methods, however, have only been justified using simulations often in strong signal regime, a scenario that does not hold in many if not most real data problems (including the one considered in this work). 

\paragraph{Concerns with model validation for CATE Estimation:}
Before deciding which estimator to choose for a given task, we would first like to know whether there is even enough signal in the data to fit a generalizable model. Again, data missingness means that there is no clear answer to this problem. The proxy loss functions are not good substitutes for quantities like $R^2$ or ROC AUC scores because they can be noisy, and furthermore they do not have an easily interpretable scale. This is especially concerning because randomized experiments often have low signal strength.\footnote{Budget constraints would dictate that they be only sufficiently powered to detect the ATE.} 

\paragraph{Subgroup analysis:} An older approach to investigating heterogeneity is through ``subgroup analysis''. The goal here is to identify subgroups of subjects in the study over which the treatment effect is significantly larger or smaller than that the population average. 
Such a conception of heterogeneity has two advantages over CATE estimation: (a) It is less ambitious, and thus promises to be more tractable given the low data regime in real settings, and (b)
it is often more aligned with the downstream tasks involving decision-making (e.g., identifying which subgroup of individuals to treat).

Traditionally, for subgroup analysis, researchers check the treatment effect over a pre-determined list of subgroups which are suggested by prior domain knowledge. Doing this, however, ignores potential unforeseen heterogeneity in the data, and there has been much recent work on how to conduct a data-driven search for subgroups. Naive searching can quickly overfit\footnote{More importantly, investigating subgroups in this manner is particularly sensitive to human failures. It opens the door to p-value hacking~\cite{intermune}, while Gelman has argued that even when researchers try to be honest, they nonetheless have a hard time accounting for ``researcher degrees of freedom''~\cite{Gelman2013}.}, so any search method has to balance aggressiveness of searching with the need to account for multiple testing. Proposed methods include using recursive partitioning~\cite{su2009subgroup,Athey2016}, Cox modeling~\cite{negassa2005tree}, controlled partitioning with significance checks using data splits ~\cite{Lipkovich2011}, and several variants~\cite{dusseldorp2014qualitative,ballarini2018subgroup}. Unfortunately, systematic analyses of these methods have usually provided unsatisfactory results in real datasettings and in low-signal simulations~\cite{ondra2016methods,huber2019comparison}.
We refer the readers to the book~\cite{carini2014clinical} (Chapter 8), and the review papers~\cite{ondra2016methods,huber2019comparison} for further discussion on these methods.


Finally, we note that some researchers have proposed using CATE estimation as a stepping stone to finding subgroups. Such a strategy was proposed by Foster et al.~\cite{foster2011subgroup} with their Virtual Twins method, namely the $T$-learner with random forests, while Chernozhukov et al.~\cite{Chernozhukov2017} recapitulate this idea in the context of a broader call to perform inference on features of the CATE function rather than the function itself. In another line of work, Shahn et al.~\cite{shahn2017latent} integrate (linear) CATE modeling with latent class mixture modeling in a Bayesian framework to allow for treatment effect heterogeneity in discrete levels. They then use the feature importance from the latent (logistic) model and the posteriors for the CATE, to estimate qualitatively, subgroups with large treatment effect.

\subsection{The PCS framework for veridical data science}

As argued in the previous section, obtaining reliable conclusions with respect to heterogeneous treatment effects is fraught with difficulty. 
On the one hand, poor signal and weak priors are prevalent, 
and on the other hand, missing potential outcomes means that test-set validation is not directly feasible. 
Methods validated on simulation studies may not work well for real data problems since their performance are often misleading. Furthermore, empirical evidence tells us that the relative and absolute performance of estimation algorithms is highly data and context-dependent~\cite{Olson_2017}.\footnote{In fact, different methods and research groups sometimes reach different conclusions on the same datasets, see the paper~\cite{Carvalho2019} and the references therein.} Given these problems, it is puzzling to see that much new methodology is being developed that is detached from solving real data problems.

In this paper, we re-analyzed the 1999-2000 VIGOR study (a 8076 patient randomized clinical trial), and had to face precisely these challenges.
To overcome them, we take advantage of the recent works on CATE estimation
~\cite{bloniarz2016lasso,Kunzel2017,Athey2016,Nie2017,Wager2018} and build on the PCS framework for veridical data science recently introduced by Yu and Kumbier~\cite{Yu2019}. As a result, we develop a methodology called Stable Discovery of Interpretable Subgroups via Calibration (StaDISC) that is generally applicable beyond this dataset.
We now briefly review the PCS framework, before turning to the overview of our contributions and StaDISC in \cref{sub:our_contributions}.

The PCS framework bridges, unifies, and expands on ideas from machine learning and statistics for the entire data science life cycle. 
The letters in PCS stand for the three core principles of data science, namely Predictability, Computability, and Stability.
In a nutshell, the PCS framework advocates using both predictability and stability analysis, argued and documented in a PCS documentation, for reliable and reproducible  scientific investigations, thereby providing a way for bridging Breiman's Two Cultures~\cite{Breiman2001a}.
More specifically, predictability emphasizes reality checks for the modeling stage, by integrating the use of data-driven validation such as out-of-sample testing favored by machine learning, and that of goodness-of-fit measures that have a rich history in traditional statistics. 
Stability, besides encompassing sampling variability, expands to other stability or robustness concerns of the contingency of modeling conclusions to researcher ``judgment calls''. These calls include the choices made by the researcher at various stages of the data science life cycle, including  data cleaning in addition to the modeling decisions such as model choices and data perturbations.
Computability reflects the need to keep computational feasibility and efficiency in mind when constructing any modern data analysis pipeline, especially those that subscribe to the first two principles, which are usually more demanding computationally.

The PCS framework addresses to a certain extent Professor Efron's concern~\cite{Efron2020} that machine learning methods (or pure prediction algorithms) are not ready to be used on scientific problems.\footnote{In Professor Efron's timely and thought-provoking revisiting~\cite{Efron2020} of the \emph{Two Cultures} debate~\cite{Breiman2001a}, it is argued that contrasting philosophies on scientific truth is a clear line that separates traditional regression methods from modern machine learning methods (or pure prediction algorithms). While the former aims at an eternal scientific truth, the latter is truth-agnostic and instead content to exploit contingent and ephemeral patterns.} The PCS framework adds a paramount consideration of stability to predictability and computability that are hallmarks of machine learning. It guides researchers in validating machine learning and statistical methods with respect to the specific task they are to be applied and extracting data conclusions that can be relied upon. As one of us has previously discussed~\cite{YuBarter2020}, even though 100\% truth is beyond reach, a useful goal is an ``accurate approximation for a particular domain, and relative to a particular performance metric," which is a more precise articulation of George Box's belief that ``all models are wrong, but some are useful.''

\subsection{Our contributions}
\label{sub:our_contributions}
This paper makes three main contributions. First, we seek subgroups with demonstrable heterogeneous treatment effects in the dataset from the 1999-2000 VIGOR study. Complementary analyses with the 2001-2004 APPPROVe study provides additional evidence for the heterogeneity in treatment effect for the found subgroups. Enroute, building on the recent CATE literature and the PCS framework, we develop a new methodology, which we call Stable Discovery of Interpretable Subgroups via Calibration (StaDISC). We provide an overview of this methodology toward the end of this section.
Finally, this paper also serves as the first articulation of the PCS framework in the context of causal inference, with StaDISC providing a template for more informative understanding of heterogeneous outcomes.

\paragraph{Organization:} The rest of the paper is organized as follows. In \cref{sec:vigor}, we start with a brief history of the VIGOR study, and then describe the dataset and data engineering, and splitting done by us. \cref{sec:framework_and_notation} reviews the Neymann-Rubin model briefly with basic notations introduced. The development of the
StaDISC methodology  (overviewed below) is carried out in \cref{sec:predictive_check_cate,sec:stability_driven_ranking,sec:finding_interpretable_subgroups} {with the final subgroups reported in \cref{sub:discussion_of_cells}. Results for the complementary analyses of the found subgroups with the APPROVe study are presented in \cref{sec:approve}.
We conclude in \cref{sec:discussion} with a recap of our results, a discussion of the relevance of our discoveries in medicine, and discuss several directions for future work with StaDISC.
Most of the figures and tables are deferred to the appendix.
Moreover, in accordance with the PCS framework's requirement for clear and careful documentation, we provide our code, data cleaning, and statistical analyses in the form of Jupyter notebooks on GitHub (\url{https://github.com/Yu-Group/stadisc}).

\paragraph{Overview of StaDISC:} 
First of all, a given data set (deemed
approximately iid) is divided into a holdout test set $\testset$ and a training set $\trainset$ (per outcome).
For hyperparameter tuning, we use 4-fold cross validation with the training data $\trainset$.\footnote{Due to the low signal in data, we decided not to split the data into training and validation sets, and instead use 4-fold cross validation on the training data.} For any set of training folds, we refer to the leftout fold as the corresponding validation fold.
The test set is used only once at the final step of checking the significance of the interpretable subgroups found by our methodology.
See \cref{sub:data_split} for more details on data splitting and \cref{sub:cate_estimators} for the fitting of CATE estimators.
With this set-up at hand, StaDISC can be summarized in three steps: a predictive reality check in \cref{sec:predictive_check_cate} based on calibration, stability-driven ranking and aggregation of CATE estimators in \cref{sec:stability_driven_ranking}, and finally the \cellsearch\ procedure for finding interpretable subgroups in \cref{sec:finding_interpretable_subgroups}. 
In \cref{sec:predictive_check_cate}, we introduce a novel calibration-based pseudo-$R^2$ score for CATE estimators denoted by $\rsquare$, which involves placing individuals (in both training and validation folds) into equally-sized bins based on their predicted CATE value, with quantiles of the predicted CATE distribution on the training folds as thresholds for the CATE estimators. 
Using such a binning and the $\rsquare$-scores, we show that 18 popular CATE estimators generalize poorly for the VIGOR data on the validation folds of the training data. However, we find that certain quantile-based bins (referred to as quantile-based top subgroups) do generalize well in the sense of having significantly stronger subgroup CATE on both training and validation folds. This provides the starting point of the next step. 
In \cref{sec:stability_driven_ranking}, we use the $t$-statistics of the treatment effect over the quantile-based top subgroups and its stability over 7 different appropriate data perturbations to rank, screen, and finally average the screened CATE estimators (the ensemble CATE estimator).
\cref{sec:finding_interpretable_subgroups} details the last step of StaDISC, where we introduce the \cellsearch\ procedure to find a stable and interpretable representation of the quantile-based top subgroup of the ensemble from the previous step, and then check its performance on the holdout test set (which was used only for final testing).

As a final overview remark, we note that we use poor performance and good/bad generalization in a slightly loose sense throughout the paper. We only use the holdout test set at the final stage, for verifying the CATE estimates of discovered subgroups. Nonetheless, we use the phrase \emph{poor generalization} to refer to worse-than-expected-performance, where the performance metric varies across results, on the validation folds.

\section{Dataset from the VIGOR study}
\label{sec:vigor}

In this paper, we are interested in finding subgroups of patients that benefit from the treatment in the dataset from the Vioxx gastro-intestinal outcomes research (VIGOR) study~\cite{Bombardier2000}.  In the process of seeking such subgroups, we develop the new StaDISC methodology.
In this section, we provide an overview of this study and the dataset, and also explain our data pre-processing and feature engineering.

\subsection{VIGOR study history and description}
\label{sub:vigor_history}

The VIGOR study was a randomized head-to-head trial comparing two drugs used to alleviate pain and inflammation for patients with rheumatoid arthritis: a ``new" cyclooxygenase-2 (COX-2) inhibitor drug Rofecoxib (Vioxx) recently approved and developed by Merck, and Naproxen, a standard nonsteroidal anti-inflammatory drug (NSAID) already in routine clinical use for many years. NSAIDs, though effective for treating pain and inflammation, cause serious gastrointestinal side effects in a small proportion of patients with frequent use. The rationale for the development of COX-2 inhibitors, such as Vioxx, was reduced gastrointestinal toxicity as compared with traditional NSAIDs. Previously conducted short term clinical studies were supportive of this hypothesis although concerns about potential cardiovascular toxicity associated with Vioxx had also been raised.

\paragraph{Aim of the study:}
The VIGOR study was designed to provide more conclusive evidence of the superior gastrointestinal safety of Vioxx.    
The study was conducted in the years 1999-2000 by Merck with the primary hypothesis that its drug Vioxx would have fewer gastrointestinal side effects than Naproxen for the treatment of rheumatoid arthritis.
The study population comprised of 8076 patients ``with rheumatoid arthritis who were at least 50 years old (or at least 40 years old and receiving long-term glucocorticoid therapy) and who were expected to require NSAIDs for at least one year''. This population was known to be at relatively high risk of gastrointestinal side effects with NSAIDs.\footnote{However, the study was conducted with a safety monitoring board: an independent committee whose purpose is to monitor the results of an ongoing trial to ensure the safety of trial participants).} The patients in the control arm were assigned the drug Naproxen, while the patients in the active treatment arm were assigned Vioxx.

\paragraph{Details and findings of the study:}
Patients were followed for a median time of 9 months, and the primary end point was time to first occurrence of a confirmed clinical upper gastrointestinal (GI) event defined as ``gastroduodenal perforation or obstruction, upper gastrointestinal bleeding, and symptomatic gastroduodenal ulcers''.
The original study report~\cite{Bombardier2000} performed a survival analysis using a Cox proportional hazard model, and estimated the relative risk for patients in the treatment arm compared with those in the control arm to be 0.5, with a confidence interval of 0.3 to 0.6.\footnote{This estimate and the other estimates reported in this paper are based on an intention-to-treat analysis. The study also performed per-protocol and sensitivity analyses and obtained similar results.} 

The study authors also conducted a subgroup analysis for the GI events, analyzing subgroups defined by gender, age, nationality, steroids, PUB history (prior history of GI events), and presence of H. pylori antibodies. The rationale was that certain patients were known to be at increased risk of GI events, and they wanted to see if the benefit of Vioxx extended to these high-risk patients. The conclusion from the subgroup analysis was that the risk ratio for every subgroup remained significant, while differences of the ratios between subgroups were not significant.

However, VIGOR demonstrated that Vioxx was associated with an increase risk of thrombotic cardiovascular events (henceforth referred to as CVT events), an aspect that was not emphasized in the original report of the study~\cite{Bombardier2000}. The study authors suggested that apparent association of Vioxx with CVT events was actually the result of Naproxen preventing CVT events. However, placebo controlled studies confirmed that Vioxx did indeed cause CVT events, and this ultimately led to the withdrawal of Vioxx from the market. We refer the reader to the articles~\cite{krumholz2007have,ross2009pooled} for more context on the VIGOR study and its consequences thereafter. 

\paragraph{Goal of our investigation into the VIGOR dataset:}
In this work, we perform analysis for both the GI and CVT events.
While the GI event was an infrequent event (experienced by around 2\% patients) in the study, the less common CVT event (around 0.6\% were reported to have a confirmed CVT event)  was considered to be more significant medically.
Since the earlier works already established that Vioxx led to an overall decrease in the GI risk but an increase in the cardio risk on the overall population of the study, an important by-product of this work is finding clinically relevant  and interpretable subgroups of interest for which Vioxx provided a significant decrease in the risk for the GI event but did not increase the risk for the CVT event. 
Interpretability of the subgroup, as well as the transparency of the search procedure is important from a clinical view point, as the doctors can then better justify their choice to favor prescribing the drug for patients in the discovered subgroup.

We present detailed results both for the GI and CVT events throughout this paper, while occasionally deferring some details to the appendix.
To perform our analysis, we created a dataset with the two outcomes---GI and CVT event---as discussed above, a treatment indicator, and 16 binary features. The data processing necessary to create this dataset is the topic of the next section.

\subsection{Feature selection and engineering}
\label{sub:data_engineering}
The VIGOR study collected an extensive range of patient data, including demographic details, prior medical history, as well as the timing and details of adverse events during the clinical experiment. From this, we extracted sixteen clinically relevant binary features, which we report in \cref{tab:covar_description} together with covariate balance details. We now describe some of the decisions we took with respect to feature engineering, as well as the meaning the selected features. 

The medical history risk factors and drug use information were all already binary, and were selected by the VIGOR study designers as being medically relevant. For instance, it is known that use of glucorticoids predisposes patients to GI events in the context of concomitant NSAID administration~\cite{hernandez2001steroids}. One feature that deserves special interest is ASPFDA. This was an indicator for patients in the study who ``met the criteria of the Food and Drug Administration (FDA) for the use of aspirin for secondary cardiovascular prophylaxis but were not taking low-dose aspirin therapy''~\cite{Bombardier2000}, and was thought to be an especially strong risk factor for cardiovascular events. Patients who were actually undergoing aspirin therapy were excluded from the study.

On the other hand, some of the demographic and lifestyle risk factors required some engineering. The goal of the feature engineering was to simplify the data using prior information, so as to avoid overfitting and to simplify downstream data analysis. While the study collected more precise data on the patient's country of residence and their race, in both cases, a single level (``US'' and ``white'' respectively) contained a large fraction of the data, and we used these to binarize the two features. We also applied a similar logic to the smoking and alcohol lifestyle risk factors. We used height and weight information to calculate the body-mass-index (BMI) for every patient, and then used a threshold value of 30 to obtain an indicator for obesity.\footnote{\url{https://www.cdc.gov/obesity/adult/defining.html}, last accessed on August 11, 2020.} Finally, we calculated the adjusted age for every patient (by multiplying their numerical age by the ratio of the life expectancy in the US to that in their country of residence), and then used a threshold value of 65 to define an indicator for being elderly. 
Finally, there was no direct indicator for patients with a prior history of GI event, so we made use of the medical history files to impute this. See \cref{sec:data_cleaning} for more details.

The dataset was fairly complete (as is the case for most RCTs), with only a single patient missing an entry for each lifestyle risk factor (we filled in this with a 1), while 35 patients were missing entries for either height or weight, leading to a missing entry for the obesity indicator (we filled this in with a 0). Furthermore, the features also have weak pairwise correlations except for the fact that the subgroup with ASPFDA=1 (321 patients) is a subset of that with ASCGRP=1 (454 patients).

\begin{table}[ht]
    \centering
    \resizebox{\textwidth}{!}{
    \rowcolors{2}{red!10}{white}
    \begin{tabular}{lcc}
    \toprule
        \bf Covariate (ABBRV) & \bf Control No. (\%) & \bf Treatment No. (\%) \\
        \midrule
        \multicolumn{1}{l}{\textbf{Overall population}} & 4029 (49.9) & 4047 (50.1)\\
        \addlinespace[0.2em]
        \multicolumn{3}{l}{\bf Demographics}\\
         \addlinespace[0.1em]
         \hspace{1.5em}\hangindent=1.5em Whether \emph{gender} is male (MALE=1) & 814 (20.2) &  824 (20.4) \\
         \addlinespace[0.1em]
         \hspace{1.5em}\hangindent=1.5em Whether \emph{race} is white (WHITE=1) &2752 (68.3) & 2764 (68.3) \\
         \addlinespace[0.1em]
         \hspace{1.5em}\hangindent=1.5em Whether \emph{country} is US (US=1) &1750 (43.4) & 1748 (43.2) \\
         \addlinespace[0.1em]
         \hspace{1.5em}\hangindent=1.5em Whether \emph{adjusted age}$^\dagger$ $> 65$ (ELDERLY=1) &1172 (29.1) & 1136 (28.1) \\
         \addlinespace[0.1em]
         \hspace{1.5em}\hangindent=1.5em Whether \emph{body-mass-index} $>30$ (OBESE=1) &1060 (26.3) &1106 (27.3) \\
        \addlinespace[0.2em]
        \multicolumn{3}{l}{\textbf{Lifestyle}}\\
         \addlinespace[0.1em]
         \hspace{1.5em}\hangindent=1.5em Whether patient \emph{smokes} $\geq 1$ cig./day (SMOKE=1) &1879 (46.6) &1919 (47.4) \\
         \addlinespace[0.1em]
         \hspace{1.5em}\hangindent=1.5em Whether patient has $\geq 1$ \emph{alcoholic drinks}/week (DRINK=1) &1045 (25.9) &1053 (26.0) \\
        \addlinespace[0.2em]
        \multicolumn{3}{l}{\textbf{Prior medical history}}\\
        \addlinespace[0.1em]
         \hspace{1.5em}\hangindent=1.5em of \emph{GI PUB events$^*$} (PPH=1) & 317 (7.9) &313 (7.7) \\
         \addlinespace[0.1em]
         \hspace{1.5em}\hangindent=1.5em of \emph{hypertension} (HYPGRP=1) & 1168 (29.0)&1217 (30.1) \\
         \addlinespace[0.1em]
         \hspace{1.5em}\hangindent=1.5em of \emph{hypercholesterolemia} (CHLGRP=1) &293 (7.3) &343 (8.5) \\
         \addlinespace[0.1em]
         \hspace{1.5em}\hangindent=1.5em of \emph{diabetes} (DBTGRP=1) & 254 (6.3) &240 (5.9) \\
         \addlinespace[0.1em]
         \hspace{1.5em}\hangindent=1.5em of \emph{atherosclerotic cardiovascular disease}
         (ASCGRP=1) &216 (5.4) &238 (5.9) \\
         \addlinespace[0.1em]
         \hspace{1.5em}\hangindent=1.5em indicating use of \emph{aspirin} under FDA guidelines (ASPFDA=1) & 151 (3.7) & 170 (4.2) \\
         \addlinespace[0.2em]
        \multicolumn{3}{l}{\textbf{Prior usage of drugs}}\\
         \addlinespace[0.1em]
         \hspace{1.5em}\hangindent=1.5em Whether used \emph{glucocorticoids/steroids} (PSTRDS=1) & 2253 (55.9) & 2244 (55.4) \\
         \addlinespace[0.1em]
         \hspace{1.5em}\hangindent=1.5em  Whether used \emph{Naproxen} (PNAPRXN=1) & 747 (18.5) & 759 (18.8)\\
         \addlinespace[0.1em]
         \hspace{1.5em}\hangindent=1.5em  Whether used \emph{NSAIDs} (PNASIDS=1) & 3341 (82.9) & 3344 (82.6)\\
        \addlinespace[0.2em]
        \multicolumn{3}{l}{\textbf{Outcomes}}\\
        \addlinespace[0.1em]
         \hspace{1.5em}\hangindent=1.5em Whether \emph{GI event} occurred (GI=1) &121 (3.0) &56 (1.4) \\
         \addlinespace[0.1em]
         \hspace{1.5em}\hangindent=1.5em Whether \emph{CVT event} occurred (CVT=1) &18 (0.4) &41 (1.0) \\
         \bottomrule
    \end{tabular}
    }
    \caption{Overview of the baseline covariates in the control and treatment arm of the VIGOR study.  $^\dagger$Adjusted age denotes age multiplied by the ratio of the life expectancy in the US to that in the individual's country of residence. $^*$PUB stands for perforations, ulcers and bleeding.}
    \label{tab:covar_description}
\end{table}

\begin{figure}[ht]
    \centering
    \includegraphics[width=0.8\textwidth]{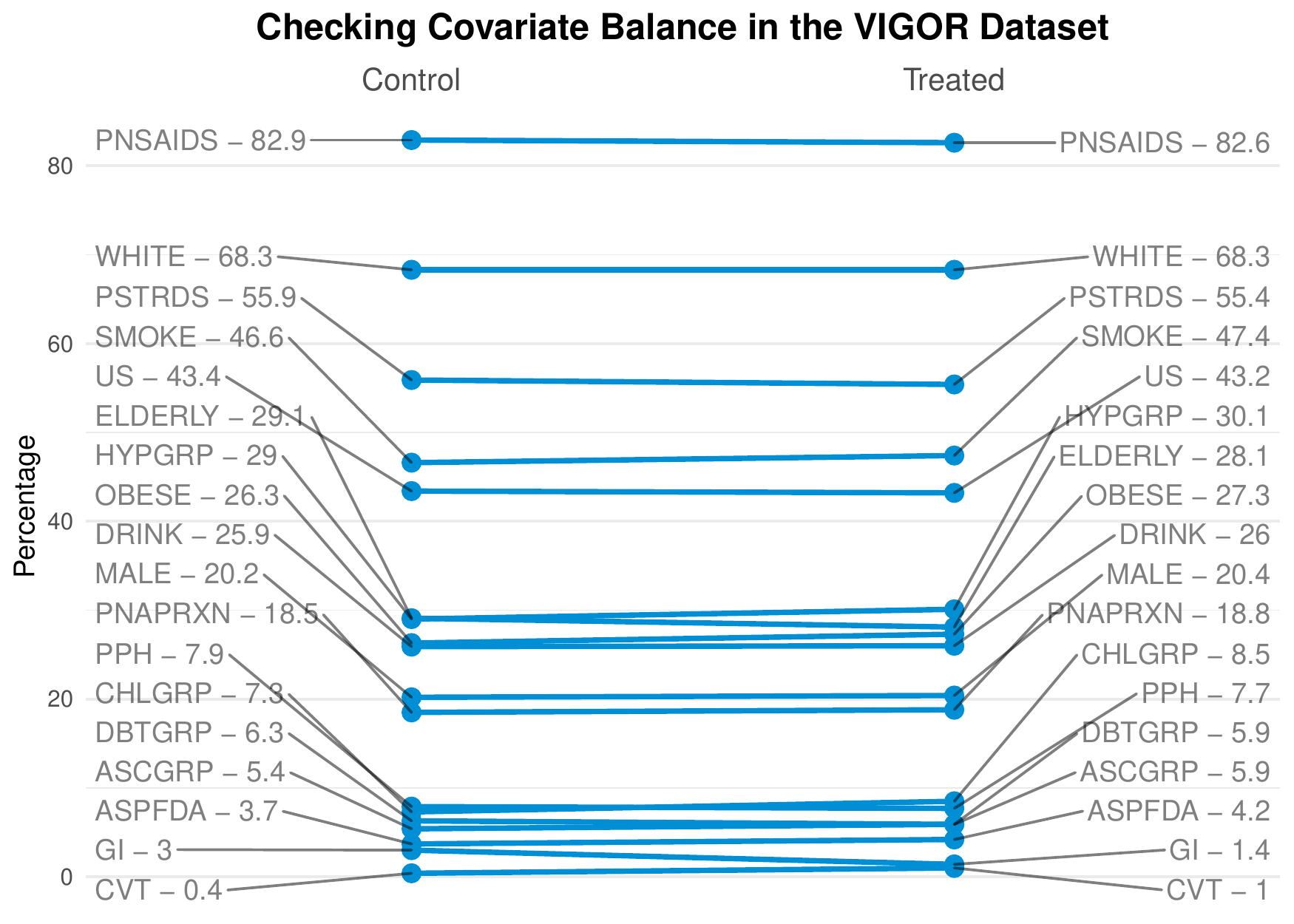}
    \caption{A visual illustration showing the covariate balance, and the outcome imbalance (GI and CVT) between the control and treatment population. The abbreviations are detailed in \cref{tab:covar_description}, the number next to the abbreviation  (ABBRV) denotes the \% of the study size taking value $1$ for that ABBRV in the respective arm. Note that the study size was 8076 total patients, and treatment and control arms comprise of 4029 (49.9\%) and 4047 (50.1\%) individuals respectively.}
    \label{fig:covar_balance}
\end{figure}

\subsection{Data splitting}
\label{sub:data_split}
As a known best practice included in the PCS framework, for each outcome, we created a holdout test set comprising 20\% of the individuals, which we did not touch in our further investigations until the very last stage of our analysis, i.e. when we wanted to verify our results. Because of the rarity of events for both outcomes, we stratified the split by both the treatment and the outcome simultaneously; such a stratification ensures that the outcome remains balanced across the test-train splits.
Let $Y$ denote the binary outcome of interest (GI or CVT event), and $T$ denote the treatment indicator.
Then such a stratification (implemented as \texttt{model\_selection.train\_test\_split} function in the sklearn library~\cite{scikit-learn}) is done by first categorizing the study subjects in 4 categories $\braces{\braces{T=0, Y=0}, \braces{T=1, Y=0}, \braces{T=0, Y=1}, \braces{T=1, Y=1}}$---once with $Y$ denoting the GI event, and once with $Y$ denoting the CVT event. Then we select a randomly sampled (without replacement) 20\% of the subjects from each category together as the test set $\testset$, with the remaining subjects form the training set $\trainset$.

Also, keeping in mind the rarity of the signals, we do not create an additional validation set, and instead we use the training data via a stratified 4-fold cross validation, where the folds are split uniformly at random, again stratified jointly according to $T$ and $Y$. For such a split, each fold has around 35 GI events and 11 CVT events among the 1615 patients.
We note that for a given outcome (say GI event), we use the same 4-fold CV split---referred to as the \emph{original split} and denoted as \cvorig---for tuning the hyperparameters for all the CATE estimators via cross-validation. We also use two \emph{additional} stratified 4-fold cross-validation (random) splits in several results throughout the paper, and denote them by \textbrace{\cvzero, \cvone}. No hyperparameter tuning is done on these additional splits, and we simply use the tuned parameters from the \cvorig\ split for fitting the estimators on different sets of training folds of these additional splits. 
Note that for any 4-fold CV split, there are 4 possible pairs of training-validation folds, denoted generically by $\trainfolds$ and $\valfold$ respectively.
Mathematically, given disjoint folds from one 4-fold CV split, namely $\braces{\trainfold[\fold]}_{\fold=1}^4$ of the training data $\trainset$ such that $\trainset=\cup_{\fold=1}^4{\trainfold[\fold]}$, the 4-pairs of training-validation folds are be denoted by $\braces{ (\trainfolds=\trainset\backslash\trainfold[\fold], \valfold=\trainfold[\fold]), \fold=1, 2, 3, 4}$.

\section{Review on Neyman-Rubin model and notation}  
\label{sec:framework_and_notation}

Throughout this paper, we will assume the standard set up for a completely randomized experiment under the Neyman-Rubin counterfactual framework. We assume that we observe a population of size $N$, in which the treatment variable $T$ is completely randomized. For each individual $i$, there are two \emph{potential outcomes}: $Y_i(0)$ when the individual $i$ is assigned to the control arm $T_i=0$, and $Y_i(1)$ when they are assigned to the treatment arm, $T_i=1$. The Individual Treatment Effect (ITE) for individual $i$ is defined as the difference of the two potential outcomes $\tau_i = Y_i(1) - Y_i(0)$.
But this quantity is unobservable since for each individual we only observe one outcome corresponding to the arm that they are assigned to, i.e,  $Y_{i,\trm{obs}} = Y_i(T_i)$ which we denote by $Y_i$ for brevity. For each individual $i$, we also observe a vector of covariates $X_i \in \fspace$. As is convention with other research into heterogeneous treatment effects, we perform inference by assuming that the samples are drawn i.i.d. from an infinite population.\footnote{Note that the standard variance estimates reported using this perspective can be taken as conservative estimates of the finite-sample variances defined in Neyman's repeated sampling framework~\cite{Ding2017}.}

We now define the various quantities of interest studied throughout this paper. Let $\group$ be a measurable subset of the feature space $\fspace$. The average treatment effect (ATE), conditional average treatment effect (CATE) and the subgroup CATE are respectively defined as 
\begin{subequations}
\label{eq:all_ate_defs}
\begin{align}
    \trm{ATE}: \ate &\coloneqq \E\brackets{Y(1)} - \E\brackets{Y(0)},  \label{eq:ate_def}   \\
    \trm{CATE}: \cate[x] &\coloneqq \E\brackets{Y(1)~\vline~ X=x} - \E\brackets{Y(0) ~\vline~ X = x},
    \qtext{for any} x \in \fspace \label{eq:cate_def} \\
    \trm{sub-group CATE}: \gate[\group] &\coloneqq \E\brackets{\tau(X)~\vline~ X \in \group}, \qtext{for measurable subset} \group \subset \fspace, \label{eq:sgcate_def}
\end{align}
\end{subequations}
where the expectation is taken with respect to the iid draws from the infinite population.

At a high-level, the goal of this work is to provide a systematic framework to find subgroups $\group \subset \fspace$, which (i) include non-trivial fraction of the observed data, (ii) are relevant and interpretable relevant for the domain problem at hand, and (iii) most importantly have significant sub-group CATE, i.e., $\gate[\group]$ has significantly larger magnitude than $\ate$.

\paragraph{Neyman difference-in-means estimates for finite samples:}
We will often use the classical Neyman difference-in-means estimator to provide plug-in estimates for the ATE and sub-group CATE values. Formally, we denote the two study arms by
\begin{subequations}
\begin{align}
\label{eq:treat_control_defn}
    \textrm{(Treatment arm) } \treat \defn \braces{i \in [n]: T_i = 1}
    \qtext{and}
    \textrm{(Control arm) } \control \defn \braces{i \in [n]: T_i = 0},    
\end{align}
Throughout this paper, we will abuse notation: for any group $\group \subset \fspace$, we will use the same symbol to refer the subpopulation of individuals that belong to it. This allows us to denote the restriction of the two arms of the study to the subgroup as follows:
\begin{align}
\label{eq:group_defn}
    \treat \cap \group \defn \treat \cap  \braces{i \in [n] : X_i \in \group}
    \qtext{and}
    \control\cap \group \defn \control \cap  \braces{i \in [n]: X_i \in \group}.
\end{align}
\end{subequations}
For a finite set $\mc A$, let $\abs{\mc A}$ denote the number of elements in the set.
With this notation at hand, the plug-in estimators for the average treatment effect $\ate$ and the sub-group average treatment effect $\gate$ are given by
\begin{subequations}
\begin{align}
    \atehat &= \frac{1}{\abs{\treat}} \sum_{i \in \treat} Y_i(1)
    -\frac{1}{\abs{\control}} \sum_{i \in \control} Y_i(0), \qquad
    \qtext{and} 
    \label{eq:ate_hat}\\
    \gatehat &= \frac{1}{\abs{\treat \cap \group}} \sum_{i \in \treat\cap \group} Y_i(1)
    -\frac{1}{\abs{\control \cap \group}} \sum_{i \in \control \cap \group} Y_i(0).
    \label{eq:gate_hat}
\end{align}
\end{subequations}

For randomized experiments, both estimates $\atehat$ and $\gatehat$ are unbiased~\cite{neyman1923translated}, and standard error estimates are available for it~\cite{imbens2015causal}. On the other hand, the precision of $\gatehat$ degrades as the size of the subgroup shrinks. 
For the same reason, a direct difference-in-means estimator for CATE~\eqref{eq:cate_def} is almost never feasible, as for most values of $x \in \fspace$ (e.g., when $\fspace$ is continuous, or combinatorially very large), there might not exist any sample with covariate equal to $x$.

\section{Calibration as a prediction (reality) check for CATE estimators}
\label{sec:predictive_check_cate}
Following the Predictability principle of the PCS framework, any statistical model must pass a test of out-of-sample prediction accuracy before we should have any trust in it. This principle is in line with the ethos of the scientific method, which correlates the strength of a hypothesis with the rigor of prior attempts to falsify it~\cite{Popper1959}. As discussed in \cref{sec:HTE}, however, no such test currently exists for CATE models. The missing potential outcomes mean we do not have a plug-in estimate for any risk function $R(\tau,\hat{\tau}) = \E\brackets{l(\cate[X],\hat{\tau}(X))}$. Furthermore, unlike $R^2$ and ROC AUC scores, the proxy loss functions proposed for model choice (see \cref{sec:HTE} and the references therein) do not have interpretable scales.

To mitigate this problem, we develop a prediction accuracy check that can be applied to any CATE estimator. This check makes use of the ideas from the calibration literature~\cite{dawid1982well,degroot1983comparison,Guo2017}, and while passing the check is not a sufficient condition for a CATE estimator to have good performance, it is at least a necessary one. 
Even though our StaDISC approach is motivated by and grounded in the analysis of CATE estimators fitted to the VIGOR study data, we believe it is a general methodology useful for other causal inference problems.

The rest of this section is organized as follows. We discuss the 18 CATE estimators used in our analysis of the VIGOR data in \cref{sub:cate_estimators}. We then introduce the calibration-based scores for prediction checks in \cref{sub:cal_based_score}, and apply it to the CATE estimators trained with VIGOR data in \cref{sub:cal_based_check_vigor}. Finally, in \cref{sub:isolate_gen_parts} we show how despite the poor performance on the overall data, the CATE estimators have good generalization locally, thereby setting the stage for identifying subgroups with subgroup CATE significantly larger than ATE in \cref{sec:stability_driven_ranking}.

\subsection{CATE estimators applied on the VIGOR dataset}
\label{sub:cate_estimators}
We now describe the 18 popular CATE estimators used in this work, 14 of which follow meta-learner strategies. Descriptions of the meta-learner strategies can be found in \cite{Kunzel2017} and ~\cite{Nie2017}. Here, we simply list our choices of base learners for each meta-learner. The base learners are all drawn from a pool comprising lasso, logistic regression, random forest (RF), and gradient-boosted trees (GB). In our statistical analyses, we used implementations of the former three algorithms from the \texttt{scikit-learn} package~\cite{scikit-learn} and the \texttt{XGBoost} implementation of the latter~\cite{Tian2014}. Furthermore, for code cleanliness, we made use of the meta-learner interface provided by the \texttt{causalml} package~\cite{chen2020causalml}. In additional to estimators based on meta-learners, we also considered two versions each of causal tree~\cite{Athey2016} and causal forest~\cite{Wager2018}. The versions differ in terms of their hyperparameter choices. We used \texttt{causalml}'s implementation of the former. For the latter, we were not able to find a well-documented python implementation of the algorithm, so we built one around \texttt{causalml}'s causal tree implementation.
\newcommand{\estimator}{\texttt}
\begin{enumerate}
    \item \emph{S-learners} (2 estimators): We used RF and GB as the base learners, denoted by. These are denoted as \estimator{s\_rf} and \estimator{s\_xgb}.
    \item \emph{T-learners} (4 estimators): We used lasso, logistic regression, RF and GB as base learners. These are denoted as \estimator{t\_lasso}, \estimator{t\_logistic}, \estimator{t\_rf} and \estimator{t\_xgb}.
    \item \emph{X-learners} (4 estimators): We used lasso, logistic regression, RF and GB as base learners for the first stage, and lasso as the only base learner for the second stage. These are denoted as \estimator{x\_lasso}, \estimator{x\_logistic}, \estimator{x\_rf} and \estimator{x\_xgb}.
    \item \emph{R-learners} (4 estimators): In the case of randomized experiments, the R-learner requires a choice of base learner for the conditional expectation of the response with the treatment variable partialed out, and a choice of base learner for the treatment effect. We use four such pairs, each member of which was chosen uniformly at random from the base learners (with logistic regression excluded due to its similarity to lasso). Doing this, we got \{lasso, lasso\}, \{lasso, GB\}, \{RF, lassso\}, and \{RF, RF\}.  These are denoted as \estimator{r\_lassolasso}, \estimator{r\_lassoxgb}, \estimator{r\_rflasso} and \estimator{r\_rfrf}.
    \item \emph{Causal Tree and Causal Forest} (4 estimators): We used 2 versions each of the causal tree and causal forest algorithms, which we have denoted as \estimator{causal\_tree\_1}, \estimator{causal\_tree\_2}, \estimator{causal\_forest\_1}, and \estimator{causal\_forest\_2}. Each pair of estimators differ in their hyperparameter choices. Specifically, \estimator{causal\_tree\_1} and \estimator{causal\_forest\_1} both use a minimum of 50 samples per leaf node, whereas \estimator{causal\_tree\_2} and \estimator{causal\_forest\_2} both use a minimum of 200 samples per leaf node. All other hyperparameter choices are standard and can be found in our documentation on GitHub.
\end{enumerate}

Here, we briefly justify our choice of the 18 CATE estimators listed above. First, we chose our pool of base learners because they are representative of the most popular supervised learning algorithms in use today, with neural networks omitted because of the poor signal and small size of the data set. The $T$-learner framework is perhaps the simplest way of fitting a CATE model and has been used and studied by many different authors. Using lasso as the base learners was proposed and analyzed by Bloniarz et al.~\cite{bloniarz2016lasso} and Imai and Ratkovic~\cite{Imai2013}. Meanwhile, ~\cite{foster2011subgroup} proposed using RF as the base learner. The X-learner~\cite{Kunzel2017} and R-learner~\cite{Nie2017} frameworks have both been used by many recent works. The former has demonstrated favorable performance over other estimators in data challenges organized by the Atlantic Causal Inference Conference, while the latter has optimality guarantees under some assumptions, and has been further supported by some follow up work~\cite{schuler2018}. We included two S-learner estimators for completion, since all four meta-learner frameworks are supported by the \texttt{causalml} package. The causal tree~\cite{Athey2016} and causal forest~\cite{Wager2018} estimators have similarly been used in much recent work, with the latter attaining the status of being a benchmark of sorts for CATE estimation methods in many simulations. 

All CATE estimators based on meta-learners had the hyperparameters of their component base learners tuned via 4-fold CV using \cvorig. A common hyperparameter grid was used for each base learner type, with details deferred to our documentation on GitHub.


\subsection{A calibration-based score for CATE estimators}
\label{sub:cal_based_score}
To develop a reality check scheme for CATE estimators, we now build on the literature of calibration of probability scores. 

A binary classifier is said to be well-calibrated if the class probabilities that it predicts for each sample point is close to the true class probabilities. This property is desirable in many situations, such as weather-forecasting, where we would like it to rain on close to 40\% of the days on which a 40\% chance of rain is forecast. Unfortunately, machine learning models are often not naturally calibrated, with neural networks in particular being overconfident in their estimated class probabilities~\cite{Guo2017}. Furthermore, because class probabilities are unobserved, we cannot directly train a model to predict these values using supervised learning. While researchers have proposed various solutions to this problem, the common theme is to \emph{bin} the observations by their \emph{predicted class probabilities}, and then use the observed class distribution over the bin to obtain plug-in estimates of the true class probabilities.

The concept of calibration has a long history~\cite{dawid1982well,degroot1983comparison}, and it has also been referred to as validity~\cite{miller1962statistical} or reliability~\cite{murphy1973new}.
Starting for evaluation of weather forecasts in the 1950s~\cite{brier1950verification}, calibration has been widely used 
as a generic scheme to compare several forecasters~\cite{degroot1983comparison}.
Related ideas have been used to calibrate a wide range of methods, including Bayesian models~\cite{dawid1982well}, SVMs, boosted trees, random forests~\cite{niculescu2005predicting,naeini2015obtaining}, and more recently deep neural networks~\cite{Guo2017}.

\paragraph{Binning via estimated CATE values:}  We now begin to define our calibration-based prediction accuracy measure for CATE estimators. While our scores---to be defined below---are easy to interpret, defining them formally requires a bit of notation which we now describe.

Consider the training set $\trainset$ and let $\trainfold, \fold=1, 2, 3, 4$ denote its $4$-fold (random) CV split.
Fix a fold $\fold$ and let $\trainfolds = \trainset \backslash \trainfold[\fold]$ denote the training folds used to fit the CATE estimator $\model: \fspace \to \real$, and let $\valfold = \trainfold[\fold]$ denote the left-out fold, which we also call as validation fold, for the estimator $\model$.
Let $\bin[\qvalue]$ denote the $q$-th quantiles of the CATE estimator $\model$ on the training folds of the data:
\begin{align}
    \bin[\qvalue] =
    \min\braces{ c~\bigg\vert~\frac{\#\{i \in \trainfolds: \model(x_i) \leq c\}}{\abs{\trainfolds}} \geq \qvalue},
    \qtext{for any} \qvalue \in (0, 1),
    \label{eq:quantile_value}
\end{align}
where by convention we set $\bin[0]=-\infty$ and $\bin[1]=\infty$.
Then given a grid of q-values denoted by $\braces{\qvalue_1 \leq \qvalue_2 \leq \cdots \leq \qvalue_{\K-1}}$ in the interval $(0, 1)$, we split the real line into $\K$ bins as follows:
\begin{subequations}
\begin{align}
    \bin[0] < \bin[\qvalue_1] \leq \bin[\qvalue_2]  \leq &\ldots \leq \bin[\qvalue_{\K-1}] < \bin[1].
    \notag
\end{align}
We use this binning to induce a partition of $\fspace$ into $K$ \emph{quantile-based subgroups} given by
\begin{align}
    \group_{\j} := \group_{\j}(\model) = \big\{x \in \fspace~\big\vert~\model(x) \in [\bin[\qvalue_{\j}], \bin[\qvalue_{\j+1}]] \big\} \qtext{for} \j=0, 1, \ldots \K-1,
    \label{eq:group_quantile}
\end{align}
Given a set of individuals $\tset$ (say, training folds $\trainfolds$ or validation fold $\valfold$), let $\overline{\model}_{\group_{\j} \cap \tset}$ denote the mean of the predicted CATE from the estimator $\model$ on the subgroups $\group_{\j} \cap \tset$ :
\begin{align}
\label{eq:model_group_cate}
    \overline{\model}_{\group_{\j} \cap \tset} := \frac{1}{\abs{\group_{\j} \cap \tset}} \sum_{i \in \group_{\j} \cap \tset} \model(X_i),
    \qtext{where}\group_{\j} \cap \tset = \braces{i \in \tset \vert X_i \in \group_{\j}},
\end{align}
Similarly, recall that $\gatehat[\group_{\j} \cap \tset]$ denotes the plug-in estimate for the subgroup CATE for the subgroup $\group_{\j}$.
\begin{align}
\label{eq:model_ate_group_cate}
    \gatehat[\group_{\j} \cap \tset] := \frac{1}{\abs{\treat \cap \group_{\j} \cap \tset}} \sum_{i \in \treat\cap \group_{\j} \cap \tset} Y_i(1)
    -\frac{1}{\abs{\control \cap \group_{\j} \cap \tset}} \sum_{i \in \control \cap \group_{\j} \cap \tset} Y_i(0).
\end{align}
\end{subequations}
\paragraph{Score definitions:} With these definitions of the sub-groups, we are now ready to define the calibration score:
\begin{subequations}
\begin{align}
\label{def:cal_score}
    \score(\tset;\model) := \sum_{\j=1}^{\K} 
    \frac{\abs{\group_{\j} \cap \tset} }{\abs{\tset}}\cdot \abs{ \overline{\model}_{\group_{\j}\cap\tset}-\gatehat[\group_{\j}\cap\tset]},
\end{align}
where we use absolute difference (and not squared difference) since the scale of the quantities $\{\overline{\model}_{\group_{\j}\cap\tset}, \gatehat[\group_{\j}\cap\tset]\}$ is pretty small for our dataset.
Nonetheless, it is still hard to interpret the absolute scale of $\score(\model)$, and hence we normalize these scores by a baseline to define a pseudo-$R^2$ score.
More precisely, we consider a  baseline calibration-score $\score(\tset;\atehat)$, obtained by replacing the the CATE estimator average $\overline{\model}_{\group_{\j}\cap\tset}$ with that of the (constant) ATE estimate $\atehat$ in \cref{def:cal_score}:
\begin{align}
\label{def:cal_score_ate}
    \score(\tset; \atehat) := \sum_{\j=1}^{\K} \frac{\abs{\group_{\j} \cap \tset} }{\abs{\tset}} \cdot \abs{ \atehat-\gatehat[\group_{\j}\cap\tset]}.
\end{align}
With \cref{def:cal_score,def:cal_score_ate} in place, we define the $\rsquare$ score as follows:
\begin{align}
\label{def:rsquare}
    \rsquare(\tset; \model) := 1- \frac{\score(\tset;\model)}{\score(\tset; \atehat)}.
\end{align}
\end{subequations}
Just like the usual $R^2$-score\footnote{While $R^2$-score was originally introduced for linear regression, several similar measures have been proposed for providing an interpretable scale to measure the model fit. The $R^2$ for linear regression takes value in [0,1] for training data, and $(-\infty, 1]$ for test data. Close to $1$ value suggests a good fit, and a smaller score implies a poor fit. 
Note that unlike the $R^2$ for linear regression, for CATE estimators, the pseudo-score $\rsquare$ is not guaranteed to take value in $[0, 1]$ even on the training data, i.e.,  $\rsquare(\trainfolds; \model) \in (-\infty, 1]$. Nonetheless, in \cref{fig:r_square}, we observe that for all the CATE estimators, this score lies in [0, 1] on the training folds, i.e., $\rsquare(\trainfolds; \model) \in [0, 1]$.}, the score $\rsquare(\tset; \model)$ can take any value between $(-\infty, 1]$, and a model can be deemed a good fit if this score is close to $1$. We interpret the score as measuring, conditioned on the partition of the feature space into bins, the degree to which the CATE estimator explains the variability of the CATE with respect to the partition, in comparison to the best constant model.

Since different models induce different partitions, the scores are not necessarily comparable across models. Furthermore, similar to how calibrated classification algorithms need not have good prediction accuracy, it is possible for a CATE model to have a good $\rsquare$ score and yet have poor overall prediction accuracy for the CATE. Nonetheless, having $\rsquare$-scores that are reasonably close to 1 across a range of data perturbations is \emph{necessary} albeit not sufficient for the CATE model to have good prediction performance.
Moreover, the variability of the score between the choices $\tset=\trainfolds$ and $\tset=\valfold$ also provides a check on the \emph{overfitting} of the CATE estimator. 

To conclude, the $\rsquare$ provides two predictive checks for the CATE estimators.
On the one hand, when $\rsquare(\trainfolds; \model)$ is much smaller than $1$, we conclude that the estimator $\model$ has a poor fit on the training data. On the other hand, a high value (close to 1) value for $\rsquare(\trainfolds; \model)$, and a relatively lower value  (close to 0 or negative) for $\rsquare(\valfold; \model)$ would necessarily indicate overfitting of the estimator $\model$.


\subsection{Calibration-based predictive check on CATE estimators for VIGOR dataset}
\label{sub:cal_based_check_vigor}
We now compute the scores defined in the previous section for the 18 popular CATE estimators when applied to the VIGOR dataset. 
We use the evenly-spaced quantile grid $\braces{0.2, 0.4, 0.6, 0.8}$ and compute the $\rsquare$-scores using the $\K=5$ bins it induces. We also consider a restricted $\rsquare$-score to measure the predictive performance of the estimators for the bottom-2 bins for the GI event, and top-2 bins for the CVT event. To compute this \emph{restricted} $\rsquare$-score, we simply replace the sum over the index $\j\in\braces{1, 2, \ldots, 5}$ in \cref{def:cal_score,def:cal_score_ate} with $\j \in \braces{1, 2}$ for the GI event and $\j \in \braces{4, 5}$ for the CVT event, and then plug this restricted sum in \cref{def:rsquare}.

In the previous section, we described how, given a CATE estimator and a fixed fold $\fold$, we obtain two (restricted) $\rsquare$-scores---one on the training folds $\trainset\backslash\trainfold[\fold]$ and one on the validation fold $\trainfold[\fold]$.
Repeating this over $4$ folds provides us with $4$ pairs of such scores. And iterating over M different types of CATE estimators yields $M \times 4$ such pairs. Furthermore, if we consider $L$ different 4-folds splits, we get $M \times 4 \times L$ such pairs of scores.

We trained 18 different CATE estimators for both the outcomes, namely the GI and CVT events. However, after fitting, the following estimators learned a zero CATE function: R-learner with XGBoost for the GI event, and S-learner with XGBoost, Causal Tree with a particular choice of hyperparameters, and R-learner with XGBoost for the CVT event. Thus, going forward we report results for the remaining 17 CATE estimators for the GI event and 15 CATE estimators for the CVT event. See \cref{sub:cate_estimators} for more details on all the estimators.
We now first discuss the details of scores presented in various plots in \cref{fig:r_square} and then discuss the conclusions in a separate paragraph.

\begin{figure}[ht]
    \centering
    \begin{tabular}{c}
    \includegraphics[width=\textwidth]{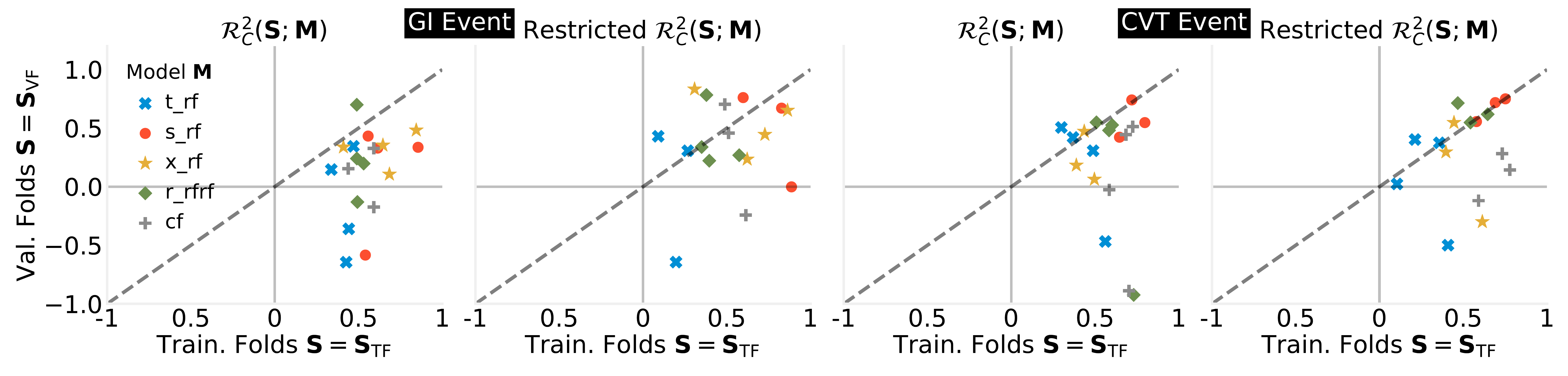} \\
    (a) \\
    \addlinespace[0.3em]
    \midrule
    \addlinespace[0.3em]
    \includegraphics[width=\textwidth]{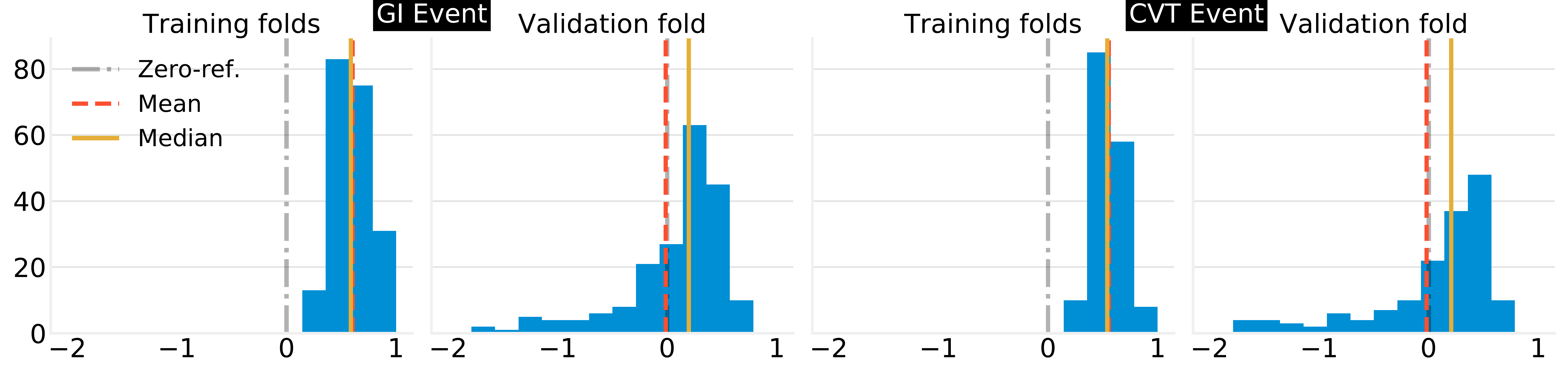}
     \\ (b)
    \end{tabular}
    \caption[Calibration-based scores]{Plots with the calibation-based $\rsquare$-scores~\eqref{def:rsquare} for various CATE estimators. \textbf{(a)} Scatter plot of $\rsquare$-scores on the training and validation folds for $5$ CATE estimators on the original 4-fold split \cvorig\ on which hyperparameters were tuned via cross-validation. Refer to the text for definition of restricted \rsquare-scores.
   \textbf{(b)} Histogram of the $\rsquare$-scores on the 12 training and validation folds, 4 each from the 3 different CV splits, namely \textbrace{\cvorig, \cvzero,\cvone} for 17 CATE estimators for GI event, and for 15 CATE estimators for CVT event.}
    \label{fig:r_square}
\end{figure}
\paragraph{Details of \cref{fig:r_square}:}
In \cref{fig:r_square}(a), we provide a scatter plot of $\rsquare(\trainfolds, \model)$ (training score) and $\rsquare(\valfold, \model)$ (validation score) for $5$ different estimators for each fold of original CV split \cvorig on the VIGOR data both for GI and CVT events.
These estimators are T\_RF, S\_RF, X\_RF, R\_RFRF and CF\_1 which denote T, S, X, R-learners with random forest as base learners, and (one of the two) Causal Forest respectively.
In addition, in the right two figures in \cref{fig:r_square}(a), we also provide the scatter plot of the corresponding restricted $\rsquare$-scores (see the first paragraph of this section for its definition) on the training and validation folds for the 5 estimators and both events.

Next, to check the \emph{stability} of our conclusion, we compute these scores for all 17 CATE estimators for the GI event, and all 15 CATE estimators for the CVT Eventon all 3 random CV splits \textbrace{\cvorig,\cvzero,\cvone}. That is, we obtain a total of 204 and 180 (training and validation) pairs of $\rsquare$-scores respectively for the GI and CVT events. In \cref{fig:r_square}(b), we plot the histogram of these scores.

\paragraph{Conclusions from \cref{fig:r_square}:}
Inspecting the scatter plots in \cref{fig:r_square}(a), we see clear evidence of overfitting, as the validation fold $\rsquare$-scores (computed as $\rsquare(\valfold, \model)$ in \cref{def:rsquare}) are systematically much smaller, and often negative, than those on the training folds (computed as $\rsquare(\trainfolds, \model)$ in \cref{def:rsquare}). Furthermore, there is substantial variability across different folds. For instance, one dot corresponding to S\_RF for GI events was not even plotted because the validation fold $\rsquare$ score exceeded the lower $y$-limit of the plot. 
These findings are supported by the histograms in \cref{fig:r_square}(b), which show that the mean of the validation fold $\rsquare$-scores is in fact a negative number for both GI and CVT events. 
While we presented histograms of the aggregated scores over all the CATE estimators, the general behavior was also true when looking at individual CATE estimators.
Next, we also note that the bottom-2-restricted \rsquare-score for the GI event and top-2-restricted \rsquare-score have slightly better generalization since the validation scores are generally positive albeit with the caveat of larger variability across the training folds. (We revisit this aspect in more detail in \cref{sub:isolate_gen_parts}.)

The poor performance on average as well as the high variability of performance both lead us to be skeptical of the conclusions from any CATE estimator on the VIGOR study data. 
Here, we remark that the variability of the scores stems from both fluctuations in the trained model as well as low SNR in the validation fold (leading to \score~ deviating from its expected value). We remind the reader that in total there are 177 GI events and 59 total CVT events, and this fact implies that for each quantile-based subgroup, we should expect to see around 7.1 and 2.3 GI and CVT events respectively in the validation fold, under the assumption of no heterogeneity. The poor performance is hence entirely to be expected, and in fact could be a general theme for RCTs, as they are often sufficiently powered for only computing the ATE.

\newcommand{\pert}{\mf D}
\newcommand{\gitopgroup}[1][\qvalue]{\wtilde{\group}_{#1}}
\newcommand{\tctopgroup}[1][\qvalue]{\wtilde{\group}_{#1}^c}
\subsection{Extracting data conclusions that can be relied upon}
\label{sub:isolate_gen_parts}
While we conclude that we cannot trust the CATE models in their entirety, it remains to be seen if we can isolate data conclusions from them that we can rely on. To this end, we take a closer look the relative ordering of scores  $\overline{\model}_{\group_{\j}\cap\tset}$~\eqref{eq:model_group_cate} and $\gatehat[\group_j \cap \tset]$~\cref{eq:model_ate_group_cate} across the quantile-based subgroups $\braces{\group_\j}_{\j=1}^5$ considered in the previous section.
Given the quantile-based definition of the groups, it is natural to test whether we have
\begin{subequations}
\label{eq:monotonicity_trends}
\begin{align}
    \overline{\model}_{\group_{1}\cap\tset}\leq \overline{\model}_{\group_{2}\cap\tset} \leq \ldots \leq \overline{\model}_{\group_{5}\cap\tset}, \quad\text{(estimator CATEs)} \qtext{and} \label{eq:mon_model}\\
    \gatehat[\group_1\cap\tset] \leq \gatehat[\group_2\cap\tset] \leq \ldots \leq \gatehat[\group_5\cap\tset], \quad\text{(subgroup CATE estimates)}\label{eq:mon_gate}
\end{align}
\end{subequations}
for a set of individuals $\tset$ comprising either the training folds or the validation fold.
In \cref{fig:ece_5_bins}, we plot these estimates for two estimators X\_RF and T\_RF for the GI event in panel (a) and the CVT event in panel (b) for one set of training and validation folds from the original split. 
In each plot, the blue error bars denotes the sample standard deviation estimate for the sample mean $\overline{\model}_{\group_{\j}\cap\tset}$ computed from $\braces{\model(X_i), i \in \group_{2}\cap\tset }$, and the red error bars denote the standard error estimate for $\gatehat[\group_\j \cap \trainfolds]$ given by \cref{eq:plugin_subgroup_cate_std}. 
We observe that generally the model CATE estimates $\braces{\overline{\model}_{\group_{1}\cap\tset}}_{\j=1}^5$ are monotonic for both events on both training folds and validation fold. However, the story with the plug-in subgroup CATE estimates  $\braces{\gatehat[\group_\j \cap \tset]}_{\j=1}^5$ is---not unexpectedly---mixed. 
For the GI event, while these estimates are monotonic on the training folds $(\tset=\trainfolds)$, they are not monotonic on the validation fold $(\tset=\valfold)$. For the rarer CVT event, the estimates $\braces{\gatehat[\group_\j \cap \tset]}_{\j=1}^5$ are not even monotonic on the training folds. This non-monotonic behavior is far from unique to the two estimators presented here. Instead, the plots are representative of what we observe for all other estimators as well, even when using alternate data splits into training and validation folds.

\begin{figure}[ht]
    \centering
    \resizebox{1\textwidth}{!}{%
    \begin{tabular}{c|c}
    \includegraphics[width=0.5\textwidth]{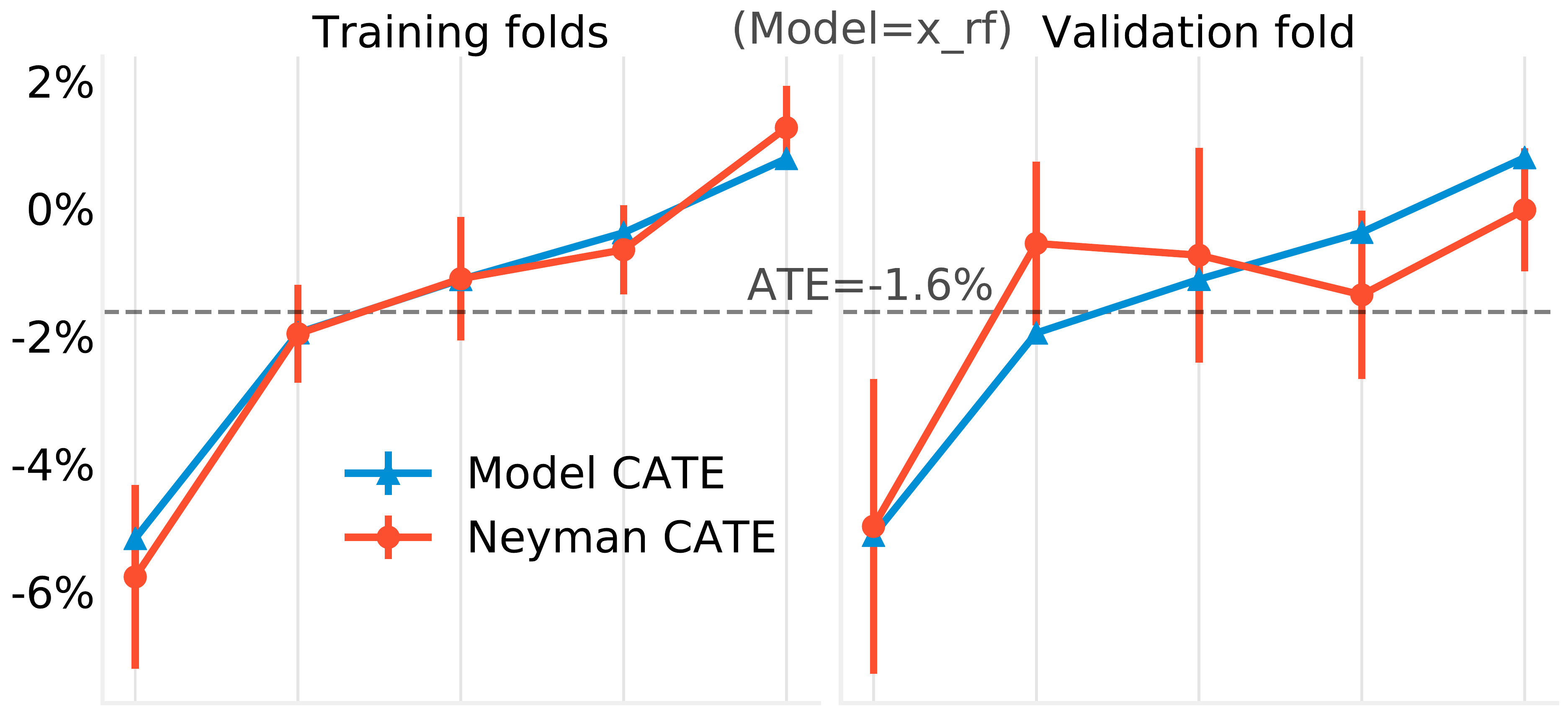} &
    \includegraphics[width=0.5\textwidth]{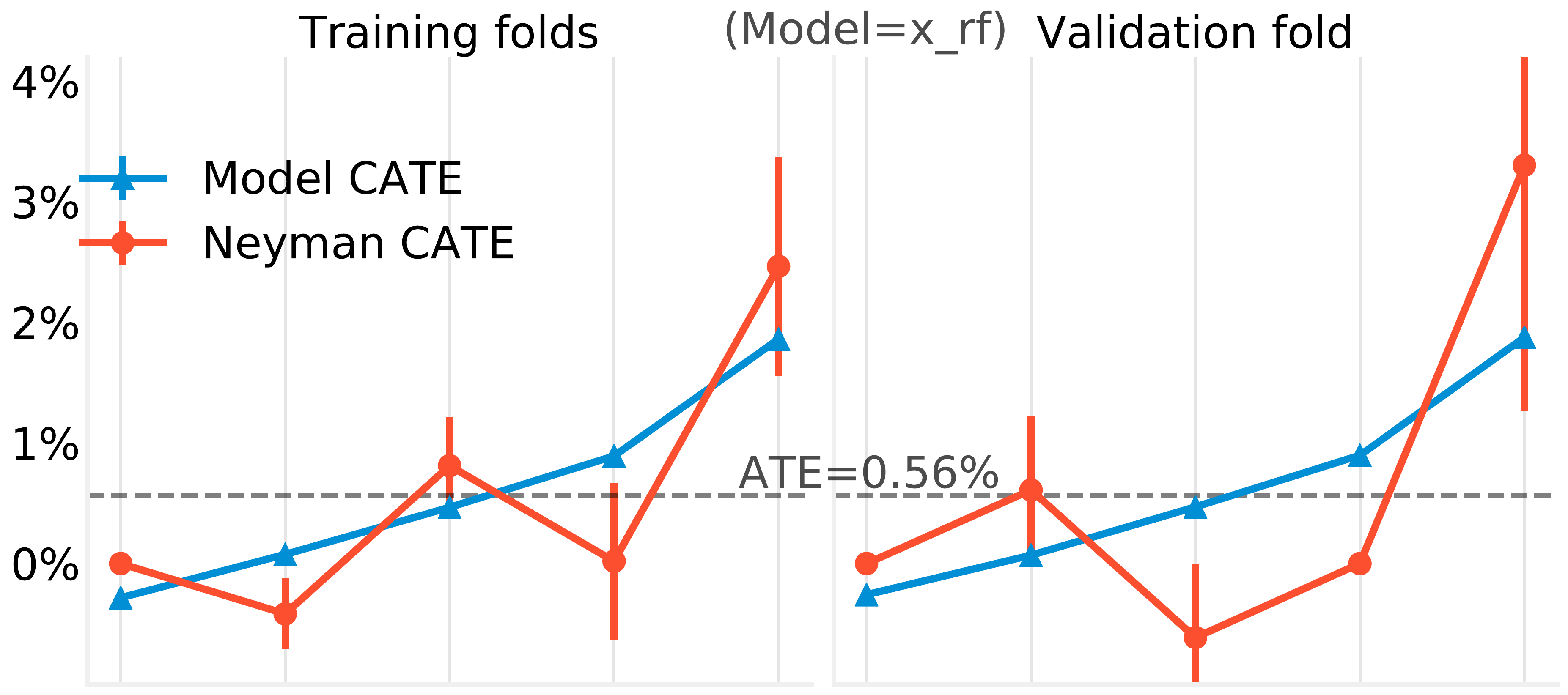}\\
    \includegraphics[width=0.5\textwidth]{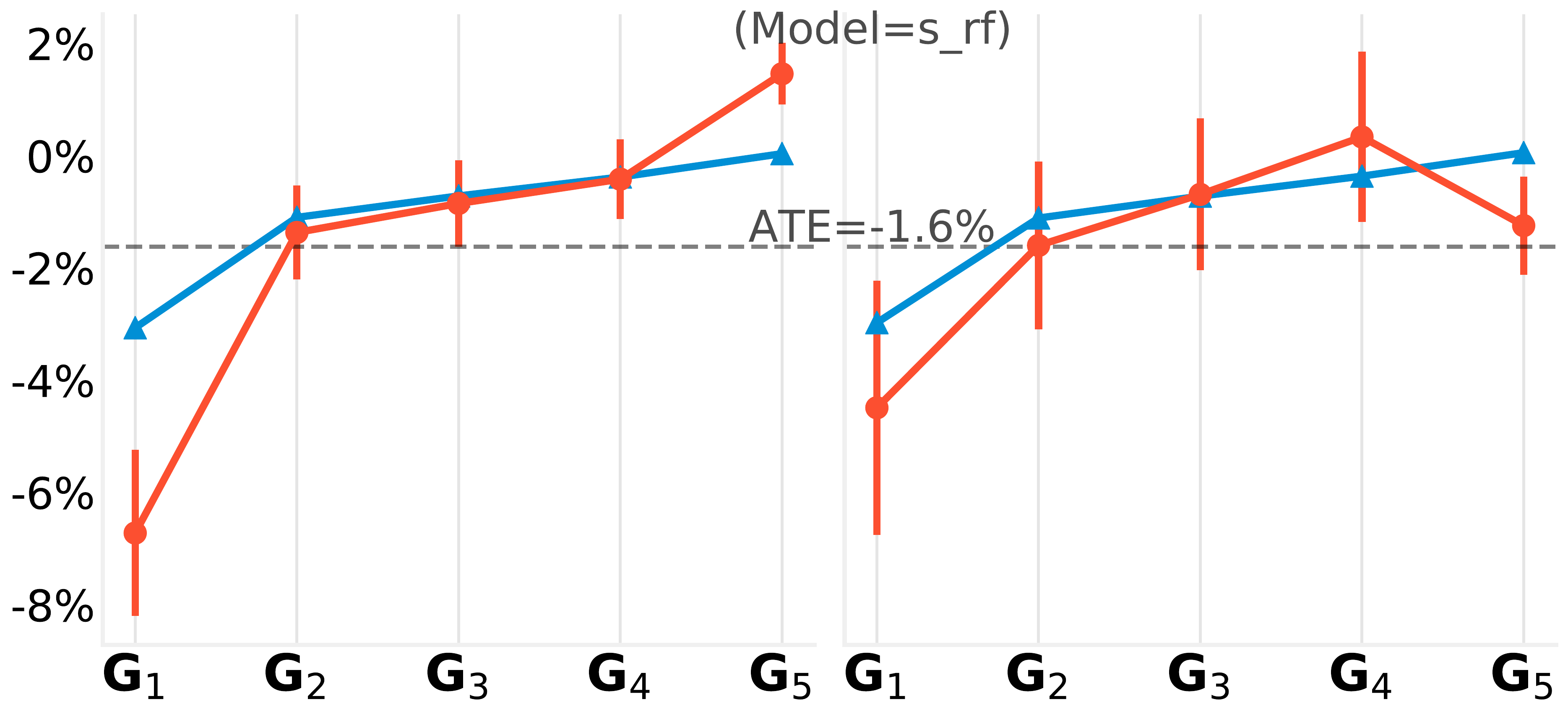} & 
    \includegraphics[width=0.5\textwidth]{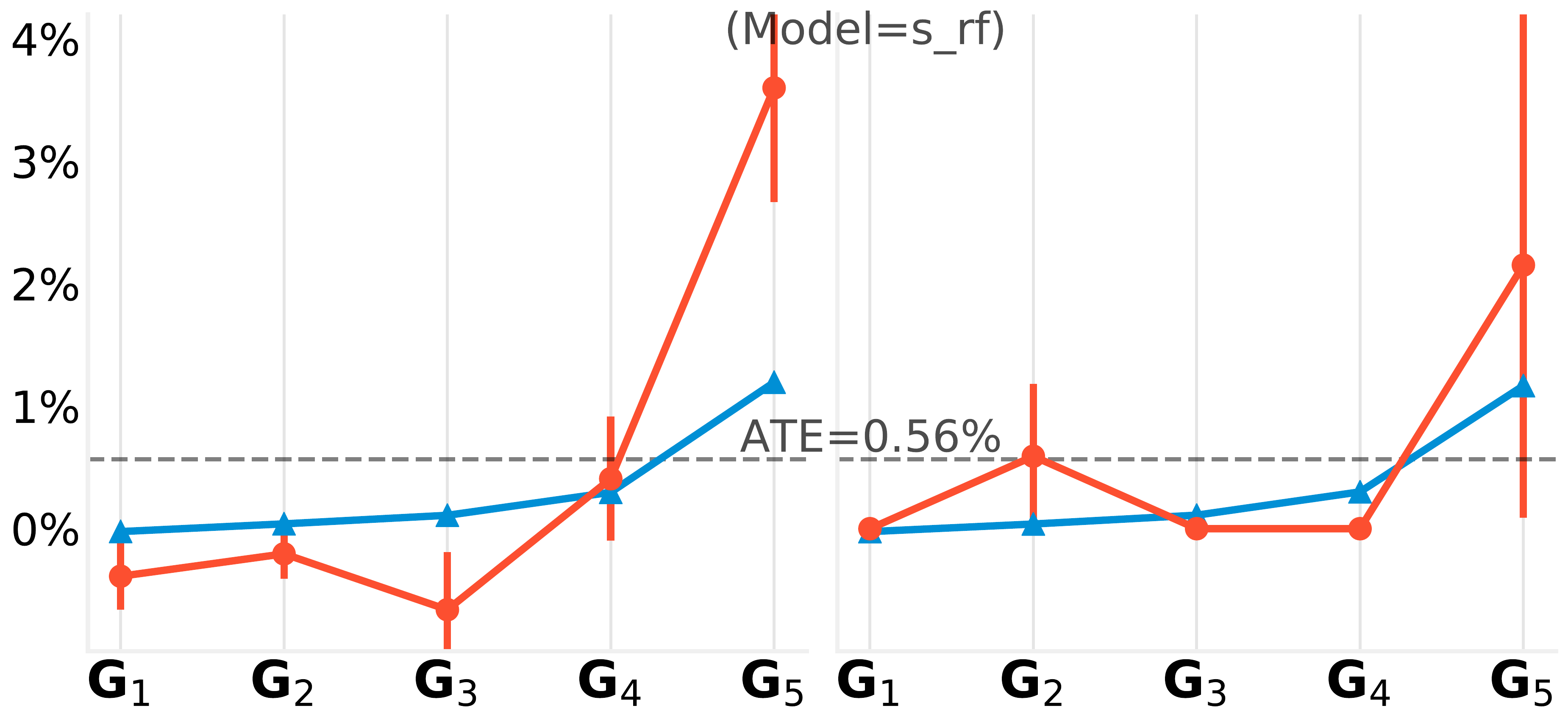}\\
    (a) GI event & (b) CVT event \\
    \end{tabular}%
    }
    \caption{Investigating the monotonicty trend (\cref{eq:monotonicity_trends}) for two CATE estimators X\_RF and S\_RF on one set of three training folds and one validation fold of the original 4-fold split \cvorig, for \textbf{(a)} the GI Event, and \textbf{(b)} the CVT Event. Here ``Model CATE" refers to the quantity $\overline{\model}_{\group_{\j} \cap \tset}$, and Neyman CATE refers to the quantity $\gatehat[{\group_{\j} \cap \tset}]$. In our notation, for training folds $\tset=\trainfolds$, and for validation fold $\tset=\valfold$. The error bars for Model CATE are the sample standard deviation for the estimated CATE values from the model, for each subgroup. For the Neyman CATE, the error bar denotes the square-root of the estimated variance~\eqref{eq:plugin_subgroup_cate_std}. Note that the subgroups $\{\group_j\}$ are defined by the CATE estimator via the training folds.}
    \label{fig:ece_5_bins}
\end{figure}

\newcommand{\bincompare}{A}
\newcommand{\qcompare}{B}
\paragraph{Pairwise comparisons:}
To summarize this phenomenon, we do a pairwise comparison of successive quantile-based subgroups and measure the frequency with which the ordering of their CATE values generalizes to the validation fold, and summarize our results in \cref{fig:monotonicity_box_plot}(a). More precisely, for a given estimator $\model$, we define the boolean indicators:
\begin{subequations}
\begin{align}
\label{eq:bin_compare}
    \bincompare_{\j,\j+1} = \mb I(\gatehat[\group_\j \cap \valfold] \leq \gatehat[\group_{\j+1} \cap \valfold])
    \qtext{for} \j=1, 2, 3, 4.
\end{align}
We then compute how often we have $\bincompare_{\j,\j+1}=1$ over the 12 validation folds 4 each from the 3 CV splits \textbrace{\cvorig,\cvzero,\cvone}, and denote this value by $\overline{\bincompare}_{\j, \j+1}$. 
Finally, we provide a box-plot of the distribution of the values $\braces{\overline{\bincompare}_{\j, \j+1}, \j=1, 2, 3, 4}$ across all 17 CATE estimators for the GI event, and 15 CATE estimators for the CVT event in panel (a) of the \cref{fig:monotonicity_box_plot}.
A value close to $1$ suggests good generalization, and conversely, a value close to 0 reflect poor generalization.
On the one hand, we see that the pairwise ordering does not generalize well for most pairs of successive quantile-based subgroups as the frequency of generalization $\overline{\bincompare}_{\j, \j+1}$ concentrates around values $\leq 0.5$ for $\j=2, 3, 4$ for the GI event, and $\j=1, 2, 3$ for the CVT event.
On the other hand, we see that values of $\overline{\bincompare}_{1, 2}$ for the GI event, and those of $\overline{\bincompare}_{4,5}$ for the CVT event are pretty close to $1$ (we present more precise numerical values in \cref{tab:mononocitiy}.)
This observation suggests that the ordering does generalize well for the subgroup with the strongest negative treatment effect for the GI event, and 
the strongest positive treatment effect for the CVT event.

\paragraph{Investigating the quantile-based ``top" subgroups:}  
We call the subgroups induced by $\group_1$ for the GI event, and $\group_5$ for the CVT event, the \emph{quantile-based top subgroup}. Note that each subgroup is specific to a choice of estimator, a choice of training-validation split, and a choice of quantile-grid. 
To further analyze the good generalization of ordering for these top subgroups, we also compare them to the other quantile-based subgroups via two boolean variables as follows:
\begin{align}
    \qtext{for GI event:} 
    \bincompare_{1, \min}&:=\mb{I}(\gatehat[\group_{1}\cap\valfold] = \min_{\j}\gatehat[\group_{\j}\cap\valfold]), \text{and}\label{eq:gi_1_min}\\
    \qtext{for CVT event:} \bincompare_{5, \max} &:= \mb{I}(\gatehat[\group_{5}\cap\valfold] = \max_{\j}\gatehat[\group_{\j}\cap\valfold]).\label{eq:tc5_max}
\end{align}
\end{subequations}
We report the distribution of the frequency of generalization $\overline{\bincompare}_{1, \min}$ (mean computed over the 12 validation folds) across the 17 CATE estimators for the GI event, and $\overline{\bincompare}_{5, \max}$ across the 15 CATE estimators for the CVT event as the rightmost entry of the corresponding figure in \cref{fig:monotonicity_box_plot}(a). 
The plots show that, on the validation fold, the \emph{quantile-based top subgroup} has the strongest treatment effect 90\% of the time for the GI outcome, and about 80\% of the time for the CVT outcome. 

Next, to better investigate the performance of quantile-based top subgroups, we compare these top subgroups directly against their complement, reporting the results in \cref{fig:monotonicity_box_plot}(b). In this plot, we also vary the $\qvalue$-value threshold used to define the quantile-based top subgroup. In particular, we consider groups of the form 
\begin{align}
\label{eq:top_quantile_group}
    \wtilde{\group}_{\qvalue} =  \braces{x \in \fspace \vert \model(x) \in (-\infty, \bin[\qvalue]]}
\end{align}
where $\bin[\qvalue]$ denotes the $q$-th quantile of the CATE estimator $\model$ on
the training folds (see \cref{eq:quantile_value} for the mathematical expression).
Note that with this notation, $\wtilde{\group}_{\qvalue}^c = \braces{x \in \fspace \vert \model(x) \in ( \bin[\qvalue], \infty}$.
In simple words, the subgroup $\wtilde{\group}_{\qvalue}$ is based on the quantile range $[0, \qvalue]$, and its complement subgroup $\wtilde{\group}_{\qvalue}^c$ is based on the quantile-range $[\qvalue, 1]$.
Then we check the ordering for between these subgroups via the following boolean indicators:
\begin{align}
\label{eq:max_compare}
    \qcompare_{\qvalue} &= \mb I\parenth{\gatehat[\displaystyle\wtilde{\group}_{\qvalue}\cap\valfold] \leq \large \gatehat[\displaystyle\wtilde{\group}_{\qvalue}^c\cap\valfold]},
    \qtext{for}
    \begin{cases}
     \qvalue \in \braces{0.1, 0.2, \ldots, 0.5} \qtext{for GI event}\\
     \qvalue \in \braces{0.9, 0.8, \ldots, 0.5} \qtext{for CVT event.}
     \end{cases}
\end{align}
Note that the subgroup of interest is $\group_{\qvalue}$ for the GI event and $\group_{\qvalue}^c$ for the CVT event. Moreover, in this new notation, the earlier subgroups (from \cref{fig:monotonicity_box_plot}(a)) would be represented as $\group_1 = \wtilde{\group}_{0.2}$ and $\group_5 = \wtilde{\group}_{0.8}^c$.
We notice that the ordering~\eqref{eq:max_compare} holds much more frequently (compared to the pairwise ordering in \cref{fig:monotonicity_box_plot}(a)). We also note from this figure that $\qvalue=0.2$ and $\qvalue=0.8$ provide the best generalization performance  for the GI and CVT events respectively.

\begin{figure}
    \centering
    \resizebox{\textwidth}{!}
    {
    \begin{tabular}{cc}
    \includegraphics[width=0.5\textwidth]{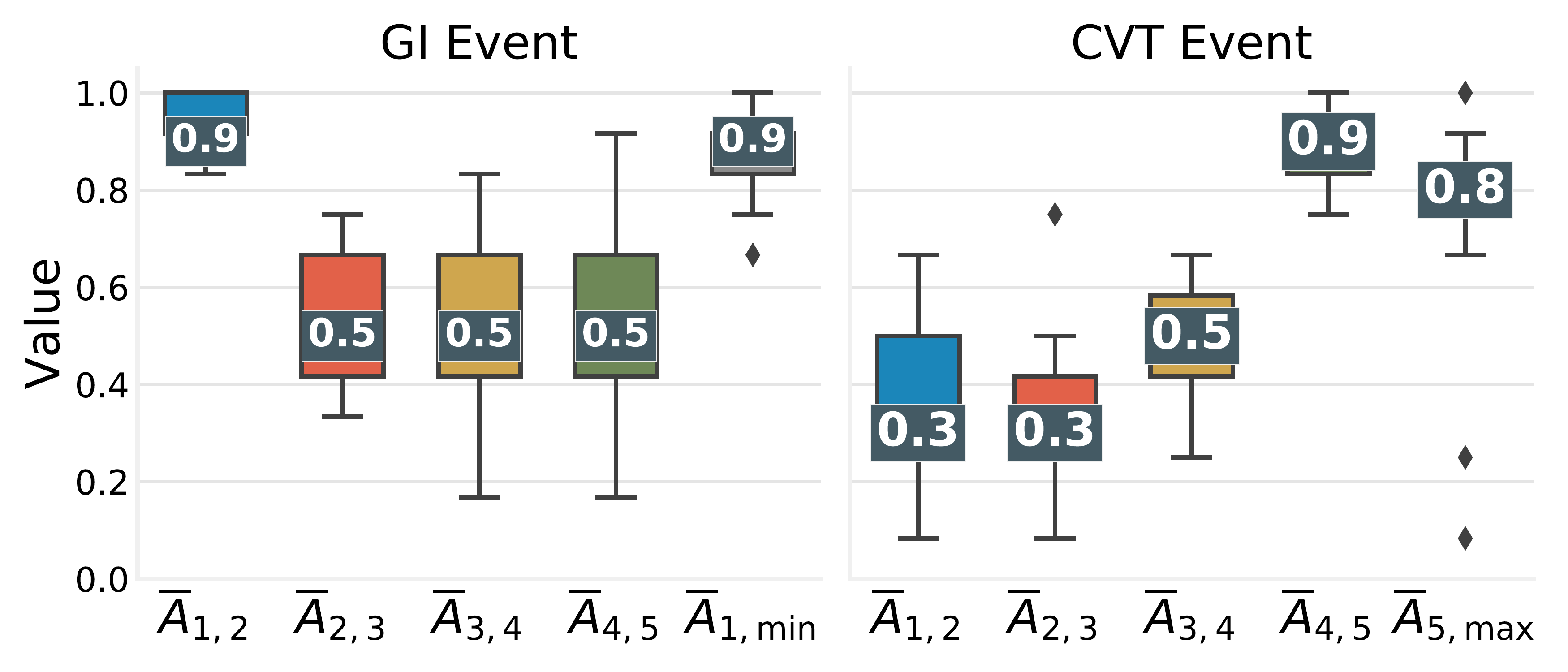} 
    & \includegraphics[width=0.5\textwidth]{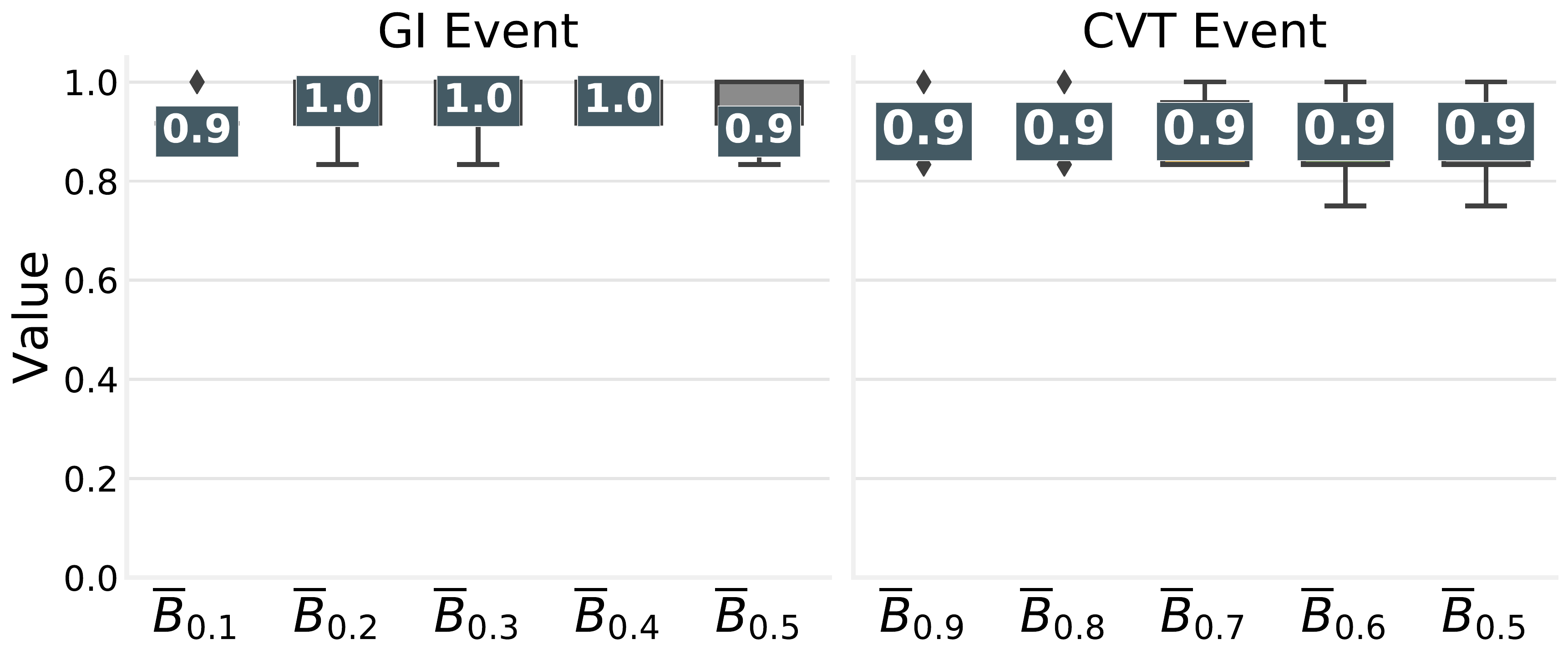}
    \\
    (a) & (b)
    \end{tabular}
    }
    \caption{Box plots for pairwise comparisons of the subgroup CATE estimates for the 5 quantile-based subgroups based on the quantile grid $\braces{0.2, 0.4, 0.6, 0.8}$.
    The boxplots in panel \textbf{(a)}, denote the distribution for the mean fraction $\overline{\bincompare}_{\j, \j+1}$~\eqref{eq:bin_compare} (where the mean is computed over the 12 validation folds, 4 each from the 3 random CV splits \textbrace{\cvorig,\cvzero,\cvone}) across various CATE estimators, for the GI event on the left, and CVT event on the right.
    In addition, we also show the boxplot of the distribution of the boolean variables $\overline{\bincompare}_{1,\min}$~\eqref{eq:gi_1_min} for the GI event, and $\overline{\bincompare}_{5, \max}$~\eqref{eq:tc5_max} in the rightmost column of respective plot.
     In panel \textbf{(b)}, we provide boxplots for the distribution of the mean value of boolean indicators $\{\overline{\qcompare}_{\qvalue}$~\eqref{eq:max_compare} across all CATE estimators, for $ \qvalue\in \braces{0.1, 0.2, \ldots, 0.5}$ for the GI event, and $ \qvalue\in \braces{0.9, 0.8, \ldots, 0.5}$ for the GI event, where the mean is computed over the and the distribution is plotted across all the CATE estimators.
      Refer to \cref{tab:mononocitiy} for estimator-wise results.}
    \label{fig:monotonicity_box_plot}
\end{figure}

In summary, we have found that at least some of the CATE estimators yield quantile-based top subgroups that have subgroup CATE that is demonstrably stronger than that of the rest of the population.
Thus, in the following sections, we use these quantile-based top subgroups, namely the subgroups $\braces{\group_{\qvalue}, \qvalue=0.1, 0.2, \ldots, 0.5}$ for the GI event, and  $\braces{\group_{\qvalue}^c, \qvalue=0.9, 0.8, \ldots, 0.5}$ for the CVT event for further analysis.

\section{Stability-driven ranking and aggregation of CATE estimators}
\label{sec:stability_driven_ranking}
Based on the discussion at the end of the last section, we believe that we can use a sub-collection of the CATE estimators to find subgroups with highly negative (in the case of the GI outcome) or positive (in the case of the CVT outcome) subgroup CATE, in the form of a quantile-based top subgroup. This observation brings us back to the question of estimator screening and choice: We seek to define a more stringent predictive test, and furthermore, out of all CATE estimators we considered, we would like to select those that are able to give us the best subgroups. While the overall goal of StaDISC is to find subgroups that are both statistically significant and interpretable, we focus in this part of paper on selecting estimators that yield the most significant subgroups, and only address interpretability in \cref{sec:finding_interpretable_subgroups}.

\subsection{Comparing estimators using $t$-statistics}
We  compare different CATE estimators using the statistical significance of their quantile-based top subgroup, measured via using standardized scores, namely $t$-statistics.
\begin{subequations}
Given a subgroup $\group$, its corresponding $\tstattext$ is given by:
\begin{align}
\label{eq:t_statistics}
    \zscore_{\group} &\defn \frac{\gatehat - \atehat}{\sqrt{\varhat\brackets{\gatehat\! -\! \atehat~\big\vert~\grouplabels}}}.
\end{align}
Here, the term in the denominator is a plug-in estimate of a conditional variance, where the conditioning is over a $\sigma$-algebra $\grouplabels$ comprising knowledge of the group labels and treatment labels for all individuals in the sample population. More precisely, the variance estimate is given by 
\begin{align}
\label{eq:plugin_subgroup_cate_std}
    \varhat\brackets{\gatehat\! -\! \atehat~\big\vert~\grouplabels} &\defn
    \parenth{\frac{ \abs{\group^c \cap \control}}{\abs{\control}}}^2 \cdot \parenth{\frac{\varhat\brackets{Y(0)~\big\vert~\group \cap \control}}{\abs{\group \cap \control} } + \frac{\varhat\brackets{Y(0)~\big\vert~\group^c \cap \control}}{\abs{\group^c \cap \control} }}\notag \\
    & + \parenth{\frac{ \abs{\group^c \cap \treat}}{\abs{\treat}}}^2 \cdot \parenth{\frac{\varhat\brackets{Y(1)~\big\vert~\group \cap \treat}}{\abs{\group \cap \treat} } + \frac{\varhat\brackets{Y(1)~\big\vert~\group^c \cap \treat}}{\abs{\group^c \cap \treat} }},
\end{align}
where for a given set $\mc A \subset \mc S$, the quantity $\varhat\brackets{Y(t)~\big\vert~\mc A}$ denotes the sample variance:
\begin{align}
    \varhat\brackets{Y(t)\big\vert\mc A} =  \frac{1}{\abs{\mc A} - 1}\sum_{i \in \mc A} \parenth{Y_i(t)- \frac{1}{\abs{\mc A}}\sum_{j\in \mc A}Y_j(t) }^2\qtext{for} t = 0, 1.
\end{align}
\end{subequations}
We show in \cref{sec:t-stat_var} that the estimator~\eqref{eq:plugin_subgroup_cate_std} is an unbiased estimator of the conditional variance of $\gatehat-\atehat$, and from the proof, it also easily follows that the estimator is consistent. As such, under the null hypothesis that $\gate - \ate = 0$, the $\tstattext$ yields an asymptotically valid $p$-value.

In this paper, we deliberately choose not to use $p$-values to report the results, so as to avoid their susceptibility to misinterpretation. For interested readers, however, we mention the mapping between $p$-values and $t$-statistics ($\zscore$). The \tstattext s presented throughout this work can be associated with one-sided $p$-values. In particular, a negative $t$-statistic with magnitude $1.65, 1.96$, and $2.33$ can be mapped to a left one-sided $p$-value of $0.05$, $0.025$ and $0.01$ respectively. The same mapping exists between positive \tstattext s and right one-sided $p$-values.

\subsection{Defining appropriate perturbations}
\label{sub:all_perturbations}
In order to guard against spurious and unreliable discoveries, the Stability principle of the PCS framework requires conclusions to be stable to reasonable or appropriate perturbations at various stages of the data science life cycle. These include modeling and data perturbations familiar to statisticians which are appropriate under the Neyman-Rubin model assumptions, and also ``judgment call'' perturbations where we reproduce or at least approximate the conclusions that would have been reached had various contingent choices been made differently. Examples of these choices include those made during data cleaning and feature engineering.\footnote{This concern is similar to that expressed by Gelman in his influential paper on \textit{The Garden of Forking Paths}~\cite{Gelman2013}.}

As mentioned earlier in the paper, we have used a random CV split in order to fit and analyze our CATE models for the VIGOR data. 
In line with our prior discussion, we do not just evaluate each estimator based on the 3 CV splits \textbrace{\cvorig,\cvzero,\cvone}, but also perform concurrent analyses of the estimator fitted and validated using four-fold splits of the data under 4 additional perturbations.  Overall, we denote the set of all 7 perturbations by $\{$\texttt{cv\_orig, cv\_0, cv\_1, cv\_time, elderly\_60, overweight, pert\_outcome}$\}$, where the 3 (random) CV splits \textbrace{\cvorig,\cvzero,\cvone} have already been used multiple times in the previous results of our paper.
For completeness and to put them in context here, we revisit them while introducing the \emph{new} perturbations \textbrace{\texttt{cv\_time, elderly\_60, overweight, pert\_outcome}} that we make use of in our subsequent analysis of the VIGOR dataset.
We remind the reader that for each perturbation, we perform the same 4-fold split for all the CATE estimators. Moreover, we continue to use the tuned hyperparameters from \cvorig\ for all other perturbations.

\paragraph{Sampling perturbations (cv\_0, cv\_1, cv\_time):} The additional CV (random) splits $\{$\texttt{cv\_0, cv\_1}$\}$, used earlier and also in the sequel, help to account for sampling variability and are pretty commonly used in statistics and machine learning.
Nonetheless, we also share Efron's concern that the use of random splits~\cite{Efron2020} does not play well with possible covariate shift, and may lead researchers to be overly optimistic about conclusions that do not have external validity. To address this, we also split the training data into four equally-sized folds by binning based on enrollment-time, denoted by $\{$\texttt{cv\_time}$\}$. This simulates possible variability in the sample population due to human choices (i.e. the date of the RCT)\footnote{In fact, such a time-based split would be even more relevant for studies based on RCTs that are \emph{online} in nature, meaning that during the trial, results from earlier stages of the trial are used to guide whether the trial would be continued further or concluded.}, and can also be seen more generally as making use of an a priori irrelevant variable to create heterogeneous folds and thus penalize ephemeral predictors.

\paragraph{Feature engineering perturbations (elderly\_60, overweight, pert\_outcome):}
We use alternative thresholds to create perturbed versions of the ELDERLY and OBESE features. Instead of thresholding the adjusted age at 65, we create an ELDERLY\_60 feature by thresholding it at 60, and instead of thresholding BMI at 30, we instead threshold it at 25 to define the feature OVERWEIGHT. In this way, we create two perturbed datasets, denoted by $\{$\texttt{elderly\_60, overweight}$\}$. Finally, for both the GI and CVT outcomes, the VIGOR study recorded for each patient both whether an event occurred, and also whether the occurred event was confirmed (meaning that it met the stringent criteria of an independent panel). In the original study, and thus far in our paper, we have used the confirmed events as the response of interest, but we now make use of the unconfirmed events to create a new response variable tracking all events. This increases the number of GI events from 177 to 190 and the number of CVT events from 59 to 84. Replacing the original responses with these one creates a further perturbed dataset for each outcome, which we denote by $\{$\texttt{pert\_outcome}$\}$. For the three perturbations $\{$\texttt{elderly\_60, overweight, pert\_outcome}$\}$, we use the original 4-fold split \cvorig\ of the patients (albeit with the perturbed features or outcomes in the data).

Performing our analyses on these perturbed datasets reveals to us what would have happened had we, or the original study authors, made different contingent decisions in feature engineering or problem formulation. Although models fit on these datasets no longer have exactly the same meaning as those fit on the original data, we still expect the estimators that perform well on the original data to also perform well on these perturbed datasets.

\subsection{Ranking and aggregation of CATE estimators}
\label{sub:ranking_aggregation}
In this section, we first rank the CATE estimators based on their performance across all data perturbations elaborated in the previous section. And, then we select the estimators that are ranked in Top-10 estimators across all the perturbations. Finally, we build a single ``ensemble CATE estimator" by taking a simple average (equal weights) of all the selected CATE estimators. Quantile-based top subgroups of the ensemble estimator form the starting point of finding interpretable subgroups in \cref{sec:finding_interpretable_subgroups}.
We now describe the details of our ranking procedure.

\paragraph{Mean \tstattext\ per data perturbation:}
For a CATE estimator $\model$, for each data perturbation $\pert \in \{$\texttt{cv\_orig, cv\_0, cv\_1, cv\_time, elderly\_60, overweight, pert\_outcome}$\}$, we compute the mean \tstattext\ averaged across all quantiles across the corresponding 4 validation folds.
In our notation, for the GI event, this mean \tstattext\ is given by
\begin{subequations}
\label{eq:mean_t_stat}
\begin{align}
\label{eq:mean_t_stat_gi}
    \overline{\zscore}_{\trm{GI}}(\pert) = 
    \frac{1}{20}\sum_{\qvalue \in \mc Q } 
    \sum_{\valfold\in \mc F} \zscore_{\displaystyle\gitopgroup\cap \valfold}
    \text{ where } \mc Q = \braces{0.1, 0.2, \ldots, 0.5},
    \mc F = \braces{\trainfold[\fold], \fold=1,2, 3, 4},
\end{align}
where the quantile-based top subgroup $\wtilde{\group}_{\qvalue}$ was defined in \cref{eq:top_quantile_group}.
Moreover, we remind the reader that the quantiles that define the subgroup $\gitopgroup$ (see \cref{eq:quantile_value,eq:group_quantile}) are computed based on the CATE estimates from the fitted $\model$ on its training folds $\trainfolds =\trainset \backslash \valfold$. On the other hand, the \tstattext\ on the RHS of \cref{eq:mean_t_stat_gi} is computed on the validation fold $\valfold$.
For the CVT event, the corresponding mean \tstattext\ is given by
\begin{align}
\label{eq:mean_t_stat_tc}
    \overline{\zscore}_{\trm{CVT}}(\pert) = \frac{1}{20}\sum_{\qvalue \in \mc Q } \sum_{\substack{\valfold\in \mc F}}\zscore_{\displaystyle\tctopgroup\cap \valfold}
    \text{ where } \mc Q = \braces{0.9, 0.8, \ldots, 0.5},
    \mc F = \braces{\trainfold[\fold], \fold=1,2, 3, 4},
\end{align}
\end{subequations}
We report the mean \tstattext\ $\overline{\zscore}(\pert)$ for each CATE estimator and all 7 data perturbations in \cref{tab:t_scores}(a) for the GI event, and \cref{tab:t_scores}(b) for the CVT event.
We also provide a visual summary of the 7 mean \tstattext\ for each estimator in the form of boxplot in \cref{fig:rank_and_tstat} in panel (a) for the GI event, and panel (b) for the CVT event.

\paragraph{Ranking the CATE estimators:}
Next, for each category $\pert$, we rank the mean \tstattext\ from lowest to highest for the GI event, and highest to lowest for the CVT event. In accordance with the Stability principle of the PCS framework, we screen for estimators that perform well across perturbations, and thereby select all estimators that rank in Top-10 across all data perturbations $\pert$. We provide the visual illustration of these ranks also in \cref{fig:rank_and_tstat} for the two events. In fact, the estimators in the \cref{fig:rank_and_tstat} are sorted based on their worst rank across the perturbations. This criterion selects (i) 2 T-learners and 4 X-learners $\{$t\_lasso, x\_rf,  t\_rf, x\_xgb, x\_lasso, x\_logistic$\}$ for the GI event, and (ii)  1 S-learner, 3 T-learners, and 1 X-learners $\{$s\_rf, t\_lasso, t\_rf, x\_xgb, t\_logistic$\}$ for the CVT event. The selected list can also be verified by a simple inspection of the rank plots from \cref{fig:rank_and_tstat}.

\newcommand{\ttt}[1]{\texttt{#1}}
\begin{table}
    \centering
    \resizebox{1.0\textwidth}{!}{
    \begin{tabular}{c}
    \rowcolors{2}{gray!10}{white}
    \begin{tabular}{lrrrrrrr}
        \toprule
        Perturbation $\pert$ &  \ttt{cv\_orig} &  \ttt{cv\_0} &  \ttt{cv\_1} &  \ttt{cv\_time} &  \ttt{elderly\_60} &  \ttt{overweight} & \ttt{pert\_outcome} \\
        Estimator $\model$ &  \multicolumn{7}{c}{$\overline{\zscore}_{\trm{GI}}(\pert)$} \\
        \midrule
      \ttt{t\_lasso}         &      -1.27 &      -1.79 &     \bf -1.52 &         -1.36 &            -1.36 &            -1.02 &         -1.24 \\
\ttt{x\_rf}            &      -1.24 &      -1.84 &      -1.37 &       \bf  -1.58 &            -1.40 &            -1.22 &         -1.38 \\
\ttt{t\_rf}            &      -1.25 &      -1.62 &      -1.39 &         -1.34 &            -1.34 &            \bf -1.24 &     \bf    -1.43 \\
\ttt{x\_xgb}           &      -1.16 &      -1.80 &      -1.44 &         -1.45 &            -1.31 &            -1.11 &         -1.10 \\
\ttt{x\_lasso}         &      -1.23 &     \bf -1.88 &      -1.49 &         -1.33 &            -1.28 &            -1.04 &         -1.15 \\
\ttt{x\_logistic}      &      -1.31 &      -1.86 &      -1.39 &         -1.26 &            -1.31 &            -0.96 &         -1.06 \\
\ttt{r\_lassorf}       &      -1.26 &      -1.34 &      -1.36 &         -1.56 &        \bf    -1.63 &            -0.95 &         -0.96 \\
\ttt{t\_logistic}      &     \bf  -1.33 &      -1.72 &      -1.56 &         -1.14 &            -1.27 &            -1.17 &         -1.19 \\
\ttt{r\_rfrf}          &      -1.24 &      -1.45 &      -1.33 &         -1.51 &            -1.50 &            -1.00 &         -0.84 \\
\ttt{causal\_forest\_2} &      -1.00 &      -1.32 &      -1.39 &         -1.23 &            -1.22 &            -0.94 &         -0.92 \\
\ttt{t\_xgb}           &      -1.02 &      -1.73 &      -1.18 &         -1.31 &            -1.38 &            -1.01 &         -1.34 \\
\ttt{r\_lassolasso}    &      -1.10 &      -1.76 &      -1.25 &         -1.19 &            -1.19 &            -1.07 &         -0.76 \\
\ttt{causal\_forest\_1} &      -0.97 &      -1.26 &      -1.25 &         -1.10 &            -1.07 &            -0.84 &         -1.32 \\
\ttt{s\_xgb}           &      -0.95 &      -1.35 &      -1.57 &         -0.99 &            -1.02 &            -0.90 &         -0.99 \\
\ttt{causal\_tree\_1}   &      -0.67 &      -1.22 &      -0.98 &         -0.50 &            -0.66 &            -0.80 &         -0.46 \\
\ttt{causal\_tree\_2}   &      -1.07 &      -0.87 &      -0.72 &         -0.96 &            -1.09 &            -0.88 &         -0.64 \\
\ttt{s\_rf}            &      -0.78 &      -1.44 &      -0.81 &         -1.19 &            -1.33 &            -0.59 &         -1.12 \\
        \bottomrule
    \end{tabular} \vspace{2mm}
    \\
    (a)  {\large GI Event}  \vspace{5mm}\\
    \rowcolors{2}{gray!10}{white}
    \begin{tabular}{lrrrrrrr}
        \toprule
         Perturbation $\pert$ &  \ttt{cv\_orig} &  \ttt{cv\_0} &  \ttt{cv\_1} &  \ttt{cv\_time} &  \ttt{elderly\_60} &  \ttt{overweight} & \ttt{pert\_outcome} \\
        Estimator $\model$ &  \multicolumn{7}{c}{$\overline{\zscore}_{\trm{CVT}}(\pert)$} \\
        \midrule
\ttt{s\_rf}            &       0.96 &      \bf 1.29 &     \bf  1.17 &     \bf     1.42 &      \bf       1.29 &             1.05 &          1.26 \\
\ttt{t\_lasso}         &       1.06 &       1.16 &       0.99 &          1.02 &             1.10 &             1.07 &          1.14 \\
\ttt{t\_rf}            &    \bf   1.10 &       1.19 &       0.90 &          1.25 &             1.24 &          \bf   1.18 &   \bf   1.45 \\
\ttt{x\_xgb}           &       1.01 &       1.15 &       0.89 &          1.03 &             1.08 &             1.04 &          1.11 \\
\ttt{t\_logistic}      &  \bf     1.10 &       1.16 &       1.03 &          1.17 &             1.17 &             0.93 &          1.02 \\
\ttt{x\_logistic}      &       0.97 &       1.11 &       0.87 &          0.94 &             1.14 &             0.92 &          1.01 \\
\ttt{x\_rf}            &       0.90 &       1.11 &       0.88 &          0.91 &             1.09 &             0.99 &          1.02 \\
\ttt{x\_lasso}         &       0.92 &       1.13 &       0.80 &          0.90 &             1.10 &             0.94 &          1.03 \\
\ttt{t\_xgb}           &       0.66 &       1.06 &       0.92 &          1.26 &             0.95 &             0.66 &          1.26 \\
\ttt{r\_rfrf}          &       0.86 &       1.12 &       0.70 &          1.01 &             0.88 &             0.96 &          0.97 \\
\ttt{r\_lassorf}       &       0.79 &       1.14 &       0.75 &          0.93 &             0.86 &             1.03 &          0.81 \\
\ttt{r\_lassolasso}    &       0.81 &       1.01 &       0.65 &          0.61 &             1.01 &             0.84 &          0.98 \\
\ttt{causal\_tree\_2}   &       0.67 &       0.88 &       0.84 &         -0.33 &             0.64 &             0.49 &          1.28 \\
\ttt{causal\_forest\_1} &       0.93 &       1.14 &       0.96 &          0.74 &             0.58 &             0.64 &          0.71 \\
\ttt{causal\_forest\_2} &       0.46 &       0.72 &       0.87 &          0.55 &             0.56 &             0.96 &          1.12 \\
        \bottomrule
    \end{tabular}\vspace{2mm}
    \\
     (b) {\large CVT Event}
     \end{tabular}
     }
    \caption{Estimator- \emph{and} perturbation-wise \tstattext~$\overline{\zscore}_{\trm{GI}}(\pert)$~\eqref{eq:mean_t_stat_gi} for the GI event in panel~\textbf{(a)}, and $\overline{\zscore}_{\trm{CVT}}(\pert)$~\eqref{eq:mean_t_stat_tc} for the CVT event in panel~\textbf{(b)}.  In each column the best (lowest for GI event, highest for CVT event) \tstattext\ is highlighted in bold. The order of the estimators in panel (a) and (b) is the same order as that in \cref{fig:rank_and_tstat}(a) and \cref{fig:rank_and_tstat}(b) respectively. }
    \label{tab:t_scores}
\end{table}

\begin{figure}
    \centering
    \resizebox{\textwidth}{!}{
    \begin{tabular}{c}
      \includegraphics[width=\textwidth]{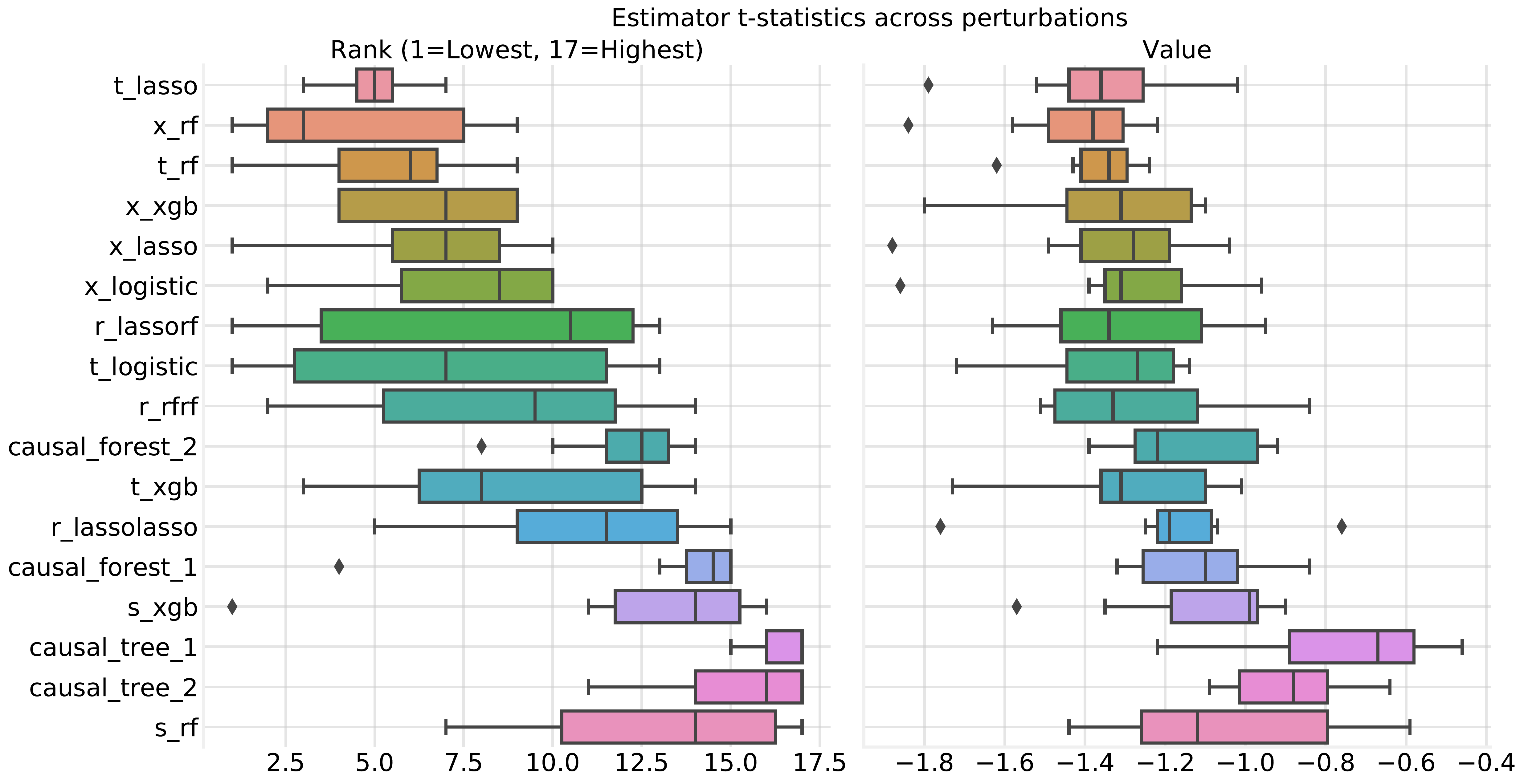}    \\
      (a) GI Event \\
      \addlinespace[1em]
      \midrule
      \addlinespace[1em]
      \includegraphics[width=\textwidth]{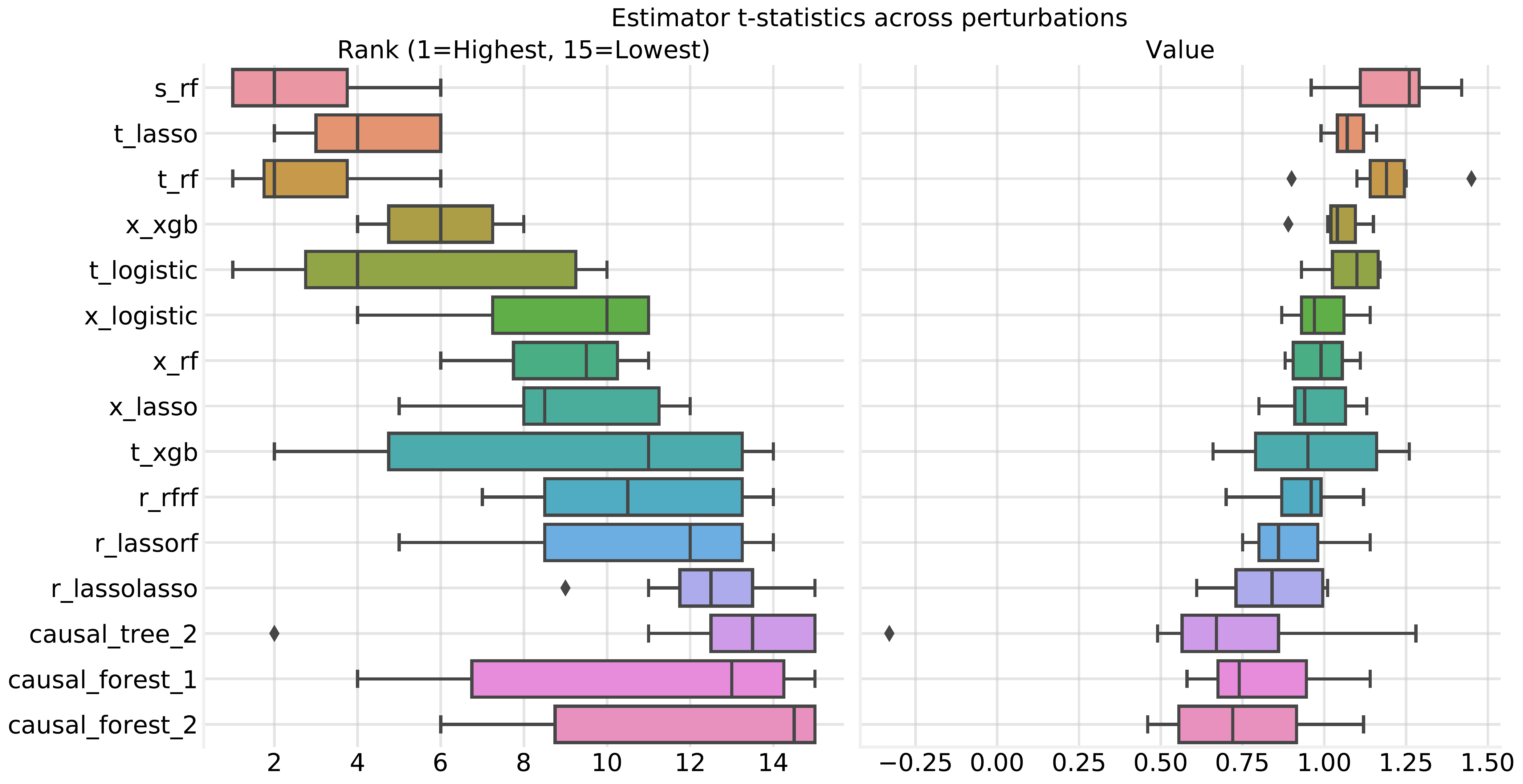}   \\
      (b) CVT Event
    \end{tabular}
    }
    \caption{Box plots of the rank and value of mean $t$-statistic scores~$\overline{\zscore}_{\trm{GI}}(\pert)$~\eqref{eq:mean_t_stat_gi}, and $\overline{\zscore}_{\trm{CVT}}(\pert)$~\eqref{eq:mean_t_stat_tc}, where the distribution is over the 7 data perturbations  $\pert \in \{$\texttt{cv\_orig, cv\_0, cv\_1, cv\_time, elderly\_60, overweight, pert\_outcome}$\}$.
    Here rank for the mean $t$-statistic score is computed per perturbation $\pert$, and all CATE estimators are ranked lowest to highest for the GI event, and highest to lowest for the CVT event. The estimator- and perturbation-wise numbers for both panels are reported in \cref{tab:t_scores}.}
    \label{fig:rank_and_tstat}
\end{figure}

\paragraph{Final step before interpreting:}
Keeping in mind the computational aspects of the next step (finding interpretable subgroups), and to increase stability, we decided to build an ensemble CATE estimator by using a simple average of the selected CATE estimators. Moreover, we also investigate the performance of the quantile-based top subgroups for this ensemble, and report the mean \tstattext\ across the 12 validation folds from \textbrace{\cvorig,\cvzero,\cvone} for $\wtilde{\group}_{\qvalue}$~\eqref{eq:top_quantile_group} for the GI event, and $\wtilde{\group}_{\qvalue}^c$ for the CVT event in \cref{tab:ensemble_quantile_t_scores}. We report the standard deviation of the \tstattext\ across these folds in parentheses. 
In addition, we also report the mean percentage overlap computed pairwise across the entire training set $\trainset$ for the 12 ensemble estimators, 4 each from the 3 CV splits \textbrace{\cvorig,\cvzero,\cvone}.
We observe that for the GI event the subgroups corresponding to $\qvalue \in \braces{0.2, 0.3}$ have relatively higher $\zscore$, and for the CVT event $\qvalue\in\braces{0.9, 0.8}$ are the top 2 choices.
The trends for overlap are as expected, with the increase in size of the group, the overlap generally increases; and remains $>70\%$ across all choices. 
In the next section, we discuss our methodology to find an interpretable representation of the quantile-based top subgroups using the ensemble CATE estimator. As a final decision before that step, we choose the groups $\wtilde{\group}_{0.2}$ and $\wtilde{\group}_{0.3}$ for the GI event, and $\wtilde{\group}_{0.9}^c$ for the CVT event, based on their high \tstattext.
We also include the group $\wtilde{\group}_{0.8}^c$ for the CVT event keeping in mind the fact that the CVT event is very rare, and thus the  low signal in the subgroup $\group_{0.9}^c$ (having only 10\% of the training data) may become a bottleneck for any reasonable inference task.
\begin{table}[ht]
    \centering
    \resizebox{\textwidth}{!}{
    \begin{tabular}{c|c}
     \begin{tabular}{ccc}
        \toprule
        {\bf   Bottom quantile} & \multicolumn{2}{c}{\bf GI Event} \\
        \bf based subgroup $\wtilde{\group}_{\qvalue}$ &$\zscore_{\wtilde{\group}_{\qvalue}}$ & Overlap
         \\
        \midrule
        $\qvalue=0.1$ &        -1.32 (0.20) &          73\% \\
        $\qvalue=0.2$ &        {\bf -1.58} (0.19) &    77\% \\
        $\qvalue=0.3$ &        -1.47 (0.16) &          82\% \\
        $\qvalue=0.4$ &        -1.02 (0.12) &          83\% \\
        $\qvalue=0.5$ &        -0.81 (0.12) &          {\bf 87}\% \\
        \bottomrule
    \end{tabular}  
    &  
    \begin{tabular}{ccc}
        \toprule
        {\bf  Top quantile } & \multicolumn{2}{c}{\bf CVT Event} \\
        \bf based subgroup $\wtilde{\group}_{\qvalue}^c$ &$\zscore_{\wtilde{\group}_{\qvalue}^c}$ & Overlap
         \\
        \midrule
        $\qvalue=0.9$ &        {\bf 1.28} (0.22) &    77\% \\
        $\qvalue=0.8$ &        1.03 (0.12) &          75\% \\
        $\qvalue=0.7$ &        0.85 (0.12) &          77\% \\
        $\qvalue=0.6$ &        0.71 (0.09) &          79\% \\
        $\qvalue=0.5$ &        0.57 (0.13) &          {\bf 82}\% \\
        \bottomrule
    \end{tabular}
    \end{tabular}
    }
    \caption{\tstattext\ for different quantile-based top subgroups of the ensemble CATE estimator. 
    ``Overlap" column reports the average \% pairwise overlap between the 12 quantile-based top subgroups on the entire training data, namely  $\gitopgroup \cap \trainset$ for the GI event, and $\tctopgroup\cap\trainset$ for the CVT event. The 12 subgroups correspond 4 each to the 3 CV splits \textbrace{\cvorig, \cvzero, \cvone}.}
    \label{tab:ensemble_quantile_t_scores}
\end{table}

\section{Finding interpretable subgroups} 
\label{sec:finding_interpretable_subgroups}

The next and final step of our investigation is to make our findings interpretable. Recall that the end goal in investigating the heterogeneous treatment effects in the VIGOR study is to inform treating physicians which subgroup of patients are likely to benefit from the reduced risk of GI events, without simultaneously incurring an increased risk of CVT events. Physicians may then favor prescribing the drug for patients in this subgroup. In situations involving high stakes decision-making such as this one, decision-makers are usually not comfortable with black-box decision rules, but instead ideally require rules to be transparent and interpretable, so as to align them with their own knowledge base, and justify them to patients and regulators.

\subsection{Interpreting using ``cells''}
\label{sub:interpreting_using_cells}

In the work by Murdoch et al.~\cite{Murdoch22071}, one of us has argued that a key element of interpretability is the notion of relevance. Interpretations need to provide  ``insight for a particular audience into a chosen domain problem.'' Since clinical decision rules usually take the form of decision trees, a decision tree is the gold standard for our problem at hand. Each leaf of a decision tree constitutes a subset of the feature space defined by constraining the values of the features occuring along the root-to-leaf path. We call such a subset of a feature space a \emph{cell}\footnote{This term is motivated by the geometric interpretation of such subsets as subcubes of the hypercube that comprises the entire feature space.}, and propose to make our quantile-based top subgroups interpretable by approximating it with a union of a few cells, which we call a \emph{cell cover}.\footnote{One may also think of this as a disjunction of conjunctions.}

Two remarks are in order. First, we find empirically that no single cell gives a good approximation of quantile-based top subgroups, so we require the additional flexibility of a union of multiple cells. Furthermore, reporting a union of cells is more flexible than reporting a decision tree, because it is not always possible to construct a tree with a given collection of cells as its leaf nodes.\footnote{For instance, leaf nodes will always involve the feature that splits the root node.} Second, by focusing on cells, we recognize the importance of interactions, or in other words, nonlinear dependence of treatment effect on the covariates. Chernozhukov et al.~\cite{Chernozhukov2017} proposed interpreting quantile-based top subgroups by estimating the differences in the ``observed characteristics'' between the quantile-based top subgroup and the subgroup that is defined to be least affected by the treatment, but this only considers the marginal importance of each feature.





\subsection{Cell-search methodology} 
\label{sub:cell_search}
In this section, we demonstrate a general framework for how to search for a cell cover that contains most of the individuals in the quantile-based top subgroup, but does not include too many individuals from outside it.

\paragraph{Feature selection:}
We start by selecting up to 10 features from the original list of 16 features. This is both to make the subsequent steps of cell search more computationally tractable, and also to act as a form of regularization.\footnote{The iterative Random Forest~\cite{basu2018iterative} algorithm for finding higher-order interactions in genomics data does soft feature selection for precisely these reasons.} To do this, we compute feature importance scores in two different ways. (i) Following Chernozhukov et al. \cite{Chernozhukov2017}, we make use of the difference between the mean of the feature values over the quantile-based top subgroup and that over its complement. 
We refer to this score as the ``Logistic" feature importance score.
(ii) We train a logistic classifier to predict membership in the quantile-based top subgroup, and make use of the coefficients. In either case, we normalize so that the absolute values of the scores sum to one. 
We refer to this score as the ``Difference" feature importance score.
We compute these two types of scores for the ensemble CATE estimators' quantile-based top subgroups selected at the end of \cref{sub:ranking_aggregation}, namely $\gitopgroup[0.2]$ and $\gitopgroup[0.3]$ for the GI outcome, and $\tctopgroup[0.9]$ and $\tctopgroup[0.8]$, across the twelve random training-validation splits (\textbrace{\cvorig, \cvzero, \cvone}).
For each outcome, we average the feature importance scores across the different splits as well as both choices of the quantile-based top subgroups. The final results are shown in \cref{fig:feature_importance}. 

Ranking the 16 features according to the two measures of feature importance, we select the features that rank among the top 8 under either measure. Note that we choose to make use of both feature importance measures because they have different meanings: While the first score measures the marginal importance of each feature, the second measures its conditional importance.
However, the choice of ``top 8" was also selected keeping in mind the fact that the top features for the two measures have a high overlap, and we end up selecting 9 and 10 features respectively for the GI and CVT events listed (alphabetically) below:
\vspace{-3mm}
\begin{align*}
    \text{\small \bf GI event: }& \text{\small CHLGRP, HYPGRP, PNAPRXN, PNSAIDS, PSTRDS, PPH, ELDERLY, OBESE, WHITE}\\
    \text{\small \bf CVT event: }& \text{\small ASCGRP, ASPFDA, CHLGRP, PPH, US, ELDERLY, MALE, OBESE, SMOKE, WHITE}
    \vspace{-3mm}
\end{align*}
Readers may refer to \cref{tab:covar_description} to remind themselves about the definitions of all the features.

\begin{figure}[ht]
    \centering
   \includegraphics[width=\textwidth]{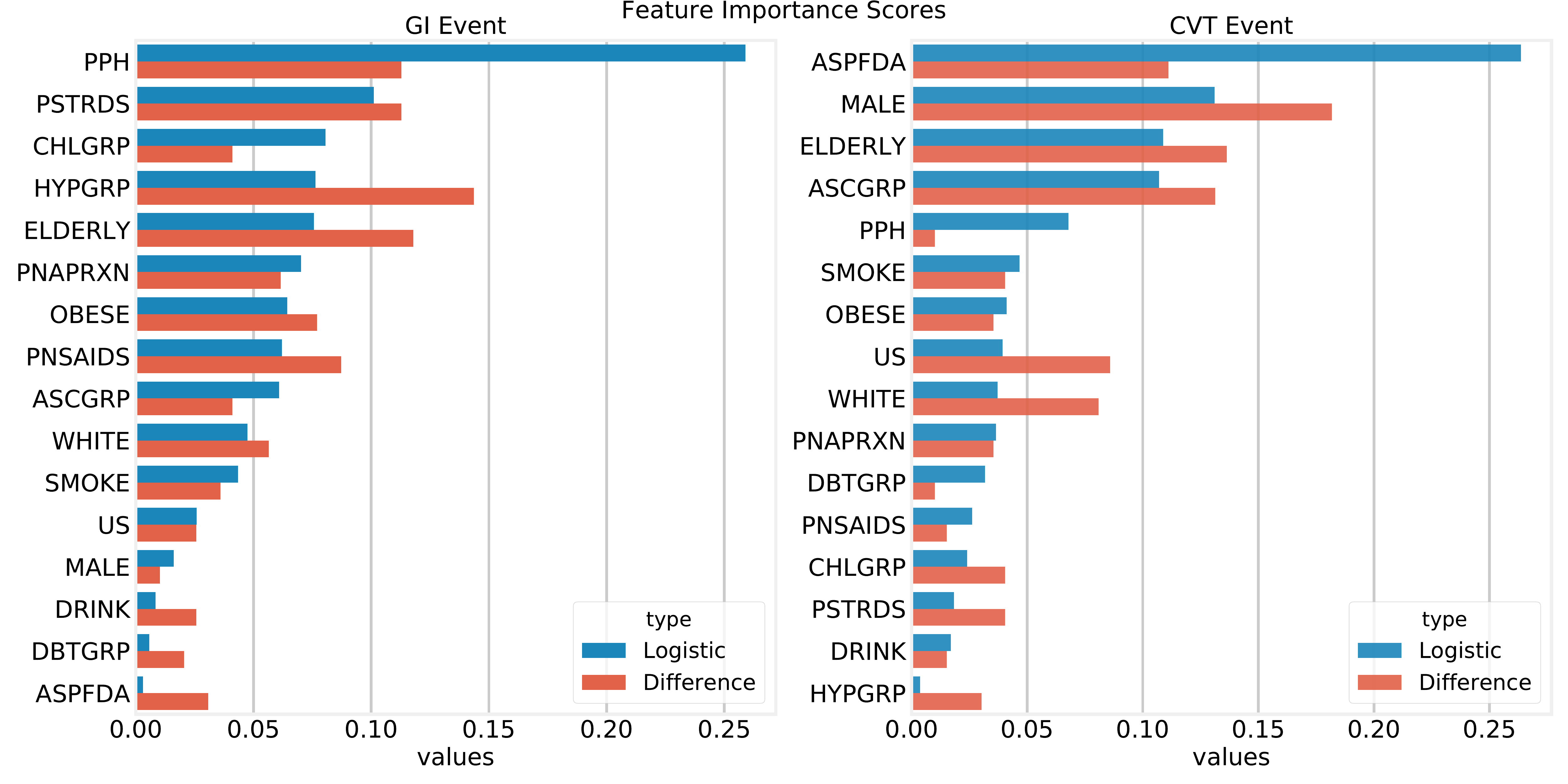}
    \caption{Mean feature importance scores  for the quantile-based top subgroups from the ensemble CATE estimator. Best seen in color. We plot both the scores next to each other for each feature with the order (top, bottom) = (logistic, difference), but separately for each outcome. The blue bars and red bars respectively denote the ``Logistic" and ``Difference" feature importance scores described in the text.}
    \label{fig:feature_importance}
\end{figure}


\paragraph{Iterative procedure:} 
We now describe the $\cellsearch$ procedure for finding the cell cover for a quantile-based top subgroup one cell at a time, with \cref{fig:cell_search} also providing a pictorial explanation. For clarity, we introduce some notation, denoting the quantile-based top subgroup by $\tbquant$, and the cell found at the $i$-th step by $\cell_i$. For GI event $\tbquant$ takes the form $\gitopgroup$, and for the CVT event $\tctopgroup$ for suitable choices of $\qvalue$. As before, we will abuse notation, using these symbols to refer to the subgroups and cells as subsets of the feature space, as well as the subpopulation of individuals that belong to them. At the first step, we consider every possible cell $\cell$ defined with $m$ features or less, where $m$ is a user-specified tuning parameter, and compute its ``true positive" (\truepos) and ``false positive" (\falsepos) values with respect to $\tbquant$ as follows:
\begin{align}
    \truepos(\cell, \tbquant) := \abs{\cell\cap\tbquant}, \text{ and }
    \falsepos(\cell, \tbquant) := \abs{\cell \cap \tbquant^c}.\footnote{We are able to compute this efficiently using the \texttt{FPGrowth} algorithm.~\cite{Han2000}}
\end{align}
Moreover, let $\Delta(\cell, \tbquant) := \truepos(\cell, \tbquant)-\falsepos(\cell, \tbquant)$ denote the difference of these values.

We rank the cells based on their difference score $\Delta(\cell, \tbquant)$, but instead of simply picking the cell achieving the largest positive value $\Delta_{\max}$, we first create a candidate list of cells for which $\Delta(\cell, \tbquant) \geq \max\paren{0,\Delta_{\max} - 0.05 \abs{\tbquant}}$, remove from cells any that are sub-cells\footnote{We say that Cell A is a sub-cell of Cell B if it is contained in Cell A when both are though as subsets of the feature space.} of other cells on this list, and then choose one of remaining cells uniformly at random. The returns on adding this layer of complexity are to favor simpler, more interpretable cells, and also (by running the procedure multiple times) to discover if two or more cells have comparable performance.\footnote{A user may wish to simply follow the greedy procedure.}

In each subsequent step of the algorithm, to find the next cell in the cell cover, we first remove from the study population all individuals belonging to the cells already found, and then repeat the above process. More rigorously, suppose cells $\cell_1, \ldots, \cell_{i-1}$ have already been determined. The true and false positive scores are now defined by
\begin{align}
    \truepos(\cell, \tbquant; \cup_{j=1}^{i-1}\cell_j) := \abs{\cell\cap\tbquant\backslash \cup_{j=1}^{i-1}\cell_j}, \text{ and }
    \falsepos(\cell, \tbquant; \cup_{j=1}^{i-1}\cell_j) := \abs{\cell \cap \tbquant^c\backslash \cup_{j=1}^{i-1}\cell_j},
\end{align}
while $\Delta_{\max}$ and the threshold are also modified accordingly. Finally, the procedure terminates if $\Delta_{\max}$ at any iteration is less than or equal to 0 or if the number of iterations has reached a pre-specified threshold (default value $3$).

\begin{figure}
    \centering
    \resizebox{\textwidth}{!}{
    \begin{tabular}{c|c}
        \includegraphics[width=0.5\textwidth]{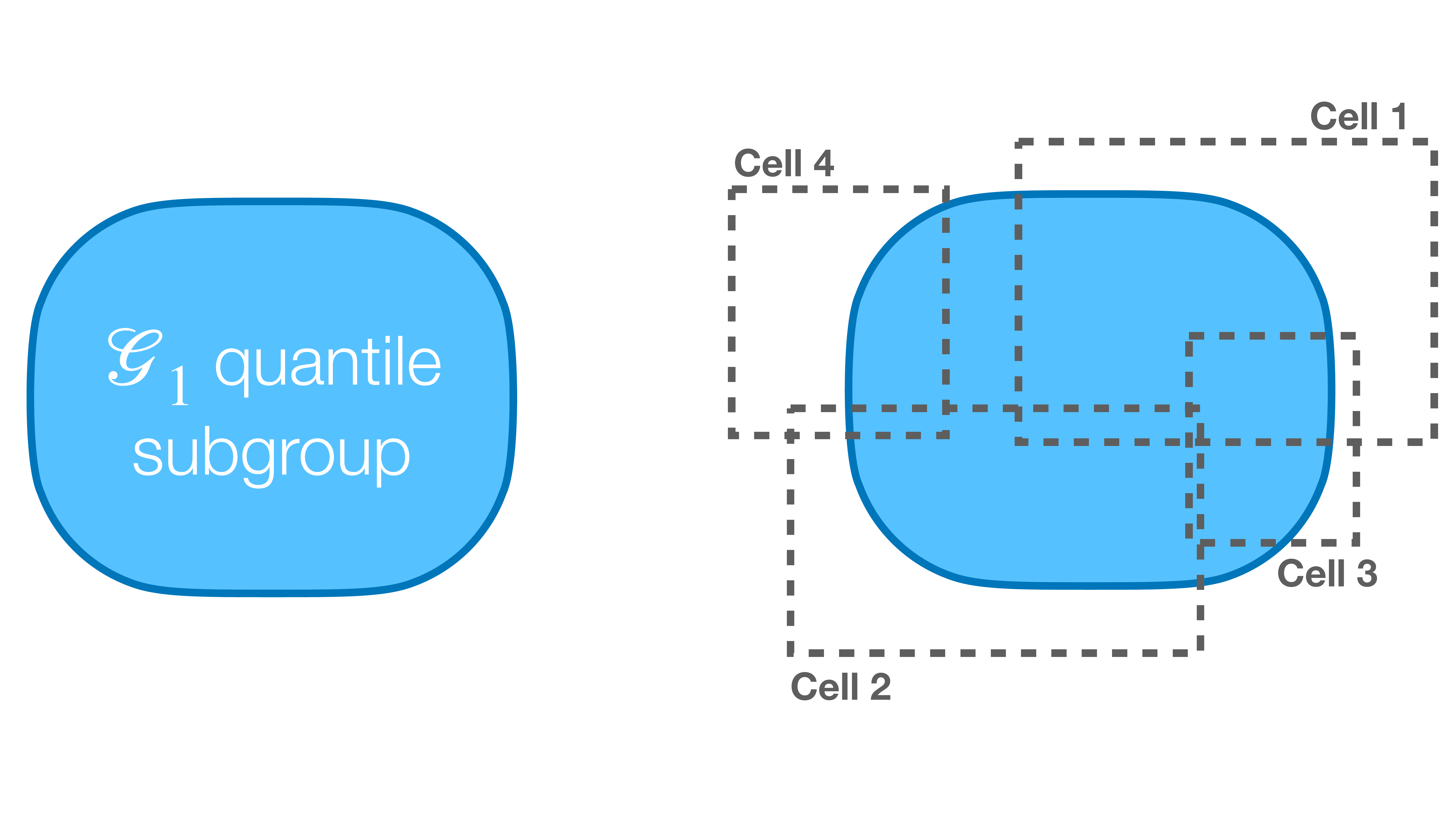} &
        \includegraphics[width=0.5\textwidth]{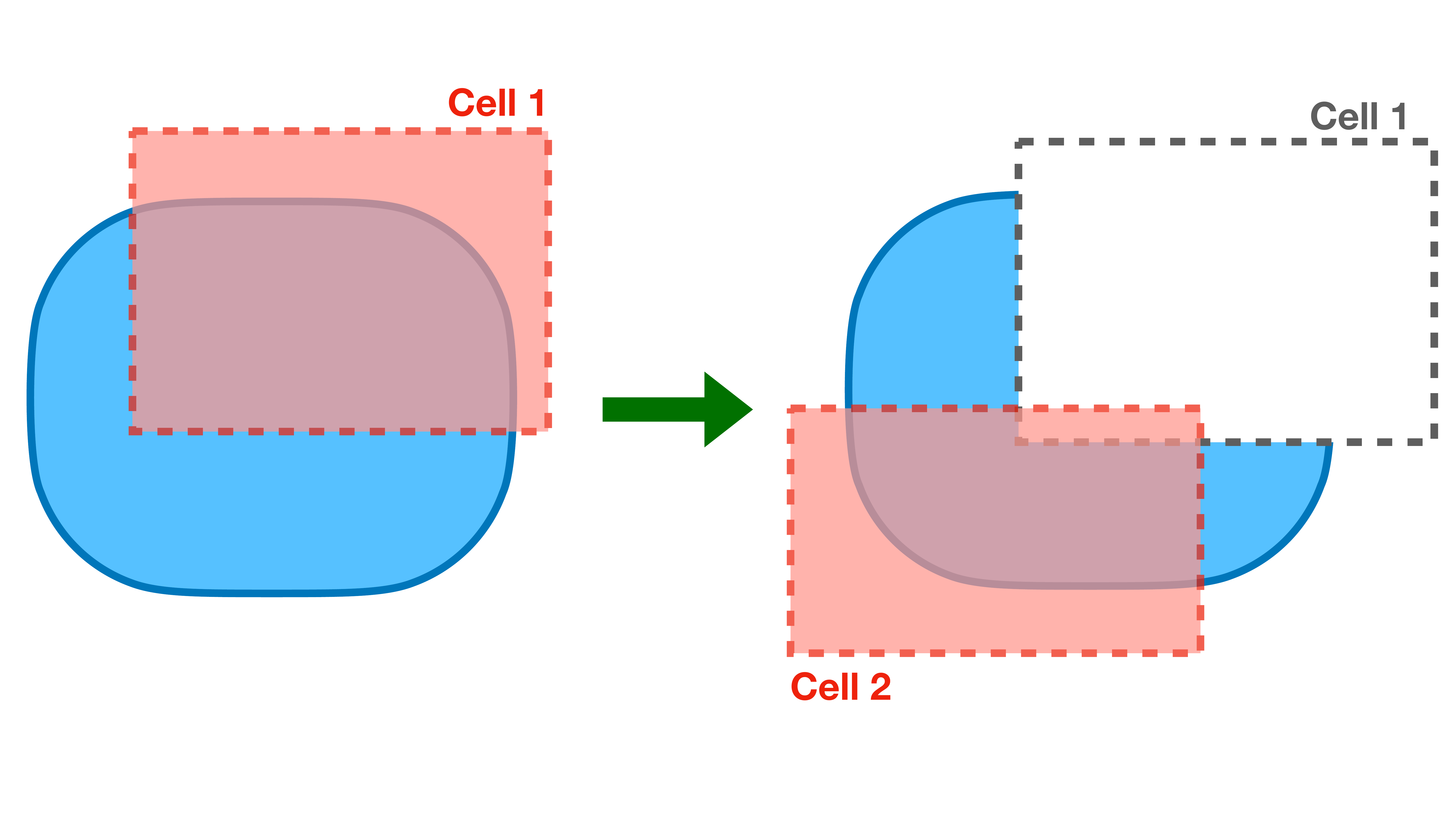} \\
        (a) Cover found by \cellsearch
        & (b) Illustration of one step of \cellsearch
    \end{tabular}
    }
    \caption{A simplified illustration of \cellsearch\ methodology for finding a cell-based cover for a given (quantile-based) subgroup.}
    \label{fig:cell_search}
\end{figure}

\paragraph{Aggregating results over multiple runs:} 
In accordance with the Stability principle, we run \cellsearch~ multiple times, and check whether the same cell cover is found. In our case, we ran it five times on each top quantile subgroup arising from 12 random training-validation splits, for a total of 60 runs. While the cell cover did not turn out to be stable, we found that certain cells or their sub-cells frequently re-appeared within each run. We thus turn our focus to individual cells, and aggregate the results over the multiple runs, calling this procedure \stabilizedcellsearch.

To describe how we aggregate the results, we first use $\mc B$ to denote the collection of all 60 runs, and for each run $b \in \mc B$, we let $\mf C_b$ denote the cover returned by the procedure, while the collection of all cells found is denoted $\mf C \coloneqq \cup_{b \in \mf B} \mf C_b$. For each cell $\cell \in \mf C_b$, we define its stability score as follows:
\begin{align}
    \stab(\cell) = \frac{1}{\abs{\mc B}}\sum_{b \in \mc B} \sum_{\cell' \in \mc C}\mathbf{1}(\cell' \in \mf C_b~\textnormal{and}~\cell'~{\textnormal{is sub-cell of}~\cell}) \frac{\abs{\cell'}}{\abs{\cell}}.
\end{align}
This score measures how frequently cell $\cell$ and its proper sub-cells are found across the different runs, with each occurrence weighted by the relative size of the sub-cell.

Finally, we rank the cells according to their stability scores, and output those for which the score exceeds a user-defined threshold. In our case, we chose the threshold to be 1/3 which results in finding 3 cells each for the GI and CVT outcomes. We discuss these cells in the next section, while the full results obtained by running \stabilizedcellsearch~ on the VIGOR data with respect to both the GI and CVT outcomes is shown in \cref{tab:cell_search_frequency}.

\subsection{Discussion of cells found and performance on test set}
\label{sub:discussion_of_cells}
In this section, we discuss the statistical significance of the cells found for both GI and CVT outcomes. First, we list the top 3 cells found for each outcome, where detailed results for top 20 cells (sroted by \stab-scores) are reported in \cref{tab:cell_search_frequency}.
For the GI outcome, the top 3 stable cells are:
\vspace{-1mm}
\begin{enumerate}[label=(\roman*)]
\itemsep0em
    \item $\cell_1$: Patients with prior history of GI Event denoted as \{PPH=1\},\vspace{-1mm}
    \item $\cell_2$: patients who (self) reported a prior (to the experiment) usage of steroids, and a history of hypertension denoted as \{PSTRDS=1, HYPGRP=1\}, and\vspace{-1mm}
    \item $\cell_3$: Elderly patients who reported a prior usage of steroid drugs denoted as \\ \{PSTRDS=1, ELDERLY=1\}.
\end{enumerate}
For the CVT outcome, they are:
\vspace{-1mm}
\begin{enumerate}[label=(\roman*)]
\itemsep0em
    \item $\wtilde{\cell}_1$: Patients for which use of Aspirin has been indicated as per FDA guidelnes \{ASPFDA=1\},\vspace{-1mm}
    \item $\wtilde{\cell}_2$: Male elderly patients \{MALE=1,ELDERLY=1\}, and\vspace{-1mm}
    \item $\wtilde{\cell}_3$: Patients that have reported prior history \{ASCGRP=1\}.
\end{enumerate}
For further details on the features appearing above, please refer back to \cref{sub:data_engineering}. In \cref{fig:final_cell_overlap}, we plot the overlap between these cells.

\begin{figure}[ht]
    \centering
    \begin{tabular}{c|c} 
        \includegraphics[width=0.43\textwidth]{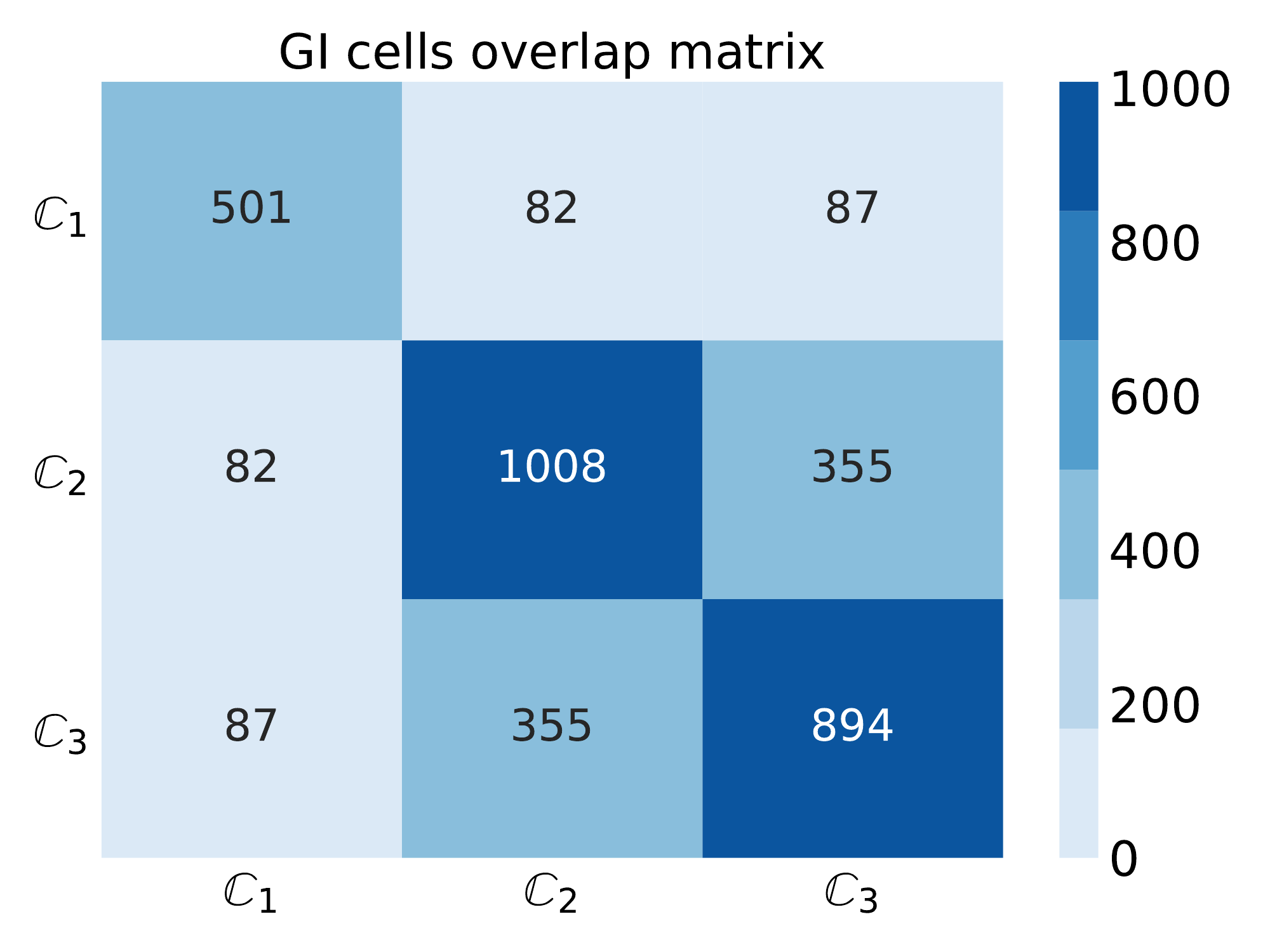} &
        \includegraphics[width=0.43\textwidth]{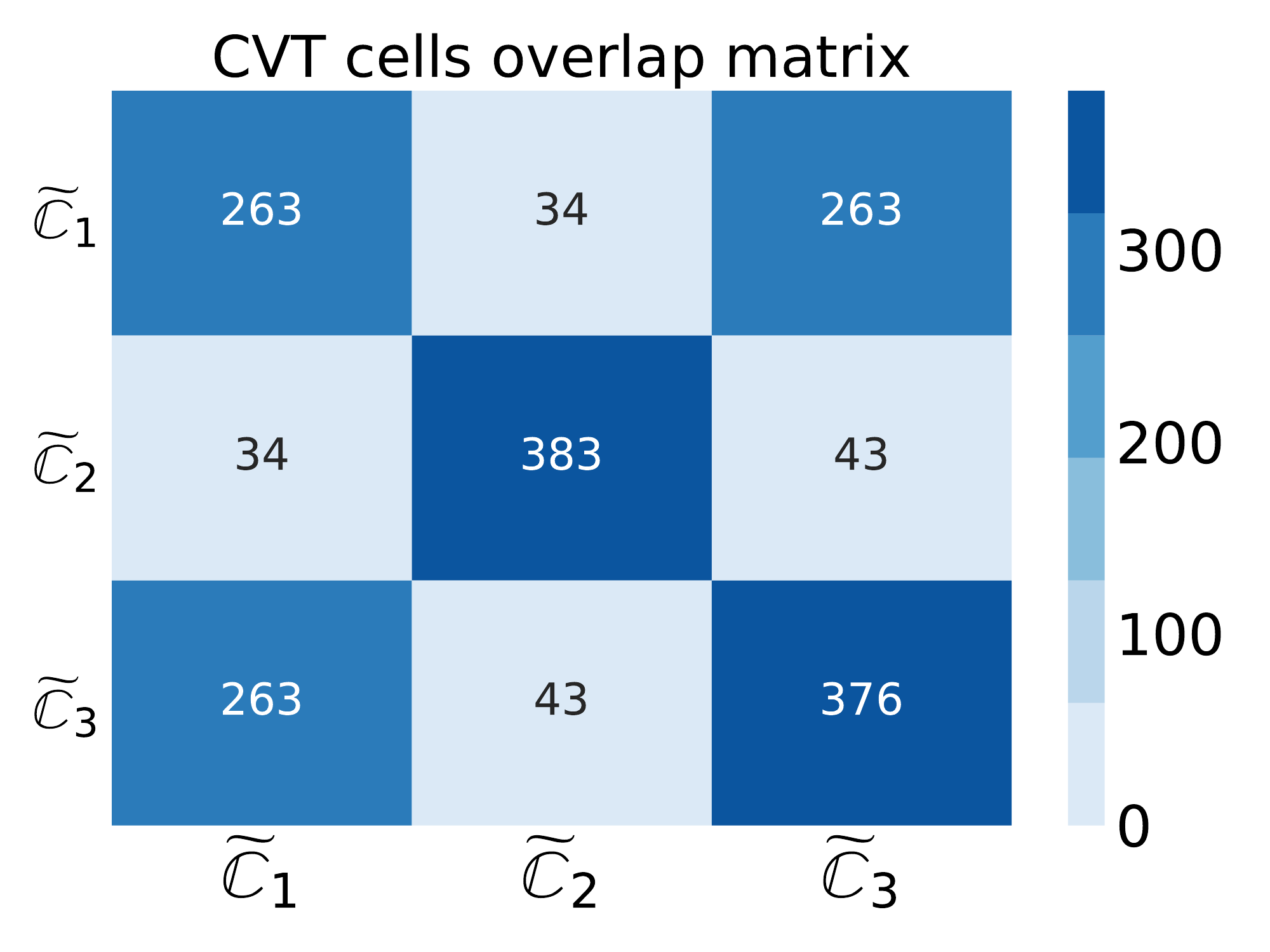}
        \\
    \begin{tabular}{l}
        $\cell_1=$\{PPH=1\}   \\
        $\cell_2=$\{PSTRDS=1, HYPGRP=1\} \\
        $\cell_3=$\{PSTRDS=1, ELDERLY=1\}
    \end{tabular}
    &
    \begin{tabular}{l}
        $\wtilde{\cell}_1=$\{ASPFDA=1\}   \\
        $\wtilde{\cell}_2=$\{MALE=1, ELDERLY=1\}   \\
        $\wtilde{\cell}_3=$\{ASCGRP=1\}
    \end{tabular} \vspace{5mm}\\
        (a) GI cells & (b) CVT cells
    \end{tabular}
    \caption{Overlap matrix for final discovered cells on the training data $\trainset$. For panel \textbf{(a)} the data split is stratified on the treatment indicator and the GI outcome, and that for \textbf{(b)} is stratified on on the treatment indicator and the CVT outcome. For instance, the number $82$ for the entry corresponding to $\cell_1$ and $\cell_2$ in panel (a) represents that the two cells had 82 patients in common on the training data.}
    \label{fig:final_cell_overlap}
\end{figure}
\paragraph{Conclusions from \cref{fig:final_cell_overlap}:} As can be seen in \cref{fig:final_cell_overlap}(a), there is little to moderate overlap among the cells $\cell_1$ and $\cell_3$, which shows that they are meaningfully different. On the other hand, there is significant overlap among the cells $\wtilde{\cell}_1,\wtilde{\cell}_3$ in \cref{fig:final_cell_overlap}(b). In particular, $\wtilde{\cell}_1$ is a subset (but not a sub-cell) of $\wtilde{\cell}_3$. 
The reason we report both cells is because of the suspected multi-scale nature of treatment effect variation for the CVT outcome, with $\wtilde{\cell}_1$ found more often for $\qvalue=0.9$, and $\wtilde{\cell}_3$ found more often for $\qvalue=0.8$.

We now compute and report several quantities for each of these 6 cells, finally making use of the holdout test dataset (20\% of the study size) for the \emph{very first time}. For cells $\cell_1,\cell_2$ and $\cell_3$, as well as the union $\cup_{j=1}^3\cell_{j}$ of these 3 cells, the results are reported in \cref{tab:test_set_GI}. 
Similar results for the cells $\wtilde{\cell}_1,\wtilde{\cell}_2$, and $\wtilde{\cell}_3$ and their union $\cup_{j=1}^3\wtilde{\cell}_{j}$ are reported in \cref{tab:test_set_CVT}. We now discuss the results from \cref{tab:test_set_GI,tab:test_set_CVT} one by one.

\begin{table}[ht]
    \centering
    \rowcolors{2}{red!10}{white}
    \resizebox{\textwidth}{!}{
    \begin{tabular}{p{2.2cm}ccccccc}
        \toprule
        &   \multicolumn{2}{c}{\bf \#evts/size} &  \multicolumn{2}{c}{\bf CATE Est. $\gatehat[\cell \cap \tset]$ (std)} &  \multicolumn{3}{c}{\bf $t$-statistic $\zscore_{\cell \cap \tset}$} \\
        \addlinespace[0.3em]
        \bf Dataset $\tset$ & $\trainset$ & $\testset$ & $\trainset$ & $\testset$ &$\trainset$ & $\testset$ & $^\dagger\valset$ \\
        \addlinespace[0.3em]
        \rowcolor{white}  \bf Cell $\cell$  \\
        \midrule
        \addlinespace[1em]
        \multicolumn{8}{l}{\large \emph{GI Event (GI-stratified split) }}\\
        \addlinespace[1em]
        \small PPH=1               &   36/501 & 8/129 & -0.057 (0.023) &  -0.055 (0.042)   & -1.89 &   -1.01 & -0.99 (0.27) \\[3mm]
        \small PSTRDS=1, HYPGRP=1 &   39/1008 & 6/238 &  -0.050 (0.012) & -0.037 (0.021)  &   -3.17 & -1.06 &  -1.57 (0.22) \\[3mm]
        \small PSTRDS=1, ELDERLY=1      &   46/894 & 9/227 &  -0.051 (0.015) & -0.063 (0.026) & -2.74 &  -2.00 & -1.38 (0.17) \\[3mm]
         Union   &  79/1905 & 19/471 & -0.038 (0.009) & -0.047 (0.018) & -3.15 & -2.22 & -1.59 (0.20) \\[3mm]
         \emph{All} &  142/6460 &  35/1616 & -0.016 (0.004) & -0.016 (0.007) &-&-&- \\
        \midrule
        \addlinespace[1em]
        \multicolumn{8}{l}{\large \emph{CVT Event (entire data) }}\\
        \addlinespace[1em]
        \small PPH=1               &  \multicolumn{2}{c}{2/630} & \multicolumn{2}{c}{-0.006 (0.004)}   &        \multicolumn{3}{c}{-2.66} \\[3mm]
        \small PSTRDS=1, HYPGRP=1 &   \multicolumn{2}{c}{11/1246} & \multicolumn{2}{c}{0.008 (0.005)}   &        \multicolumn{3}{c}{0.44} \\[3mm]
        \small PSTRDS=1, ELDERLY=1      &   \multicolumn{2}{c}{16/1121} & \multicolumn{2}{c}{0.015 (0.007)}   &        \multicolumn{3}{c}{1.42} \\[3mm]
        Union   &   \multicolumn{2}{c}{21/2376} & \multicolumn{2}{c}{0.007 (0.004)}   &        \multicolumn{3}{c}{0.55} \\[3mm]
         \emph{All} &  \multicolumn{2}{c}{ 59/8076} & \multicolumn{2}{c}{ 0.006 (0.002)}   &        \multicolumn{3}{c}{-} \\
        \bottomrule
        \end{tabular}
        }
    \caption{Results for the final cells selected after \stabilizedcellsearch\ for the GI event, namely $\cell_1=\braces{\text{PPH=1}}$, $\cell_2=\braces{\text{PSTRDS=1,HYPGRP=1}}$ and $\cell_3=\braces{\text{PSTRDS=1,ELDERLY=1}}$ from \cref{sub:discussion_of_cells}. We also report the results for the other outcome, namely CVT event, on the entire data (all 8076 patients).
    In the column $^\dagger\valset$, we report the mean $t$-statistics and standard deviation in parentheses, across the 12 different folds of the training data $\trainset$ obtained from the 3 random CV splits \textbrace{\cvorig, \cvzero, \cvone}.
    }
    \label{tab:test_set_GI}
\end{table}

\paragraph{Results from \cref{tab:test_set_GI}:}
In the first three rows of \cref{tab:test_set_GI}, we examine the subgroup treatment effect for these cells with respect to the GI outcome. In the second and third columns, we report two versions of the Neyman estimate for the cell CATE $\gatehat[\cell\cap \tset]$, one computed on the training set $\trainset$ as well as one computed on the test set $\testset$. Likewise, in the next two columns, we report the $t$-statistic $\zscore_{\group\cap\tset}$, one computed on the training set $\trainset$, and on the test set $\testset$. Finally, in the last column with header $^\dagger\valset$, we report the mean (and standard deviation in parenthesis) of the $t$-statistics $\zscore$ computed on the 12 different folds of $\trainset$ from the 3 random CV splits \textbrace{\cvorig, \cvzero, \cvone}. Overall, the test set results are promising, with test set CATE estimates being much more negative than the estimated ATE, and comparable to their training set counterparts. While we do not report $p$-values because they can be easily misunderstood, we note that the test set $t$-statistic values for the GI outcome are $\cell_3$, and the union $\cup_{j=1}^3\cell_{j}$, are both significant at the 0.025 level for a one-sided $z$-test.

The starting point of our investigation of the VIGOR dataset was the hope to identify a subgroup for which Vioxx simultaneously has a strong negative treatment effect for GI risk and a low positive treatment effect for CVT risk. Consequently, in the last three rows of \cref{tab:test_set_GI}, we report the treatment effect results for the cells $\braces{\cell_{j}}_{j=1}^3$ and their union, with respect to the CVT outcome. While $\cell_2$ and $\cell_3$ experience increased CVT risk, $\cell_1=\braces{\text{PPH}=1}$ in fact shows reduced CVT risk, which makes it especially promising for further clinical investigation. We note that for the CVT outcome we report the CATE estimates and the \tstattext\ on the entire data as this outcome had no role to play in the entire StaDISC pipeline with the GI outcome, and hence the entire data can be treated as a ``valid" test set for estimating heterogeneous treatment effect of Vioxx with the CVT outcome. 

\begin{table}[ht]
    \centering
    \rowcolors{2}{red!10}{white}
    \resizebox{1\textwidth}{!}{
    \begin{tabular}{p{2.2cm}ccccccc}
        \toprule
         &   \multicolumn{2}{c}{\bf \#evts/size} &  \multicolumn{2}{c}{\bf CATE Est. $\gatehat[\cell \cap \tset]$ (std)} &  \multicolumn{3}{c}{\bf $t$-statistic $\zscore_{\cell \cap \tset}$} \\
        \addlinespace[0.3em]
        \bf Dataset $\tset$ & $\trainset$ & $\testset$ & $\trainset$ & $\testset$ &$\trainset$ & $\testset$ & $^\dagger\valset$ \\
        \addlinespace[0.3em]
        \rowcolor{white}  \bf Cell $\cell$  \\
        \midrule
        \addlinespace[1em]
        \multicolumn{8}{l}{\large \emph{CVT Event (CVT-stratifed split)}}\\
        \addlinespace[1em]
        \small ASPFDA=1  &  13/263 & 5/58 &  0.062 (0.025) &    0.103 (0.074) & 2.28 &  1.38 & 1.09 (0.20) \\[3mm]
        \small MALE=1, ELDERLY=1 &   12/383 &    0/111 &  0.040 (0.017) & 0 (0) &    2.09 & -1.16 & 0.85 (0.24) \\[8mm]
        \small ASCGRP=1  &  15/376 & 6/78 &  0.044 (0.020) & 0.047 (0.060)   & 2.05 & 0.74 & 1.04 (0.23)\\[3mm]
         Union &  24/716 & 6/175 &  0.042 (0.013) & 0.024 (0.028) & 3.09 &  0.77 & 1.55 (0.13) \\[3mm]
         \emph{All} &  \bf 47/6460 &  12/1616 & 0.006 (0.002) &   0.005 (0.004) &-&-&- \\
        \midrule
        \addlinespace[1em]
        \multicolumn{8}{l}{\large \emph{GI Event (entire data) }}\\
        \addlinespace[1em]
        \small ASPFDA=1             &  \multicolumn{2}{c}{6/321} & \multicolumn{2}{c}{-0.027 (0.016)}   &        \multicolumn{3}{c}{-0.71} \\[3mm]
        \small MALE=1, ELDERLY=1 &   \multicolumn{2}{c}{17/494} & \multicolumn{2}{c}{-0.045 (0.016)}   &        \multicolumn{3}{c}{-1.85} \\[3mm]
        \small ASCGRP=1      &   \multicolumn{2}{c}{8/454} & \multicolumn{2}{c}{-0.028 (0.013)}   &        \multicolumn{3}{c}{-0.96} \\[3mm]
         Union   &   \multicolumn{2}{c}{25/891} & \multicolumn{2}{c}{-0.040 (0.011)}   &        \multicolumn{3}{c}{-2.27} \\[3mm]
        \emph{All} & \multicolumn{2}{c}{  177/8076} & \multicolumn{2}{c}{ -0.016 (0.003)}   &        \multicolumn{3}{c}{-} \\
        \bottomrule
        \end{tabular}
    }
    \caption{Results for the final cells selected after \stabilizedcellsearch\ for the CVT event, namely $\wtilde{\cell}_1=\braces{\text{ASPFDA=1}}$, $\wtilde{\cell}_2=\braces{\text{MALE=1,ELDERLY=1}}$ and $\wtilde{\cell}_3=\braces{\text{ASCGRP=1}}$ from \cref{sub:discussion_of_cells}. We also report the results for the other outcome, namely GI event, on the entire data (all 8076 patients).
    In the column $^\dagger\valset$, we report the mean $t$-statistics and standard deviation in parentheses, across the 12 different folds of the training data $\trainset$ obtained 4 each from the 3 random CV splits \textbrace{\cvorig, \cvzero, \cvone}.}
    \label{tab:test_set_CVT}
\end{table}

\paragraph{Results from \cref{tab:test_set_CVT}:}
In \cref{tab:test_set_CVT}, we report the analogous results for cells $\wtilde{\cell}_1,\wtilde{\cell}_2$, and $\wtilde{\cell}_3$, and their union $\cup_{j=1}^3\wtilde{\cell}_{j}$, first for the CVT outcome, and then the GI outcome. For these cells, the generalization to the holdout test set is weaker, with only $\wtilde{\cell}_1$ and $\wtilde{\cell}_3$ having test set CATE values that remain substantially positive. Furthermore, the test set $t$-statistic values are smaller. All these observations are unsurprising given the rarity of the CVT outcome---in particular, only 12/1616 individuals in the test set $\testset$ experienced an event. Nonetheless, the test set CVT-CATE estimates for $\wtilde{\cell}_1$ and $\wtilde{\cell}_3$ support the view that the treatment effect is stronger on these subgroups, while the GI-CATE estimates do not suggest that these subgroups benefit especially strongly from the treatment with Vioxx.


\section{Complementary analysis with the APPROVe study}
\label{sec:approve}

It is well-documented that RCTs have problems with \emph{external validity}~\cite{juni2001assessing,rothwell2005external,fortin2006randomized,krauss2018all}, which is defined by Rothwell to be ``whether the results can be reasonably applied to a definable group of patients in a particular clinical setting in routine practice''~\cite{rothwell2005external}. This phenomenon arises primarily because RCTs have carefully defined enrollment criteria, so conclusions in such studies may not apply to patients who do not conform to these criteria. In more mathematical language, the ATE, subgroup CATE, and other estimands of interest are all defined in terms of expectations with respect to a particular distribution of patients, a particular outcome, and a particular treatment, and hence do not directly apply when any of these change. We refer the interested reader to the excellent articles by Rothwell~\cite{rothwell2005external,rothwell2006factors} for a further discussion on these topics.


Despite its importance for clinical relevance, external validity has been relatively neglected by researchers and institutions overseeing the conduct of RCTs)~\cite{rothwell2005external,krauss2018all}. One way to argue for external validity is to attempt \emph{external validation}, i.e. to reproduce the results obtained on one data set on a different but related data set. Recent voices that urge the community to give external validiation a higher priority across many domains~\cite{debray2015new,krauss2018all,norgeot2020minimum} are very much in accordance with Yu and Kumbier's~\cite{Yu2019} call to statisticians to broaden the scope of their concern from data-modeling to the entire data science life cycle as part of the PCS framework. This can be seen not only as one more predictive and stability check under the PCS framework, but also as a special case of ``transfer learning" where the desiderata is the transferability of the conclusions or findings from one dataset to other related datasets.

These reasons motivate the following complementary analysis of the APPROVe study~\cite{baron2008cardiovascular}, another RCT investigating Vioxx. 
More precisely, we compute the subgroup CATEs with respect to both the GI and CVT outcomes over this new data set, and show that the \emph{qualitative} conclusions obtained by applying StaDISC to the VIGOR study also generalize to this data set for four out of the six subgroups from \cref{fig:final_cell_overlap}; the other two subgroups were too small in size and did not have any GI events.
We now start with a background on the APPROVe study followed by a discussion of the results on subgroup CATEs.

\subsection{Background for the APPROVe study}
In this section, we provide only a brief background for the APPROVe study and refer the readers to the original paper~\cite{baron2008cardiovascular} for additional details.

The Adenomatous Polyp Prevention on Vioxx (APPROVe) study was another randomized trial sponsored by Merck, but unlike VIGOR, it was placebo-controlled. Conducted in 2001-2004, it was designed to assess whether Vioxx could ``reduce the risk of adenomatous polyps in individuals with a recent history of these tumours''~\cite{baron2008cardiovascular}. The study population comprised 2587 patients who had colon adenomatous polyps removed during a 12 week period before being entered into the study, and who had no known polyps remaining. After discovering that Vioxx had significant cardiovascular toxicity, the study was terminated two months early in September 2004, but all individuals were followed-up for at least a year afterwards off-treatment. 

The data files of the APPROVe study followed a very similar format to that of the VIGOR study albeit with two major differences: (i) GI event was not directly labeled in the dataset, and (ii) the risk factor file was not available. As a result, outcomes related to the GI event, and features (including but not limited to) ASPFDA, ASCGRP, HYPGRP, PSTRDS---which were used to define the final subgroups obtained in the previous section---were not directly available for APPROVe. However, with the data available to us, we were able to impute the GI outcome and the missing relevant features (used for the cells reported in \cref{tab:test_set_GI,tab:test_set_CVT}). The data cleaning and imputation were done before looking at the final results.  The details for this data cleaning are provided in Appendix~\ref{sec:data_cleaning}, and the distribution of the selected features and the two outcomes is reported in \cref{tab:approve_covar_description}. Once we have the features and the outcomes, we compute the subgroup CATE~\eqref{eq:gate_hat} and $t$-statistics~\eqref{eq:t_statistics} and report the results in \cref{tab:approve}.

\subsection{Results with the APPROVe study}
\label{sub:approve_results}

Before presenting the quantitative results, we make a few remarks. In direct analogy with the problems with external validity mentioned earlier, there are several ways in which the causal estimands in APPROVe differ from those in VIGOR. First, the ``control'' arm of both studies were of entirely different natures: while VIGOR was a comparison between Vioxx and Naproxen, APPROVe compared Vioxx with a placebo. Second, the lengths of both studies were different, which is important because our estimands are defined in terms of accumulated risk over the duration of the study. Patients in VIGOR were followed for a median time of 9 months, whereas most patients in APPROVe were tracked for at least 4 years. Furthermore, while GI events were adjudicated in VIGOR, this was not the case for APPROVe. Lastly, the study populations are different. As elaborated earlier in \cref{sec:vigor}, the VIGOR study comprised patients who were diagnosed with rheumatoid arthritis. On the other hand, APPROVe comprised patients with a recent history of colon polyps. Furthermore, unlike VIGOR, APPROVe excluded patients likely needing regular NSAID treatment, but allowed for concomitant low-dose aspirin therapy.


\begin{table}[ht]
    \centering
    \rowcolors{2}{red!10}{white}
    \begin{tabular}{lccc}
        \toprule
          \bf Cell $\cell$ &   {\bf \#evts/size} &  {\bf CATE Est. $\gatehat[\cell \cap \tset]$ (std)} & {\bf $t$-statistic $\zscore_{\cell \cap \tset}$} \\
        \addlinespace[0.3em]
        \midrule
        \addlinespace[1em]
        \multicolumn{4}{l}{\large \emph{GI Event with $\tset =$ all data}}\\[1mm]
        \addlinespace[1em]
        \small PPH=1  &  6/184 & 0.066 (0.026) & 2.012 \\[1mm]
        \small PSTRDS=1, HYPGRP=1  &  0/30 & - & - \\[1mm]
        \small PSTRDS=1, ELDERLY=1   & 0/21 & - & - \\[1mm]

        \emph{All} &   33/2587 &  0.016 (0.004) & - \\
        \midrule
        \addlinespace[1em]
        \multicolumn{4}{l}{\large \emph{CVT Event with $\tset =$ all data}}\\[1mm]
        \addlinespace[1em]
        \small ASPFDA=1  &  13/151 & 0.107 (0.043) & 2.128 \\[1mm]
        \small MALE=1, ELDERLY=1  &  30/416 & 0.069 (0.025) & 2.251 \\[1mm]
        \small ASCGRP=1   &  17/250 & 0.068 (0.031) & 1.664 \\[1mm]
         Union (of 3 cells above) &  41/588 & 0.065 (0.021) & 2.650 \\[1mm]
        \small PPH=1   &  4/184 & 0.022 (0.022) & 0.119 \\[1mm]
        \emph{All} &  89/2587 &  0.020 (0.007) & - \\[1mm]
        \bottomrule
        \end{tabular}
    \caption{Results for the subgroups found with StaDISC on VIGOR, for the APPROVe dataset. Note that unlike VIGOR, the patients in the control arm for the APPROVe study were treated with a placebo, which makes the quantitative results reported here not directly comparable with that reported in \cref{tab:test_set_GI,tab:test_set_CVT}. Refer to the text for further discussion. The armwise statistics of the features and outcomes for the APPROVe study are provided in \cref{tab:approve_covar_description}.}
    \label{tab:approve}
\end{table}

\cref{tab:approve} describes the quantitative results for the final subgroups (from \cref{sub:discussion_of_cells}) for the APPROVe study. For the reasons explained in the previous paragraph, we do not expect the subgroup CATE estimands to be the same across the two studies. However, comparing the results across \cref{tab:test_set_GI,tab:test_set_CVT,tab:approve}, it is reassuring that the subgroups we found for the VIGOR study continue to be meaningful for APPROVe in illustrating the heterogeneity of treatment effects. We now discuss the results first for the CVT outcome followed by that for the GI outcome as the interpretation of the results for the latter is a bit more subtle.

\paragraph{Results for the CVT outcome:}
We note that the three subgroups \{ASPFDA=1\}, \{MALE=1, ELDERLY=1\}, \{ASCGRP=1\}, and the union of these 3 subgroups all had subgroup CATEs that were much larger than the ATE, with $t$-statistics that were significant at the 0.05 level for a one-sided $z$-test, even after accounting for multiple-testing (refer to end of this section for further discussions related to multiple-testing.) Overall these results provide evidence for the heterogeneous treatment effects of Vioxx for the CVT outcomes over these subgroups, namely that Vioxx disproportionately increases the CVT event risk for these subgroups when compared to either Naproxen or a placebo. To be consistent with the earlier results in \cref{tab:test_set_GI}, we also computed the subgroup CATE for \{PPH=1\} for the CVT outcome and (like the VIGOR study) did not find any evidence for a disproportionate increase in the risk for the CVT event compared with the entire population.

Recall that, the found increase in risk for VIGOR was relative to Naproxen. This observation alone may suggest a possibility that Vioxx was not the cause of the observed increase in CVT events, and the positive ATE could have resulted due to a protective effect of Naproxen reducing them. Merck, the manufacturer of Vioxx, interpreted the CVT signal in VIGOR as being a consequence of a hitherto unknown protective effect of Naproxen, rather than a deleterious consequence of Vioxx. The CVT signal in the APPROVe study associated with Vioxx relative to placebo conclusively confirmed that Vioxx can have deleterious consequences. Moreover, both VIGOR and APPROVe study suggest that Vioxx has significant heterogeneity in how it increases the risk for CVT events for different subgroups.


\paragraph{Results for the GI outcome:}
As noted above, additional care is required to interpret the CATE results for the GI outcome. Whereas Naproxen was known to have GI toxicity, and was shown in VIGOR to increase the risk of GI events more than Vioxx, a placebo by definition does not have any toxicity. As such, our finding that treatment with Vioxx had a positive estimated ATE (1.6\%) with the GI outcome in the APPROVe study does not contradict our earlier reporting of a negative ATE with respect to the GI outcome (-1.6\%) in the VIGOR study. In fact, this discovery is surprising insofar as Vioxx was initially believed to have minimal if any, GI toxicity whatsoever~\cite{laine1999randomized}.

We found the subgroup \{PPH=1\} to have a large positive estimated subgroup CATE (6.6\%) resulting in a $t$-statistic score significant at the 0.025 level for a one-sided $z$-test (without correcting for multiple-testing.) As discussed above, this result does not contradict the negative CATE value of -5.7\% (or -5.5\% for the test set) estimated for the VIGOR study (see \cref{tab:test_set_GI}). We furthermore note that the GI event rates over both arms in VIGOR, and the Vioxx arm in APPROVe were all elevated compared to the entire population. The corresponding rates for the placebo in the APPROVe study were fairly similar (0\% for \{PPH=1\} and  0.4\% on average.)

We summarize our finding across the two studies as follows. (i) VIGOR study: Vioxx, in comparison to Naproxen, reduced the GI Toxicity disproportionately for the subgroup \{PPH=1\} when compared to the the average. (ii) APPROVe study: Vioxx, in comparison to the placebo, increases the GI Toxicity disproportionately for the subgroup \{PPH=1\} when compared to the average. Nonetheless, the conclusion that the estimated subgroup CATE for \{PPH=1\} was significantly different than the estimated ATE is \emph{consistent} across the two studies.

Finally, due to the difference in the study population, two out of the three subgroups for the GI event reported in \cref{tab:test_set_GI}, namely \{PSTRDS=1, HYPGRP=1\} and \{PSTRDS=1, ELDERLY=1\}, were too small in size and had no GI events.\footnote{Indeed, comparing \cref{tab:covar_description,tab:approve_covar_description}, we can attribute the discrepancy in these subgroups' sizes between the two studies to the smaller population of patients (74/2587) with a history of using glucocorticoids (PSTRDS = 1) in the APPROVe study versus that of the much larger population of such patients (4479/8076) in the VIGOR study.} Consequently, it does not make sense to quantify the subgroup CATE for these subgroups.

\paragraph{Multiple-testing with FWER control:} Given enough data points in the APPROVe study, we also perform corrected multiple hypothesis testing using Holm-Bonforreni procedure controlling family wise error rate (FWER) at level 0.05. Overall, we test 5 null hypotheses, that the subgroup CATE is equal to the average treatment effect for the following cases: (i) $\cell_1=$ \{PPH=1\} for the GI event, (ii) $\widetilde{\cell}_1=$\{ASPFDA=1\}, (iii) $\widetilde{\cell}_2=$\{MALE=1, ELDERLY=1\}, (iv) $\widetilde{\cell}_3=$\{ASCGRP=1\}, and (v)~the union $\cup_{i=1}^3\widetilde{\cell}_i$---where the treatment effect in subgroups (ii)-(v) corresponds to the CVT event. 
The t-statistics for these hypotheses (sorted by magnitude) as reported in \cref{tab:approve} are 2.650, 2.251, 2.128, 2.012 and 1.664, and thereby the corresponding one-sided p-values are 0.004, 0.012, 0.0167, 0.022 and 0.048. The  corrected procedure for significance level 0.05 compares these sorted p-values with the cut-offs 0.01, 0.0125, 0.0167, 0.025 and 0.05. On doing so we find that all five hypotheses are rejected, and thus we conclude all the subgroups (i)-(v) have statistically significant heterogeneous treatment effect.\footnote{Note that, for the APPROVe study, we did not test for heterogeneity in the subgroups \{PSTRDS=1, HYPGRP=1\}, and \{PSTRDS=1, ELDERLY=1\} due to their small size in this study. Since the size of the subgroup can only be observed once we know the group membership of the patients, our testing procedure and the associated discoveries can be considered as being conditional on observing the group membership, and treatment variable for all the patients in the APPROVe study. In other words, the statistical significance is over the randomness in the outcome, and the conditional randomness in the covariates given the group membership indicators.}

\section{Discussion}
\label{sec:discussion}
In this work, we have made three major contributions: (I) We have re-analyzed a dataset from the 1999-2000 VIGOR study, an RCT of 8076 patients, and found three clinically relevant subgroups each for the GI outcome (total size 29.4\%), and the CVT outcome (total size 11.0\%), for which the treatment drug Vioxx has significantly large estimated treatment effect when compared to that from the estimated ATE. We provided external evidence for the significance of the heterogeneous treatment effects for four out of the six subgroups through a complementary analysis of the 2001-2004 APPROVe study, another RCT of 2587 patients. (II) Our work is an illustration of how clinical trial data can be analyzed to provide a basis for differential treatment decisions in subgroups in order to optimize outcomes, and how the findings can be validated with another study. We call this novel methodology StaDISC, and develop it by building on the PCS framework~\cite{Yu2019}, the calibration literature, and recent developments in CATE estimation. (III) Our work introduces the PCS framework to the causal inference community, and provides a template for a more informative understanding of heterogeneous treatment effects.

An important point to note is that the notions of estimated treatment effects ATE, CATE and subgroup CATE (defined in \cref{eq:all_ate_defs}) used in this work and more broadly in CATE estimation, measure the \emph{difference} in the adverse event risk in the treatment group to that in the control group. However, when investigating the efficacy of medical interventions, medical professionals are often more interested in relative risk, which measures the \emph{ratio} of the two risks. This alternate conception of treatment effect in terms of relative risk changes the meaning of heterogeneity. For instance, the subgroup $\cell_1$~\{PPH=1\} has a relative risk of 0.43 with respect to GI events, which is barely any different than the population relative risk of 0.46. On the other hand, because the baseline risk of individuals in this subgroup is far higher than the rest of the population, the subgroup CATE is similarly inflated.

We do not attempt to debate which notion of heterogeneity is better since it is context-dependent. Nevertheless, given the popularity of relative risk in the medical literature, in our future work we plan to develop a formal framework for subgroup discovery with respect to relative risk by adapting generic CATE estimation methods, and consequently extend StaDISC for relative risk estimation.

There are several other extensions of StaDISC that remain interesting future directions. First, StaDISC is currently motivated and defined for randomized experiments. We intend to formulate a statistical framework that would also make it applicable to observational studies. Second, the cell search step of StaDISC only works with binary features. One can either propose to incorporate continuous features through either careful binary encoding using quantile-thresholding, or through amending the cell search procedure. Third, we have thus far applied StaDISC to the GI and CVT outcomes in the VIGOR study one at a time and a joint investigation with multiple outcomes, even more generally, is an interesting future direction.

\section*{Acknowledgements}
BY acknowledges the support from ARO Army Research Office grant W911NF-17-1-0005, Office of Naval Research Grant N00014-17-1-2176, the Center for Science of Information (CSoI), a US NSF Science and Technology Center, under grant agreement CCF-0939370, UCSF fund N7251, National Science Foundation grants NSF-DMS-1613002, 1953191, IIS 1741340, and a research award from the Amazon Web Services (AWS Servers were used for running all our Jupyter Notebooks). BY is a Chan Zuckerberg Biohub investigator. The research was also partly supported by National Science Foundation grant NSF-CCF-1740855.
The authors would like to thank Peng Ding, Sam Pimental, Chandan Singh, Tiffany Tang and our anonymous reviewer for their helpful comments.

\appendix



\begin{table}
    \centering
    \resizebox{1\textwidth}{!}{
    \centering
    \begin{tabular}{cc}
    \rowcolors{2}{gray!10}{white}
         \begin{small}

    \begin{tabular}{lccccc}
    \toprule
    Estimator $\model$
    &  $\overline{\bincompare}_{1, 2}$ &  $\overline{\bincompare}_{2, 3}$ 
    &  $\overline{\bincompare}_{3, 4}$ &  $\overline{\bincompare}_{4, 5}$ 
    & $\overline{\bincompare}_{1, \min}$\\
    \midrule
\ttt{t\_logistic}      &                                   \bf     1.00 &                                      0.67 &                                  \bf      0.83 &                                      0.25 &              \bf  1.00 \\
\ttt{causal\_forest\_2} &                                    \bf    1.00 &                                      0.50 &                                   \bf     0.83 &                                      0.17 &              \bf  1.00 \\
\ttt{x\_lasso}         &                                    \bf    1.00 &                                      0.50 &                                      0.58 &                                      0.67 &             \bf  1.00 \\
\ttt{x\_rf}            &                                    \bf    1.00 &                                      0.42 &                                      0.42 &                                      0.67 &               \bf 1.00 \\
\ttt{t\_lasso}         &                                   \bf     1.00 &                                      0.42 &                                      0.50 &                                      0.58 &              0.92 \\
\ttt{x\_logistic}      &                                    \bf    1.00 &                                      0.33 &                                      0.50 &                                      0.75 &              0.92 \\
\ttt{s\_xgb}           &                                    \bf    1.00 &                                      0.67 &                                      0.58 &                                      0.58 &              0.92 \\
\ttt{r\_lassolasso}    &                                      0.92 &                                      0.42 &                                      0.42 &                                      \bf  0.92 &              0.92 \\
\ttt{r\_rfrf}          &                                      0.92 &                                      0.50 &                                      0.42 &                                      0.50 &              0.92 \\
\ttt{r\_lassorf}       &                                      0.92 &                                      0.42 &                                      0.42 &                                      0.42 &              0.92 \\
\ttt{causal\_forest\_1} &                                      0.92 &                                      0.67 &                                      0.75 &                                      0.50 &              0.83 \\
\ttt{x\_xgb}           &                                      0.92 &                                      0.33 &                                      0.50 &                                      0.83 &              0.83 \\
\ttt{t\_xgb}           &                                      0.92 &                                      0.42 &                                      0.67 &                                      0.17 &              0.83 \\
\ttt{t\_rf}            &                                      0.92 &                                      \bf  0.75 &                                      0.50 &                                      0.33 &              0.83 \\
\ttt{causal\_tree\_2}   &                                      0.92 &                                      \bf  0.75 &                                      0.25 &                                      0.42 &              0.75 \\
\ttt{s\_rf}            &                                      0.83 &                                      0.58 &                                      0.67 &                                      0.42 &              0.75 \\
\ttt{causal\_tree\_1}   &                                      0.83 &                                      0.58 &                                      0.17 &                                      0.67 &              0.67 \\
    \bottomrule

    \end{tabular}\vspace{2mm}
            \end{small}
    
    &
    \rowcolors{2}{gray!10}{white}
     \begin{small}
    \begin{tabular}{lrrrrr}
    \toprule
    Estimator $\model$
    &  $\overline{\bincompare}_{1, 2}$ &  $\overline{\bincompare}_{2, 3}$ 
    &  $\overline{\bincompare}_{3, 4}$ &  $\overline{\bincompare}_{4, 5}$ 
    & $\overline{\bincompare}_{5, \max}$\\
    \midrule
\ttt{t\_lasso}         &                                      0.33 &                                      0.42 &                                      0.42 &                                     \bf   1.00 &           \bf    1.00 \\
\ttt{x\_xgb}           &                                      0.33 &                                      0.50 &                                      0.58 &                                      0.92 &             0.92 \\
\ttt{x\_logistic}      &                                      0.50 &                                      0.50 &                                      0.42 &                                      0.92 &             0.92 \\
\ttt{r\_rfrf}          &                                      0.25 &                                      0.42 &                                      0.50 &                                      0.92 &             0.83 \\
\ttt{s\_rf}            &                                      0.42 &                                      0.42 &                                      0.42 &                                      0.92 &             0.83 \\
\ttt{x\_lasso}         &                                      0.50 &                                      0.33 &                                      0.50 &                                      0.83 &             0.75 \\
\ttt{t\_rf}            &                                      0.33 &                                      0.25 &                           \bf             0.67 &                                      0.83 &             0.75 \\
\ttt{x\_rf}            &                                      0.50 &                                      0.33 &                                      0.58 &                                      0.83 &             0.75 \\
\ttt{t\_logistic}      &                                      0.33 &                                      0.25 &                                      0.58 &                                      0.83 &             0.75 \\
\ttt{r\_lassorf}       &                                      0.17 &                                      0.42 &                                      0.42 &                                      0.92 &             0.75 \\
\ttt{causal\_forest\_1} &         \bf                               0.67 &                                      0.33 &                             \bf           0.67 &                                      0.92 &             0.75 \\
\ttt{causal\_forest\_2} &                                      0.50 &                                      0.08 &                                      0.33 &                                      0.92 &             0.75 \\
\ttt{r\_lassolasso}    &                                      0.17 &                                     \bf   0.75 &                                      0.50 &                                      0.75 &             0.67 \\
\ttt{causal\_tree\_2}   &                                      0.25 &                                      0.08 &                                      0.33&                                      0.83 &             0.25 \\
\ttt{t\_xgb}           &                                      0.08 &                                      0.08 &                                      0.25 &                                      0.75 &             0.08 \\
    \bottomrule
     \end{tabular} \vspace{2mm}
    \end{small}
    \\
    (a) GI Event &
    (b) CVT Event
    \end{tabular}
    }
    \caption{Estimator-wise values of the mean scores $\overline{\bincompare}_{\j, \j+1}$~\eqref{eq:bin_compare} for $\j=1,2,3, 4$ for both GI and CVT events, $\overline{\bincompare}_{1, \min}$~\eqref{eq:gi_1_min} for the GI event, and $\overline{\bincompare}_{5, \max}$~\eqref{eq:tc5_max} for the CVT event, where the mean was taken over the 12 validation folds, 4 each from the 3 random CV splits \textbrace{\cvorig,\cvzero,\cvone}.
    In each column the maximum score is highlighted in bold. The estimators are listed in the order sorted by the value in last column. Recall that each column was plotted earlier as a boxplot in \cref{fig:monotonicity_box_plot}(a).}
    \label{tab:mononocitiy}
\end{table}

\begin{table}[ht]
    \centering
    \resizebox{1\textwidth}{!}{
    \begin{tabular}{cc}
   \begin{small}
   \rowcolors{2}{gray!10}{white}
    \begin{tabular}{lccc}
        \toprule
        &\multicolumn{3}{c}{\Large $\stab(\cell)$-score in \% }\\[2pt]
        &\multicolumn{3}{c}{ \Large  with $\tbquant=\gitopgroup$ }\\[0.3cm]
        \rowcolor{white}\Large Cell $\cell$  for GI event                     &  \Large $\qvalue=0.2$ &  \Large $\qvalue=0.3$ & \large \bf Mean\\[2pt]
        \midrule
\{PPH=1\}                   &\large  \bf 92 &\large   \bf  92 &\large   \bf 92 \\[2pt]
\{PSTRDS=1, HYPGRP=1\}                    &\large    \bf 36 &\large   \bf  54 &\large   \bf 45 \\[2pt]
\{PSTRDS=1, ELDERLY=1\}            &\large    \bf 37 &\large    \bf 48 &\large   \bf 42 \\[2pt]
\{PNAPRXN=0, PSTRDS=1, ELDERLY=1\} &\large     23 &\large     18 &\large    21 \\[2pt]
\{PNAPRXN=0, HYPGRP=1, PSTRDS=1\}         &\large     25 &\large      8 &\large    17 \\[2pt]
\{PSTRDS=1, PNSAIDS=0\}                   &\large      8 &\large     23 &\large    15 \\[2pt]
\{WHITE=0, PSTRDS=1, ELDERLY=1\}   &\large     18 &\large      3 &\large    11 \\[2pt]
\{CHLGRP=1, HYPGRP=1\}                    &\large     17 &\large      2 &\large    10 \\[2pt]
\{OBESE=1, WHITE=0, PSTRDS=1\}            &\large     10 &\large      8 &\large     9 \\[2pt]
\{PNAPRXN=0, ELDERLY=1\}           &\large      0 &\large     18 &\large     9 \\[2pt]
\{OBESE=1, WHITE=0\}                      &\large      0 &\large     17 &\large     8 \\[2pt]
\{HYPGRP=1, PNSAIDS=0\}                   &\large     16 &\large      0 &\large     8 \\[2pt]
\{WHITE=0, PNSAIDS=0\}                    &\large     14 &\large      0 &\large     7 \\[2pt]
\{OBESE=1, WHITE=0, PNAPRXN=0\}           &\large      3 &\large     10 &\large     7 \\[2pt]
\{OBESE=1, PSTRDS=1, HYPGRP=1\}           &\large      5 &\large      8 &\large     7 \\[2pt]
\{PSTRDS=1, HYPGRP=1, ELDERLY=1\}  &\large     12 &\large      0 &\large     6 \\[2pt]
\{WHITE=0, PSTRDS=1, PNSAIDS=0\}          &\large     10 &\large      2 &\large     6 \\[2pt]
\{CHLGRP=1\}                              &\large      0 &\large     11 &\large     6 \\[2pt]
\{PNAPRXN=0, HYPGRP=1\}                   &\large      0 &\large     10 &\large     5 \\[2pt]
\{OBESE=1, PNSAIDS=0\}                    &\large      4 &\large      6 &\large     5 \\[2pt]
    \bottomrule
    \end{tabular}
    \end{small}      
    \vspace{1cm}
    &
    \begin{small}
    \rowcolors{2}{gray!10}{white}
    \begin{tabular}{lccc}
        \toprule
        &\multicolumn{3}{c}{\Large $\stab(\cell)$-score in \% }\\[2pt]
        &\multicolumn{3}{c}{\Large with $\tbquant=\tctopgroup$ }\\[0.3cm]
       \rowcolor{white} \Large Cell $\cell$    for CVT event   & \Large $\qvalue=0.9$ & \Large  $\qvalue=0.8$ & \large \bf Mean \\[2pt]
        \midrule
\{ASPFDA=1\}                                      &\large    \bf 82 &\large     50 &\large \bf    66 \\[2pt]
\{MALE=1, ELDERLY=1\}                      &\large     \bf 70 &\large    \bf 57 &\large   \bf 64 \\[2pt]
\{ASCGRP=1\}                                      &\large \bf    32 &\large  \bf    54 &\large \bf    43 \\[2pt]
\{MALE=1\}                                        &\large      0 &\large    \bf 62 &\large    31 \\[2pt]
\{ELDERLY=1, SMOKE=1\}                    &\large     22 &\large     27 &\large    25 \\[2pt]
\{MALE=1, ELDERLY=1, US=1\}                &\large     30 &\large      0 &\large    15 \\[2pt]
\{MALE=1, US=1\}                                  &\large      0 &\large     26 &\large    13 \\[2pt]
\{OBESE=1, ELDERLY=1\}                     &\large      0 &\large     21 &\large    10 \\[2pt]
\{MALE=1, WHITE=1, ELDERLY=1\}             &\large     20 &\large      0 &\large    10 \\[2pt]
\{MALE=1, ASCGRP=1\}                              &\large     18 &\large      0 &\large     9 \\[2pt]
\{WHITE=1, OBESE=1, ELDERLY=1\}            &\large      0 &\large     15 &\large     8 \\[2pt]
\{MALE=1, PPH=0, ELDERLY=1\} &\large     13 &\large      0 &\large     7 \\[2pt]
\{MALE=1, WHITE=1\}                               &\large      0 &\large     12 &\large     6 \\[2pt]
\{PPH=0, US=1, ASCGRP=1\}           &\large      2 &\large      8 &\large     5 \\[2pt]
\{WHITE=1, ELDERLY=1, SMOKE=1\}           &\large      7 &\large      3 &\large     5 \\[2pt]
\{ELDERLY=1, US=1, SMOKE=1\}              &\large      7 &\large      3 &\large     5 \\[2pt]
\{MALE=1, PPH=0\}                   &\large      0 &\large      9 &\large     4 \\[2pt]
\{ELDERLY=1, US=1, CHLGRP=1\}              &\large      0 &\large      8 &\large     4 \\[2pt]
\{CHLGRP=1, ASCGRP=1\}                            &\large      8 &\large      0 &\large     4 \\[2pt]
 \{MALE=1, ELDERLY=1, SMOKE=1\}            &\large      7 &\large      0 &\large     3 \\[2pt]
    \bottomrule
    \end{tabular}
    \end{small}  
    \vspace{-2mm}
    \\
        \Large (a) GI Event & 
        \Large (b) CVT Event
    \end{tabular}
    }
    \caption{ $\stab(\cell)$-scores (in \% rounded to nearest integer) for the top 20 cells $\cell$ found by \cellsearch-methodology for quantile-based top subgroups $\tbquant$ of the ensemble CATE estimator. The cells are sorted by the ``Mean" column of $\stab(\cell)$-scores, which in turn denote the average of the the scores in second and third columns. For each score column, cells corresponding to top-3 scores are displayed in bold. The choices $\qvalue=0.2, 0.3$ for the GI event in panel \textbf{(a)}, and $\qvalue=0.8, 0.9$ for the CVT event in panel \textbf{(b)} were made based on the results reported in \cref{tab:ensemble_quantile_t_scores} and the discussion around it.}
    \label{tab:cell_search_frequency}
\end{table}

\section{Derivation of variance formula in $t$-statistic} 
\label{sec:t-stat_var}

In this section, we derive a formula for the variance of $\gatehat - \atehat$, thereby justifying the formula for the plug-in estimator used in the definition of the $t$-statistic, which we repeat here for convenience.
\begin{align}
    \zscore_{\group} &\defn \frac{\gatehat - \atehat}{\sqrt{\varhat(\gatehat - \atehat)}},
\end{align}
We first group terms to get
\begin{align*}
    \gatehat - \atehat & = \parenth{\frac{1}{\abs{\group \cap \treat}}\sum_{i\in \group \cap \treat} Y_i(1) - \frac{1}{\abs{\group \cap \control}} \sum_{i\in \group \cap \control}Y_i(0)} - \parenth{\frac{1}{\abs{\treat}}\sum_{i\in \treat} Y_i(1) - \frac{1}{\abs{\control}} \sum_{i\in \control}Y_i(0)} \\
    & = \alpha_1\sum_{i\in \group \cap \treat} Y_i(1) + \alpha_0\sum_{i\in \group \cap \control} Y_i(0) + \beta_1\sum_{i\in \group^c \cap \treat} Y_i(1) + \beta_0\sum_{i\in \group^c \cap \control} Y_i(0)
\end{align*}
where 
\begin{align*}
    \alpha_1 = \parenth{\frac{1}{\abs{\group \cap \treat}} - \frac{1}{\abs{\treat}}},
    \quad
    \alpha_0 = -\parenth{\frac{1}{\abs{\group \cap \control}} - \frac{1}{\abs{\control}}},
    \quad
    \beta_1 = -\frac{1}{\abs{\treat}},
    \qtext{and}
    \beta_0 = \frac{1}{\abs{\control}}.
\end{align*}
Next, observe that even after we condition on $\grouplabels$, the collection of random variables $\{Y_i(1), Y_i(0) \colon 1 \leq i \leq N\}$ are fully independent, and furthermore, the terms within each sum are identically distributed. Applying the linearity of variance thus gives us
\begin{align*}
    \Var\brackets{\gatehat - \atehat~\vert~\grouplabels} & = \alpha_1^2 \abs{\group \cap \treat}\cdot \Var\brackets{Y(1)~\big\vert~\group\cap\treat} + \alpha_0^2 \abs{\group \cap \control}\cdot \Var\brackets{Y(0)~\big\vert~\group\cap\control} \\
    & + \beta_1^2 \abs{\group^c \cap \treat}\cdot \Var\brackets{Y(1)~\big\vert~\group^c\cap\treat} + \beta_0^2 \abs{\group^c \cap \control}\cdot \Var\brackets{Y(0)~\big\vert~\group^c\cap\control}
\end{align*}
where $\Var\brackets{Y(1)~\big\vert~\group\cap\treat}$ denotes the variance of $Y(1)$ when conditioned on $X \in \group$ (recall our abuse of notation described in \cref{sec:framework_and_notation}) and $T = 1$, with the other terms defined similarly.
Simplifying this formula leads to 
\begin{align*}
    \Var\brackets{\gatehat - \atehat~\vert~\grouplabels} & = 
    \parenth{1-\frac{ \abs{\group \cap \control}}{\abs{\control}}}^2 \cdot \frac{\Var\brackets{Y(0)~\big\vert~\group \cap \control}}{\abs{\group \cap \control} }
    + \parenth{1-\frac{ \abs{\group \cap \treat}}{\abs{\treat}}}^2 \cdot \frac{\Var\brackets{Y(1)~\big\vert~\group \cap \treat}}{\abs{\group \cap \treat} }
    \notag \\
    &+ \parenth{\frac{ \abs{\group^c \cap \control}}{\abs{\control}}}^2 \cdot \frac{\Var\brackets{Y(0)~\big\vert~\group^c \cap \control}}{\abs{\group^c \cap \control} }
    + \parenth{\frac{ \abs{\group^c \cap \treat}}{\abs{\treat}}}^2 \cdot \frac{\Var\brackets{Y(1)~\big\vert~\group^c \cap \treat}}{\abs{\group^c \cap \treat} } \\
    & = \parenth{\frac{ \abs{\group^c \cap \control}}{\abs{\control}}}^2 \cdot \parenth{\frac{\Var\brackets{Y(0)~\big\vert~\group \cap \control}}{\abs{\group \cap \control} } + \frac{\Var\brackets{Y(0)~\big\vert~\group^c \cap \control}}{\abs{\group^c \cap \control} }} \\
    & + \parenth{\frac{ \abs{\group^c \cap \treat}}{\abs{\treat}}}^2 \cdot \parenth{\frac{\Var\brackets{Y(1)~\big\vert~\group \cap \treat}}{\abs{\group \cap \treat} } + \frac{\Var\brackets{Y(1)~\big\vert~\group^c \cap \treat}}{\abs{\group^c \cap \treat} }}.
\end{align*}

\section{Details on data cleaning with VIGOR and APPROVe}
\label{sec:data_cleaning}
Here we collect additional details deferred from the main paper. First, we provide the details on how we identified the patients with prior history of GI event (PPH=1) for the VIGOR study. Although this subgroup was analyzed in the original study, the data files we had did not contain a membership indicator, not were there specific constructions on how to construct this subgroup. We applied a similar procedure to determine the patients with PPH=1 for the APPROVe study. Following that we describe the steps we followed to impute the GI ouctome as well as the features \{ASPFDA, ASCGRP, HYPGRP, PSTRDS\} for the APPROVe study. We also note that the other features, namely MALE and ELDERLY, reported in \cref{tab:approve} could be readily identified from the demographics dataset for the APPROVe study, where the ELDERLY feature uses normalized age as detailed in \cref{sub:data_engineering}.


\begin{table}[ht]
    \centering
    \resizebox{\textwidth}{!}{
    \rowcolors{2}{red!10}{white}
    \begin{tabular}{lcc}
    \toprule
        \bf Covariate (ABBRV) & \bf Control No. (\%) & \bf Treatment No. (\%) \\
        \midrule
        \multicolumn{1}{l}{\textbf{Overall population}} &  1300 (50.3) &1287 (49.7)\\
        \addlinespace[0.2em]
        \multicolumn{3}{l}{\bf Demographics}\\
         \addlinespace[0.1em]
         \hspace{1.5em}\hangindent=1.5em Whether \emph{gender} is male (MALE=1) & 805 (61.9) &  804 (62.4) \\
         \addlinespace[0.1em]
         \hspace{1.5em}\hangindent=1.5em Whether \emph{adjusted age}$^\dagger$ $> 65$ (ELDERLY=1) & 338 (26.0) & 329 (25.6) \\\addlinespace[0.2em]
        \multicolumn{3}{l}{\textbf{Prior medical history}}\\
        \addlinespace[0.1em]
         \hspace{1.5em}\hangindent=1.5em of \emph{GI PUB events$^*$} (PPH=1) & 93 (7.2) & 91 (7.1) \\
         \addlinespace[0.1em]
         \hspace{1.5em}\hangindent=1.5em of \emph{hypertension} (HYPGRP=1) & 446 (34.3) & 463 (36.0) \\
         \addlinespace[0.1em]
         \hspace{1.5em}\hangindent=1.5em of \emph{atherosclerotic cardiovascular disease}
         (ASCGRP=1) & 121 (9.3) & 129 (10.0) \\
         \addlinespace[0.1em]
         \hspace{1.5em}\hangindent=1.5em indicating use of \emph{aspirin} under FDA guidelines (ASPFDA=1) & 70 (5.4) & 81 (6.3) \\
         \addlinespace[0.2em]
        \multicolumn{3}{l}{\textbf{Prior usage of drugs}}\\
         \addlinespace[0.1em]
         \hspace{1.5em}\hangindent=1.5em Whether used \emph{glucocorticoids/steroids} (PSTRDS=1) & 40 (3.1) & 34 (2.6) \\
        \addlinespace[0.2em]
        \multicolumn{3}{l}{\textbf{Outcomes}}\\
        \addlinespace[0.1em]
         \hspace{1.5em}\hangindent=1.5em Whether \emph{GI event} occurred (GI=1) & 6 (0.46) & 27 (2.1) \\
         \addlinespace[0.1em]
         \hspace{1.5em}\hangindent=1.5em Whether \emph{CVT event} occurred (CVT=1) & 32 (2.5) & 57 (4.4) \\
         \bottomrule
    \end{tabular}
    }
    \caption{Overview of the selected baseline covariates in the control and treatment arm of the APPROVe study. The treatment arm was given Vioxx, while the control arm was given placebo. $^\dagger$Adjusted age denotes age multiplied by the ratio of the life expectancy in the US to that in the individual's country of residence. $^*$PUB stands for perforations, ulcers and bleeding.}
    \label{tab:approve_covar_description}
\end{table}

\paragraph{PPH for both studies:}
To identify patients with a history of GI events, we identified a list of medical terms associated with such events, namely gastroduodenal perforation, obstruction, ulcer, or upper GI bleeding, from the medical history file. (We used REPTTERM field for this part as PREFTERM was not available in the medical history file for the VIGOR dataset.) Using this procedure, we identified 313 patients in the control arm, and 317 patients in the treatment arm who had a prior history of GI events (identified as PPH = 1). These number are off by $1$ when compared to the 314 and 316 patients reported with PPH = 1 for the control and treatment arms respectively, by Bombardier et al in their paper on the VIGOR study~\cite{Bombardier2000}.

To identify the patients with PPH = 1 for the APPROVe study, we used the medical terms identified above, with some adjustment for different spellings. Doing this gives us a subgroup of 184 patients. For this dataset, we used the PREFTERM in the medical history file for identification (since PREFTERM uses standardized terminology). Note that the paper on APPROVe study by Baron et al~\cite{baron2008cardiovascular} does not report any information about the PPH feature.

\paragraph{GI outcome and other features for APPROVe study:}
On the VIGOR dataset, we identified all possible medical terms (PREFTERM field in the adverse event file) that were relevant and possibly associated with GI events during the treatment period. To be consistent with our procedure on VIGOR dataset, we excluded pre-treatment events, and included events that occurred during the treatment and post-study periods. A confirmed CVT event was a designated end point of the study, so these labels were directly provided to us in the study's data files. Such a process, i.e., using only a relevant list of medical terms in the adverse event file, correctly identified 166 out of 177 patients with GI events. Despite our best efforts, this procedure also falsely identified 12 out of the remaining 7899 patients who did not have a confirmed GI event. 
Next, we found that 33 patients in APPROVe had recorded adverse events with PREFTERM contained in the list of terms identified above (with some adjustment of different spellings) during the treatment or the post-study periods. We declared these 33 patients to have had a GI event. Because the APPROVe study did not aim to study GI toxicity, the paper on the study by Baron et al~\cite{baron2008cardiovascular} does not report any information about the GI event as well as the risk factor features that we discuss next.

We followed a similar strategy to develop a mapping using the medical terms from the medical history file to the risk factor indicators for \{ASPFDA, ASCGRP, HYPGRP\} for the VIGOR study. Doing so, we correctly identified (i) 320/321 patients with ASPFDA = 1 (indication of aspirin usage by FDA due to their medical history), (ii) 453/454 patients with ASCGRP = 1 (history of atherosclerosis), and (iii) all 2385/2385 patients with HYPGRP = 1 (history of hypertension). For all three features (\{ASPFD, ASCGRP, HYPGRP\}), we did not have any false inclusion, i.e., using just the selected list of medical terms did not incorrectly impute a value of $1$ for any patient. Finally to identify the patients with prior usage of glucortocoids (PSTRDS=1), we developed a mapping between the information from the concomitant therapy file and the PSTRDS indicator from the risk factor file. Our mapping correctly identified all 4479/4479 patients with PSTRDS = 1 but also falsely identified an additional 248 (out of the remaining 3597) patients.

The mappings described above were then used to impute the GI outcome and relevant missing features in the APPROVe study, thereby allowing us to report the ``transfer" results for the subgroups found by StaDISC on the VIGOR study (\cref{tab:test_set_GI,tab:test_set_CVT}) to the APPROVe study (\cref{tab:approve}).

\FloatBarrier
\bibliographystyle{abbrv}
\bibliography{vioxx}

\end{document}